\documentclass{ar-1col-S2O}
\usepackage[comma]{natbib}
\usepackage{url}
\usepackage[bottom=1.6in]{geometry}

\usepackage{aas_macros}
\usepackage{amsmath,amssymb, graphics}
\usepackage{bm}
\usepackage{graphicx}
\usepackage{subfigure}
\usepackage{epstopdf}
\usepackage{float}
\usepackage{mathrsfs}

\newcommand{\bTb}{{\bm T}_{\rm b}}

\newcommand{\bl}{\bm {\hat l}}
\newcommand{\blb}{\bm {\hat l}_{\rm b}}
\newcommand{\beb}{\bm {e}_{\rm b}}

\newcommand{\eb}{e_{\rm b}}

\newcommand{\der}{\text{d}}
\newcommand{\br}{{\bf r}}

\jname{Annu. Rev. Astron. Astrophys.}
\jvol{AA}
\jyear{2022}
\doi{10.1146/((please add article doi))}

\def\gtrsim{\mathrel{\raise.3ex\hbox{$>$}\mkern-14mu
             \lower0.6ex\hbox{$\sim$}}}
\def\lesssim{\mathrel{\raise.3ex\hbox{$<$}\mkern-14mu
             \lower0.6ex\hbox{$\sim$}}}

\begin{document}

\markboth{D. Lai \& D.J. Mu\~noz}{Circumbinary Accretion}

\title{Circumbinary Accretion: From Binary Stars to Massive Binary Black Holes}
\author{Dong Lai$^1$ and Diego J. Mu\~noz$^{2,3,4}$
\affil{$^1$Department of Astronomy, Center for Astrophysics and Planetary Science, Cornell University, Ithaca, New York, USA; email: dong@astro.cornell.edu}
\affil{$^2$Center for Interdisciplinary Exploration and Research in Astrophysics, Department of Physics \& Astronomy, Northwestern University, Evanston, Illinois, USA}
\affil{$^3$Facultad de Ingenier\'ia y Ciencias, Universidad Adolfo Ib\'a\~nez, Pe\~nalol\'en, Santiago, Chile}
\affil{$^4$Millennium Institute for Astrophysics, Chile}
}


\begin{abstract}
We review recent works on 
the dynamics of 
circumbinary accretion, including 
time variability, angular momentum transfer between the disk and the binary, and the secular evolution of accreting binaries.
These dynamics can 
impact
stellar binary formation/evolution, circumbinary planet formation/migration, 
and the evolution of (super)massive black-hole binaries. We discuss the
dynamics and evolution of inclined/warped circumbinary disks 
and connect with recent observations of protoplanetary disks. 
A special kind of 
circumbinary accretion involves binaries embedded in ``big'' disks,
which may contribute to the mergers of stellar-mass black holes in AGN disks. Highlights include:

\vspace{0.1in}
\begin{minipage}[l]{0.65\textwidth}
\begin{itemize}
\item[{\scriptsize$\blacksquare$}] Circumbinary accretion
is highly variable, being modulated at $P_{\rm b}$ (the binary period) or $\sim 5P_{\rm b}$, depending on the binary eccentricity $e_{\rm b}$ and mass ratio $q_{\rm b}$.
\item[{\scriptsize$\blacksquare$}] The inner region of the circumbinary disk can develop coherent eccentric structure, which may modulate the accretion and affect the physical processes (e.g. planet migration) taking place in the disk.

\item[{\scriptsize$\blacksquare$}] Over long timescales, circumbinary accretion steers binaries toward equal masses, and it does not always lead to binary orbital decay, as is commonly assumed.
The secular orbital evolution depends on the binary parameters
($e_{\rm b}$ and $q_{\rm b}$), and on the thermodynamic properties of the accreting gas.
\item[{\scriptsize$\blacksquare$}] A misaligned disk around a low-eccentricity binary tends to evolve toward coplanarity due to viscous dissipation. But when $e_{\rm b}$ is significant, the disk can evolve toward ``polar alignment'', with the disk plane perpendicular to the binary plane.
\end{itemize}
\end{minipage}
\end{abstract}

\maketitle

\vspace{-0.1in}
\tableofcontents

\section{INTRODUCTION}
\label{sec:1}

Circumbinary disk (CBD) accretion plays an important role in the
evolution of many types of astrophysical systems, ranging from young binary
stars, main-sequence and post-main-sequence binaries to supermassive binary black holes.
Figure 1 illustrates the basic features of such accretion: Gas from large distances
gradually spirals toward the binary in a CBD driven by
viscous dissipation; the disk is truncated at a few binary separations by the varying gravitational force from the binary (the ``egg beater''), forming a
cavity; the gas at the inner edge falls inward through accretion streams
towards individual stars (or black holes), forming circum-single disks (CSDs, or ``mini-disks'') and eventually accreting onto each stars.

Circumbinary accretion has long been suggested to exist around binary
massive black holes (MBHs) following galaxy mergers 
\citep[e.g.,][]{Begelman:1980,Milosavljevic:2001,Escala:2005,Milosavljevic:2005,Dotti:2007,Cuadra:2009,Chapon:2013}.
This has been demonstrated in many numerical simulations over the years. An example can be found in
\citet{Mayer:2007}: two galaxies (each containing a MBH), initially separated at $\sim 100$~kpc, collide with each other, and eventually end up with two
MBHs separated by $\sim 10$~pc and surrounded by an extended ($\sim 100$~pc) disk/torus at the center of the merged galaxy.

One of the key questions concerning circumbinary accretion is: Does the binary lose
or gain angular momentum and how does the binary orbit evolve?
The first discussion of this issue appeared in \citet{Begelman:1980}:
\begin{quote}
  $\cdots$ infall of gas onto the binary can also lead to some orbital evolution. Gas may be flung out of the system, acquiring energy (and
  angular momentum) at the expense of the binary; alternatively, gas
  may accrete onto the larger hole, causing orbital contraction as the
  product $Mr$ is adiabatically invariant. In either case, the evolution time scale is
$$  t_{\rm gas}\sim 10^8M_8\left(\dot M/1~M_\odot\,{\rm yr}^{-1}\right)^{-1}\,\,{\rm yr}$$
\end{quote}
As we discuss in this review (see Section~\ref{sec:3}), this issue has been controversial and the prevailing view
has been challenged by recent studies. The evolution of SMBH binaries undergoing
gas accretion may directly impact the low-frequency gravitational wave signatures
probed by space interferometers such as LISA \citep{AmaroSeoane:2017}
and Pulsar Timing Arrays \citep{Burke-Spolaor:2019} (see Section~\ref{sec:mbhb}).

\begin{figure}[h]
\includegraphics[scale=0.31]{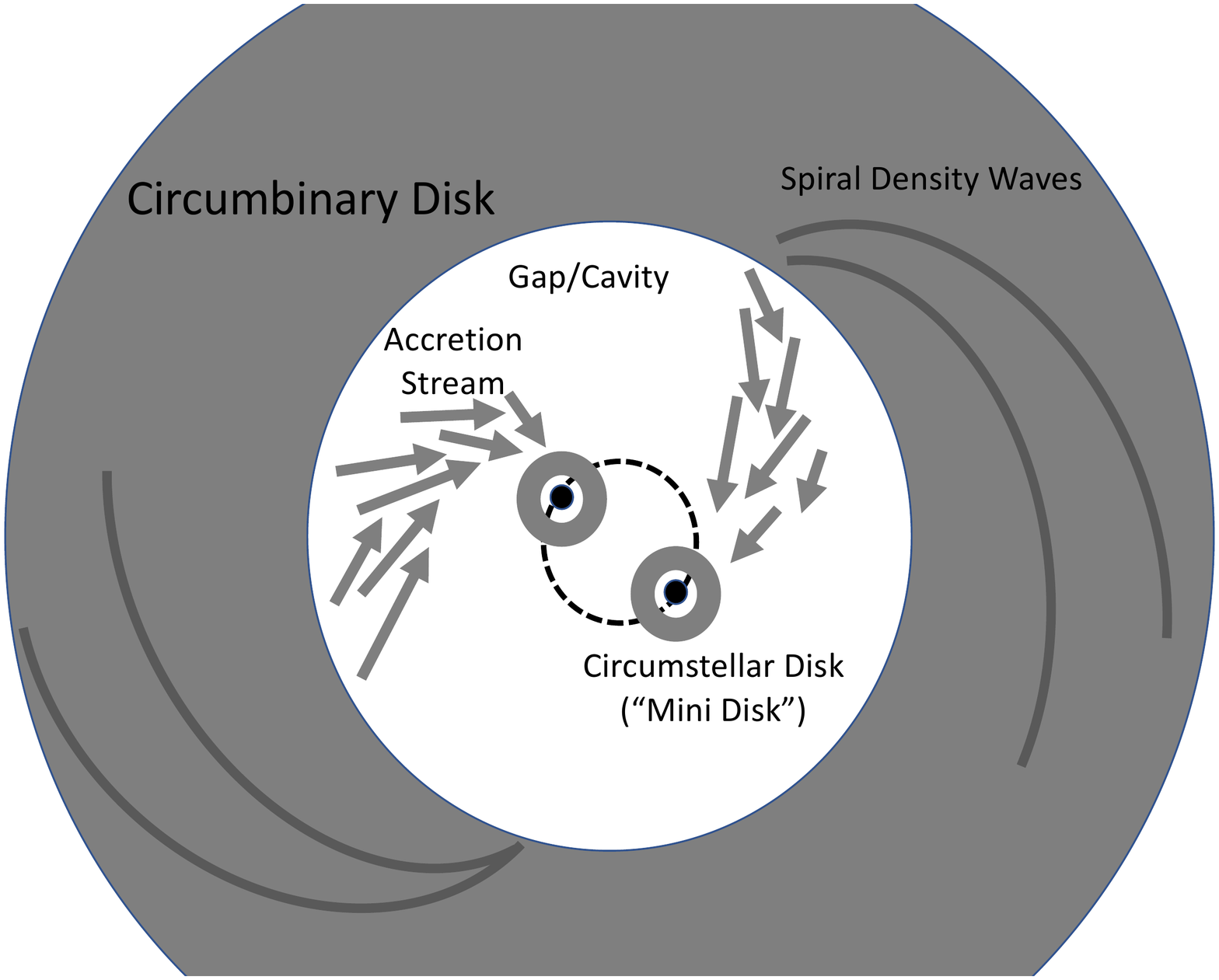}                                        
\caption{An illustration of the basic features of circumbinary accretion.
Figure credit: Ryan Miranda (2017).}
\label{fig1}
\end{figure}

\begin{marginnote}
\entry{CBD}{circumbinary disk, the disk around (and exterior to) the binary} 
\entry{CSD}{circum-single disk (or circum-stellar disk, mini-disk), the disk around each binary component}
\entry{MBH}{massive black hole (BH); SMBH: supermassive BH}
\end{marginnote}

Circumbinary accretion disks are a natural byproduct of binary star
formation via disk fragmentation \citep[e.g.,][]{Bonnell:1994,Kratter:2008,Offner:2022}.
 A number of these disks have been
observed around Class I/II young stellar binaries -- well-known examples include GG Tau, DQ Tau, and UZ Tau E \citep[e.g.,][]{Dutrey:1994,Mathieu:1997,Phuong:2020} -- and recently even around much younger Class 0 objects like L1448 IRS3B \citep{Tobin:2016}
and IRAS 16293-2422 A \citep{Maureira:2020}.
With ALMA, many CBDs have been discovered through direct imaging
\citep[e.g.,][]{Czekala:2021}.
Recent observations have revealed that the disk can be highly misaligned with the central binary \citep{Kennedy:2019} -- an issue we will address in
Section~\ref{sec:5}.  The observed properties of ``mature'' stellar binaries (such as the mass ratio distribution) may have been shaped by circumbinary accretion at the earlier (proto-stellar) phase (see Section~\ref{sec:stellar_binaries}).

Circumbinary disks have been also found around many 
post-AGN binaries \citep[e.g.,][]{vanWinckel:2018}. These "second-generation" disks, likely formed as a result of binary interactions during the AGB phase, can impact the evolution of the systems (see Section~\ref{sec:post-MS}).

Starting from NASA's Kepler mission, planets have been found around
stellar binaries using the transit method \citep[e.g.,][]{Doyle:2011}.
So far, more than a dozen of such systems are known \citep[see][and references therein]{Welsh:2018,Kostov:2020}.
An interesting features of these circumbinary systems is that many of
the planets are found very close (within a factor of 1.5) to the
stability limit, i.e., if the semi-major axis of the planet is a bit smaller (e.g. by $10\%$ in the case of Kepler-16b), the planet would be ejected from the system. These planets are unlikely to have formed in-situ, but must have migrated from far away in the protoplanetary disk. The dynamics of CBDs can strongly influence the formation and migration of the planets (see Section~\ref{sec:circumbinary_planets}).

Recently, a special type of circumbinary accretion has gained interest, in connection with the gravitational wave sources detected by LIGO/VIRGO. It has been suggested that merging stellar black-hole binaries can be produced in AGN disks \citep[e.g.,][]{Bartos:2017,Stone:2017,McKernan:2018,Tagawa:2020}.
The hydrodynamical flows generated by binaries embedded in AGN disks have several distinct features compared to normal circumbinary accretion disks (see Section~\ref{sec:6}).

\section{THEORY AND SIMULATION OF CIRCUMBINARY DISK ACCRETION: BASIC CONCEPTS AND OVERVIEW}\label{sec:2}

We first review some of the key theoretical concepts related to circumbinary accretion. The analytic aspect of binary-disk interaction through gravitational forcing 
is relatively straightforward, at least in the linear regime
(i.e., when the perturbation of the disk by the binary is weak).
The gravitational potential
produced by the binary (with total mass $M_{\rm b}$, semi-major axis $a_{\rm b}$ and eccentricity $e_{\rm b}$) on the disk 
(assumed aligned with the binary plane) at the position $\br=(r,\phi)$ 
(measured from the binary's center of mass) can be written as 
\citep{Goldreich:1980}
\begin{equation}
\Phi(\br,t)    =\sum_{m=0}^{\infty}\sum_{n=-\infty}^{\infty}\Phi_{mn}(r)\cos\left[m\phi-(m\Omega_{\rm b}+n\kappa_{\rm b})t\right],
\end{equation}
where $\Omega_{\rm b}=(GM_{\rm b}/a_{\rm b}^3)^{1/2}$ is the mean angular frequency (or ``mean motion'') of the
binary and $\kappa_{\rm b}$ is the radial epicyclic frequency.
For binaries in Keplerian orbits, $\kappa_{\rm b}=\Omega_{\rm b}$.
The potential component $\Phi_{mn}$ depends on $e_{\rm b}$ and $a_{\rm b}/r$; to the leading order in $e_{\rm b}$, we have $\Phi_{mn}\sim e_{\rm b}^{|n|}\Phi_{mm}$. For $m>0$, the $(mn)$-potential rotates with the pattern frequency 
\begin{equation}
\omega_{mn}
={m\Omega_{\rm b}+n\kappa_{\rm b}\over m}
={N\Omega_{\rm b}\over m},
\label{eq:omega_mn}\end{equation}
(where $N=m+n$ and the second equality assumes $\kappa_{\rm b}=\Omega_{\rm b}$), and excites spiral density waves at the Lindblad resonances (LRs), where 
\begin{equation}
\omega_{mn}-\Omega(r)=\pm {\kappa(r)\over m},
\end{equation}
with the upper (lower) sign corresponding to the outer (inner) LR.
\begin{marginnote}
\entry{$\Omega_{\rm b}$}{$=(GM_{\rm b}/a_{\rm b}^3)^{1/2}$,
mean angular frequency (mean motion) of the binary; $P_{\rm b}=2\pi/\Omega_{\rm b}$ is the binary orbital period}
\entry{$\Omega(r)$}{$(GM_{\rm b}/r^3)^{1/2}$, angular frequency of the CBD}
\entry{LR}{Lindblad resonance; OLR: outer LR}    
\end{marginnote}  
As the CBD is approximately Keplerian, $\kappa\simeq \Omega\simeq (GM_{\rm b}/r^3)^{1/2}$, the LRs are located at 
\begin{equation}
\frac{\Omega\left(r_{\rm LR}\right)}{\Omega_\mathrm{b}}\simeq \frac{N}{m \pm 1}\quad{\rm or}\quad {r_{\rm LR}\over a_{\rm b}}\simeq \left({m\pm 1\over N}\right)^{2/3}.
\end{equation}
The torque on the disk at a LR is 
\citep{Goldreich:1979}
\begin{equation}
\label{eq:lr_torque}
T_{mn}^{\mathrm{LR}} = -m\pi^2 \left[\Sigma \left(\frac{\mathrm{d}D}{\mathrm{d}\ln r}\right)^{-1}\lvert\Psi_{mn} \rvert^2\right]_{r_\mathrm{LR}},
\end{equation}
where $\Sigma$ is the disk surface density, $D = \kappa^2 - m^2\left(\Omega-\omega_{mn}\right)^2$, and
\begin{equation}
\Psi_{mn} = \frac{\mathrm{d} \Phi_{mn}}{\mathrm{d}\ln r} + \frac{2 \Omega}{\Omega - \omega_{mn}} \Phi_{mn}.
\end{equation}
At the outer LRs (which are most relevant for CBDs), 
$\left(\mathrm{d}D/\mathrm{d}\ln r\right)= - {3N^2}\Omega_{\rm b}^2/(m+1)$, we find 
$T_{mn}^{\mathrm{OLR}}>0$, i.e., the disk particles gain angular momentum from the binary through resonant gravitational torques.
This is a general result: A rotating potential always transports angular momentum from higher to lower angular velocity \citep{Lynden-Bell:1972,Goldreich:2003}. The associated energy transfer rate to the disk through the LR is given by 
\begin{equation}
{\der E_{\rm d}\over \der t}=
\omega_{mn}\,{\der J_{\rm d}\over \der t}
=\omega_{mn}\, T_{mn}^{\rm LR}.
\end{equation}
These expressions are useful for determining how gravitational binary-disk interaction affects the binary orbit (see Section \ref{sec:post-MS}).

On the other hand, the disk particles lose angular momentum through viscous torque. 
Assuming the $\alpha$-ansatz for the kinematic viscosity coefficient, $\nu = \alpha c_\mathrm{s}^2/\Omega$ (where $c_s$ is the disk sound speed), the viscous torque is given by \citep[e.g.,][]{Pringle:1981}
\begin{equation}
T_\nu = 3\pi \nu\Sigma \Omega r^2 =3\pi \alpha h^2\Sigma \Omega^2 r^4,
\end{equation}
where $h = H/r$ is the disk aspect ratio.
A gap is opened at the $(mn)$-LR if $T_{mn} \geq T_\nu$
\citep{Artymowicz:1994}.  In this picture, 
the radius $r_{\rm cav}$ of the inner cavity of a CBD
is determined by largest radius at which a gap can
be cleared. 

As an example, for a circular binary, the dominant potential has $m=2,\,n=0$, with $\Phi_{mn}\simeq -3G\mu_ba_{\rm b}^2/(4r^3)$ (assuming $r\gg a_{\rm b}$, where $\mu_{\rm b}$ is the reduced mass of the binary).
The OLR is located at $r_{\rm LR}\simeq (3/2)^{2/3}a_{\rm b}$, and the LR torque is $T_{20}^{\rm LR}\simeq (49\pi^2/2)\Sigma (\Phi_{20}/\Omega_{\rm b})^2$.  Gap opening at the OLR requires $T_{20}^{\rm LR}\gtrsim T_\nu$, i.e.
\begin{equation}
\alpha h^2\lesssim 0.14 \left({4\mu_{\rm b}\over M_{\rm b}}\right)^2,
\end{equation}
a condition easily satisfied for binaries with comparable component masses. For a small but finite $e_{\rm b}$, gap is cleared by the $m=2$, $n=-1$ potential (with $\Phi_{mn}\propto e_{\rm b}$), and the corresponding OLR is at 
$r_{\rm LR}\simeq 3^{2/3}a_{\rm b}$.

\citet{Miranda:2015} considered eccentric
binaries with general mass ratios and binary-disk inclination angles. For typical disk $c_s$ and viscosity parameter, 
the inner radius $r_{\rm cav}$ of the CBD is found to be (2-3)$a_{\rm b}$ and depends on $e_{\rm b}$ in a discrete manner. 
Misaligned disks generally have smaller inner radii than aligned disks. In any case, 
it is important to recognize that such theoretical calculation of the inner disk radius has obvious limitations, as real disks are expected to 
have a fuzzy and dynamical inner boundary, with gas streaming into the cavity. Numerical simulations are needed to capture the whole complexities of binary-disk interactions, especially near the inner truncation radius.

More subtle resonant binary-disk interactions can also play a role in the dynamics of CBDs, such as 
the parametric instability associated with 
Lindblad resonances (see Section~\ref{sec:disk_eccentricity})
that can excite disk eccentricity \citep{Hirose:1990,Lubow:1991a,Lubow:1991b}.
In addition, for an eccentric binary, secular (orbital-averaged) interaction may affect the disk eccentricity evolution \citep[][see Section~\ref{sec:disk_eccentricity}]{Miranda:2017,Lubow:2022}

Because of the importance of circumbinary accretion in various
astrophysical contexts, many numerical simulations have been carried
out over the years. 
Some works were based on Smoothed Particle
Hydrodynamics (SPH)
\citep[e.g.,][]{Bate:1995,Artymowicz:1996,Escala:2005,Cuadra:2009,Roedig:2012,Pelupessy:2013,Ragusa:2016}.
Others used grid-based
Eulerian hydrodynamical methods: these include, among others, 
\citet{Gunther:2002}  (hybrid grid),
\citet{MacFadyen:2008} ({\footnotesize FLASH}, polar grid with inner cavity excised), \citet{Hanawa:2010} (nested cartesian grid),
\citet{Devalborro:2011} (cartesian grid),
\citet{DOrazio:2013} ({\footnotesize FLASH}, polar grid with inner cavity excised),
\citet{Lines:2015} ({\footnotesize FARGO}, polar grid with inner cavity excised),
\citet{Miranda:2017} ({\footnotesize PLUTO}, polar grid with inner cavity excised).
Most simulations solved
hydrodynamical equations with parameterized viscosity, but some
short-duration (essentially) Newtonian MHD simulations have also been carried out 
\citep{Noble:2012,Shi:2012,Shi:2015,Bowen:2019}
and even GRMHD simulations, in which a time-dependent, sewn-together metric \citep[e.g.,]{Matzner:1998} can be prescribed in order to  to take into account the Kerr metric in the vicinity of each black hole \citep{Combi:2021, Combi:2022}.

\begin{figure*}[t]
\includegraphics[width=\textwidth]{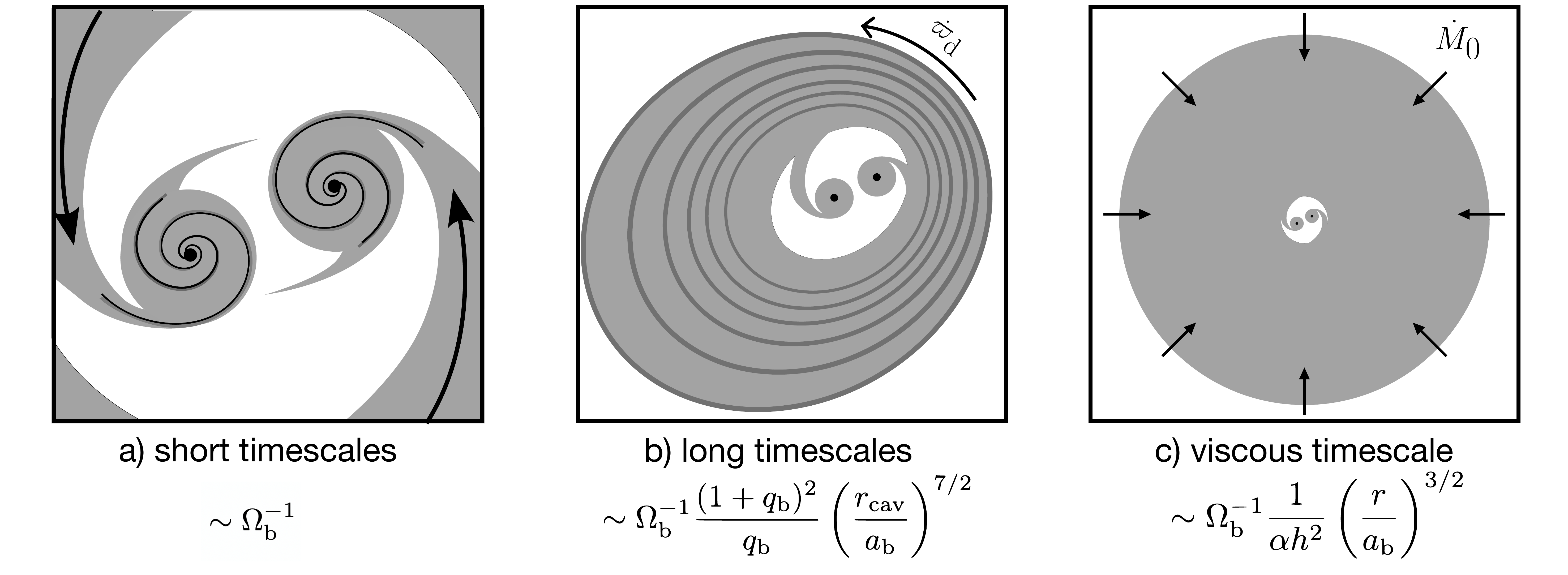}
\caption{Some relevant timescales involved in circumbinary accretion. a) Dynamical timescales: variability is measured in the accretion rate on timescales of order the binary orbital period. b) Secular timescales: variability is measured in the accretion rate on timescales of hundreds of binary orbital periods in tandem with the secular apsidal precession of the inner disk. c) Viscous timescales: for quasi-steady state to be reached, the disk must be viscously relaxed. Depending on the initial conditions, quasi-steady state can be reached on a few viscous times as measured at the cavity or tens of times longer if the initial condition is far from the steady state.
\label{fig:cbd_timescales}}
\end{figure*}

Numerical simulations of circumbinary accretion are challenging because
of the wide spatial range involved, and the multiples timescales on which variability takes place (Figure~\ref{fig:cbd_timescales}). The accreting gas flows
from a large, viscously-evolving disk surrounding the
binary, transitions into plunging accretion streams, which then feed the circum-single disks
around each individual binary
component, which accretes mass at spatial scales typically much smaller than the binary separation. In addition, since the flow in the disk and near the
binary is highly dynamical, to determine the long-term effect of the
flow on the binary, sufficiently long simulations with careful
averaging must be carried out. 

In recent years, several finite-volume moving mesh codes have been
used to study circumbinary accretion. The first is {\footnotesize DISCO} that
ultilizes a moving ring grid
\citep{Farris:2014,Dorazio:2016,Tang:2017, Duffell:2020}.
 Another is {\footnotesize AREPO}
\citep{Munoz:2016,Munoz:2019,Munoz:2020a}.
The general-purpose grid-based Godunov code {\footnotesize ATHENA++} has also been used
for long-term simulations of CBDs 
\citep{Moody:2019,Wang:2022a,Wang:2022b}.

\section{SIMULATIONS OF CIRCUMBINARY ACCRETION: KEY RESULTS}\label{sec:3}

In this section, we summarize the key findings
from recent simulations. Our discussion will be guided by our own simulation results
using {\footnotesize AREPO} \citep{Munoz:2016,Munoz:2019,Munoz:2020a,Siwek:2022}
and related work using {\footnotesize PLUTO} \citep{Miranda:2017}, but we 
will compare with the results from other simulations when relevant, and discuss more recent 
progress.
{\footnotesize AREPO} \citep{Springel:2010,Pakmor:2016} is a quasi-Lagrangian Godunov-type moving-mesh code. It has
unstructured moving grid with adaptive resolution, and with hydrodynamical equations
solved in the moving frame. AREPO was been adapted for viscous accretion disk
simulations by
\citet{Munoz:2014,Munoz:2015a}.
In our works on circumbinary accretion, we
simulate 2D Newtonian viscous flow, with a locally isothermal equation of
state $P=\Sigma c_s^2$, where the sound speed $c_s(r)$ is a prescribed
function of $r$ -- we assume the disk 
has a constant disk aspect ratio $h\equiv H/r$. We use the
Shakura–Sunyaev $\alpha$ prescription for the viscosity $\nu(r)$.
Our typical simulations resolve accretion onto individual binary components down to $0.02a_{\rm b}$.

In the following, we first consider binaries with mass ratio $q_{\rm b}\sim 1$
and disks with $H/r\sim 0.1$ and viscosity parameter $\alpha=0.05-0.1$.
We examine both ``infinite'' disks (with a fixed mass supply rate at $r_{\rm out}\gg a_{\rm b}$ as
well as finite disks.
Later in this section, we discuss how the results vary for different parameters as well as various complications.

\begin{figure*}[h]
\centering
\vskip -0.5truecm  
\includegraphics[width=1.0\textwidth]{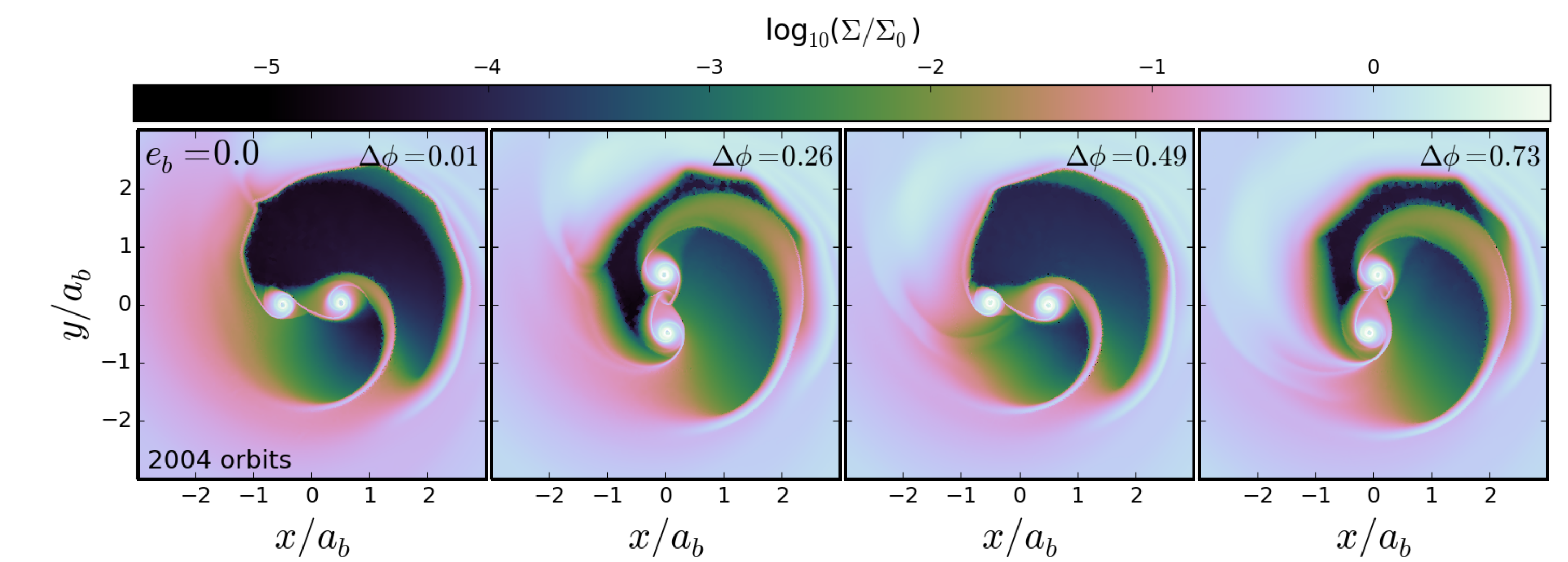}%
\includegraphics[width=0.8\textwidth]{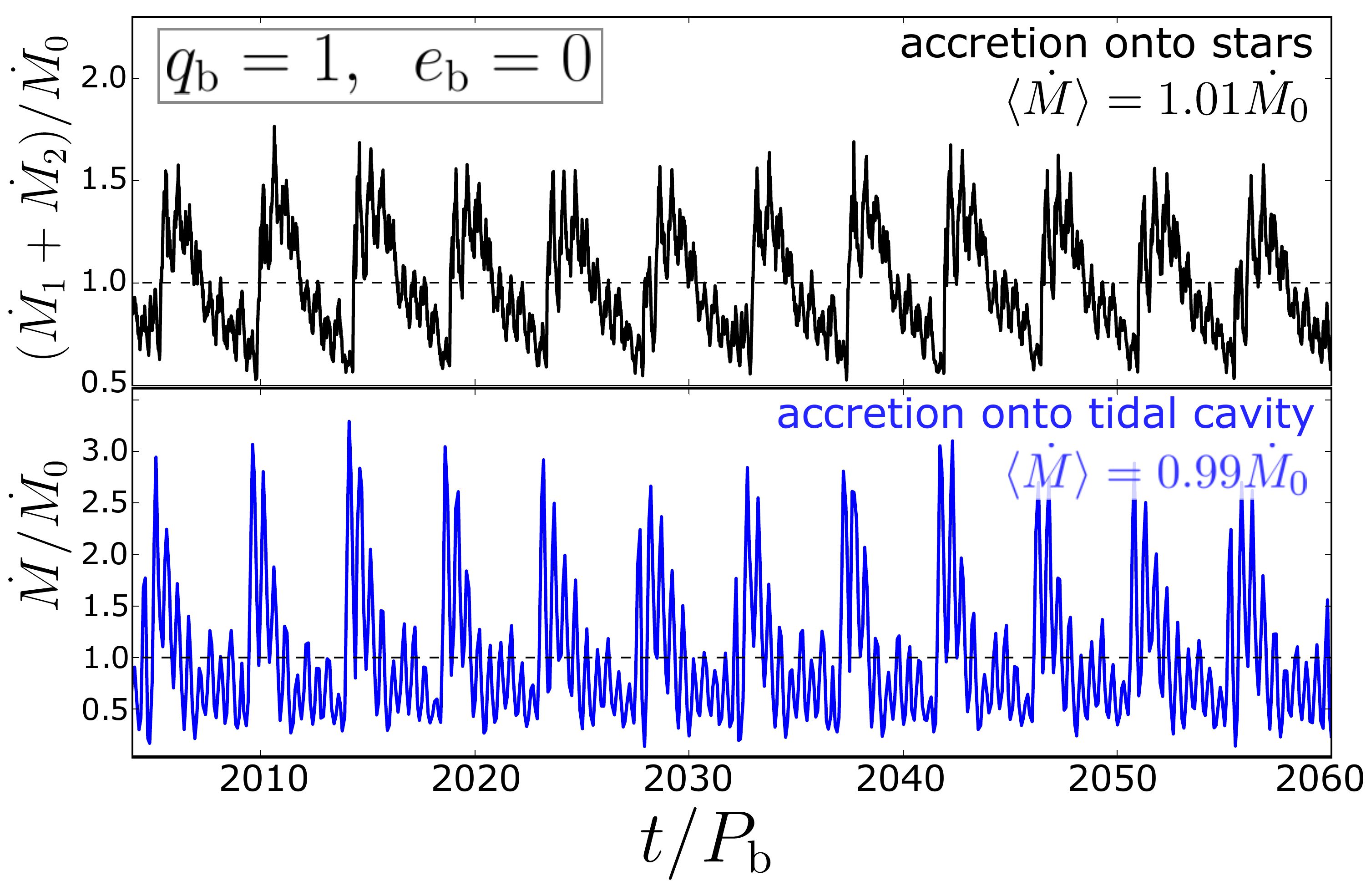}%
\caption{Short-term variability of circumbinary accretion 
for an equal-mass ($q_{\rm b}=1$) circular binary ($e_{\rm b}=0$) with $h=0.1,\,\alpha=0.1$. Upper panel: Surface density field evolution within timescales of about one binary orbit at
$2004 + \Delta\phi$ orbits. In this case, the pattern repeats every half orbit.
Lower two panels: Accretion rate onto the binary and at the inner edge of the
CBD; both exhibit bursty behaviors with a dominant period of about $5P_{\rm b}$.
Adapted from \citet{Munoz:2016} \copyright AAS. Reproduced with permission.
\label{fig:circular_binary}}
\end{figure*}

\begin{figure*}[h]%
\centering
\includegraphics[width=1.0\textwidth]{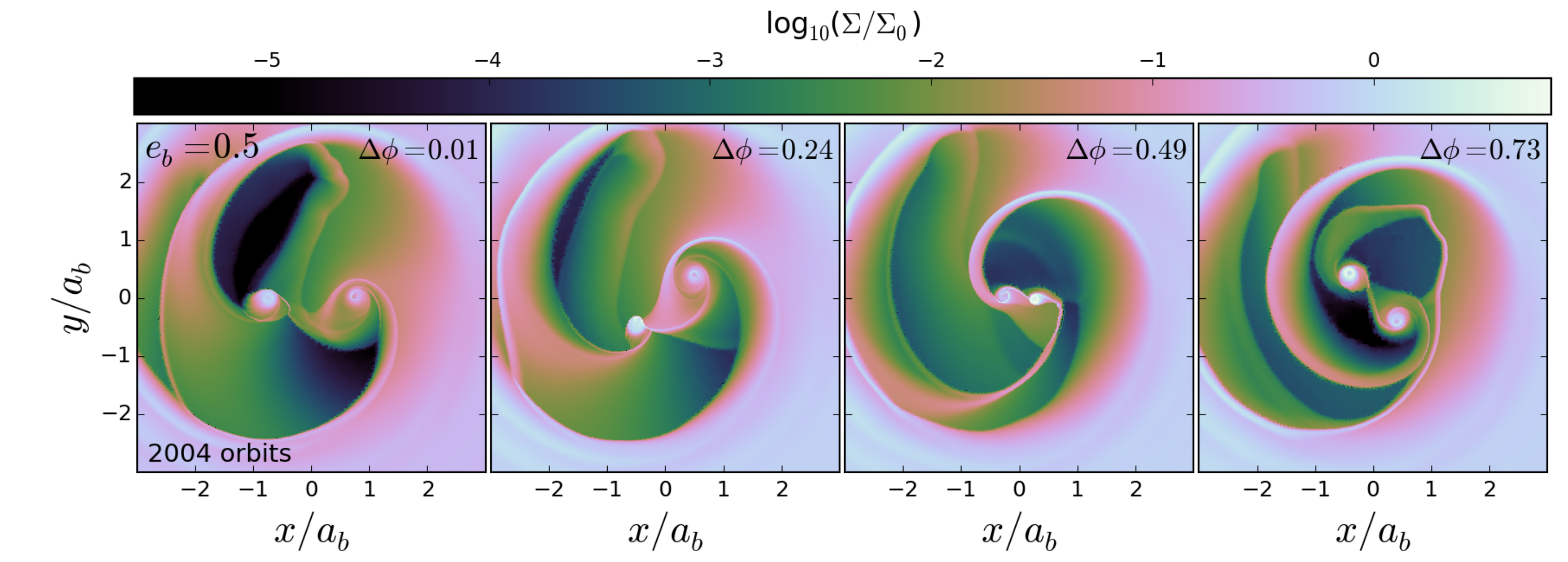}%
\includegraphics[width=0.8\textwidth]{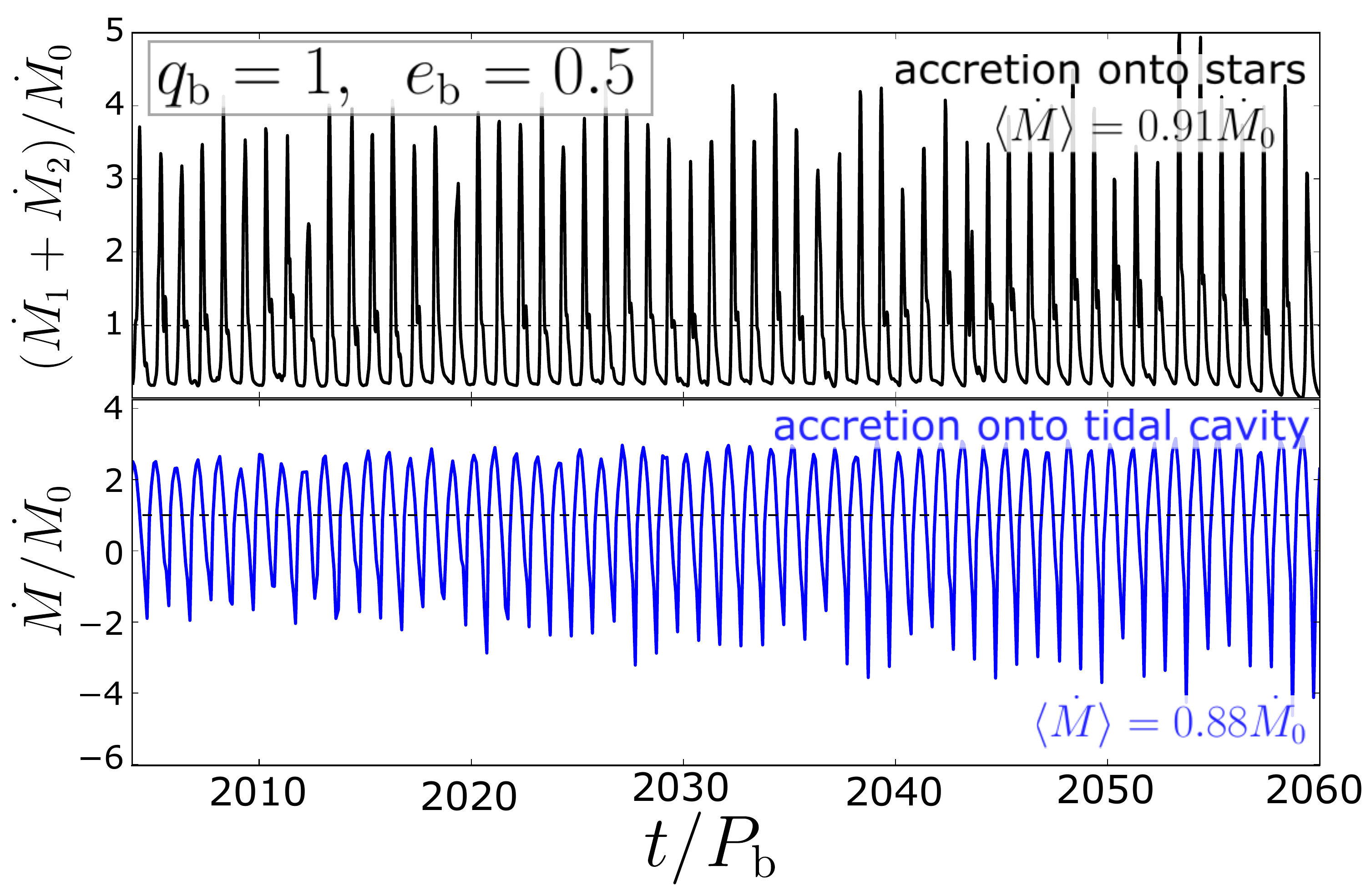}%
\caption{Same as Fig.~2, except for an eccentric binary with 
$e_{\rm b}=0.5$. The accretion rate exhibits pulsation with a dominant period of $P_{\rm b}$.
  Note that in the upper panel, there is an asymmetry in the density field. This 
  is related to the way gas is funneled into the central cavity. At a given time,
  an eccentric inner disk could favor one star over the other; but the accretion rate
  asymmetry is reversed after hundreds of binary orbits. Adapted from \citet{Munoz:2016}
   \copyright AAS. Reproduced with permission.
\label{fig:eccentric_binary}}
\end{figure*}

\subsection{Short-Term Accretion Variability}

For an extended disk, when the gas supply rate $\dot {M}_0$
at large distances is constant, the whole binary-disk
system can reach a quasi-steady state, in the sense that the time-averaged
accretion rate across the CBD and total mass accretion rate onto the binary,
$\langle\dot M_{\rm b}\rangle=\langle\dot M_1\rangle+\langle\dot M_2\rangle$,
are equal to $\dot {M}_0$.
However,  the flow rates onto the inner binary cavity and onto individual binary components
are highly variable on the binary orbital timescale \citep{Munoz:2016,Miranda:2017}:
\begin{itemize}
\item[{\scriptsize$\blacksquare$}]  For nearly circular binaries (with $e_{\rm b}\lesssim 0.05$, the accretion rates $\dot M_1$ and
$\dot M_2$ vary with a dominant period about $5P_{\rm b}$ (where $P_{\rm b}$ is the binary orbital period),
a result that was already found in earlier simulations 
\citep[e.g.,][]{MacFadyen:2008,DOrazio:2013,Shi:2012}. This dominant period
corresponds to the Kepler period of gas at the inner edge of the
disk, $r_{\rm cav}\simeq 3a_{\rm b}$. The reason for this is that for circular binaries,
the accretion onto the inner cavity arises from the development of $m=1$ lumps at the inner
disk edge and the lump has a pattern speed equal to the local Kepler velocity (see Figure~\ref{fig:circular_binary}).

\item[{\scriptsize$\blacksquare$}]  For $e_{\rm b}\gtrsim 0.05$, the dominant variability of $\dot M_1$ and $\dot M_2$
has a period of $P_{\rm b}$. The reason is that for eccentric binaries, the mass transfer onto the cavity
mainly occurs when the binary is at apocenter, where the binary component is closest to the disk
inner edge and ``grabs'' the gas from the disk and funnels it inwards (see Figure~\ref{fig:eccentric_binary}).
\end{itemize}

The different variability timescales ($P_{\rm b}$ vs $5P_{\rm b}$) are important when trying to 
infer the orbital period of binary massive black holes from the observed variability 
of dual AGNs (see Section 4).
These short-term variabilities can be also used to confront observations of
T Tauri binaries \citep[e.g.][]{Tofflemire:2017a,Tofflemire:2017b,Tofflemire:2019}.
In some cases, the effective sizes (which can be different from the stellar radii because
of the magnetic fields) 
of the individual accretors can be constrained.

\subsection{Longer-term Accretion Variability}

\begin{figure*}[h]%
\centering
\begin{minipage}[l]{0.495\textwidth}
\includegraphics[width=0.99\textwidth]{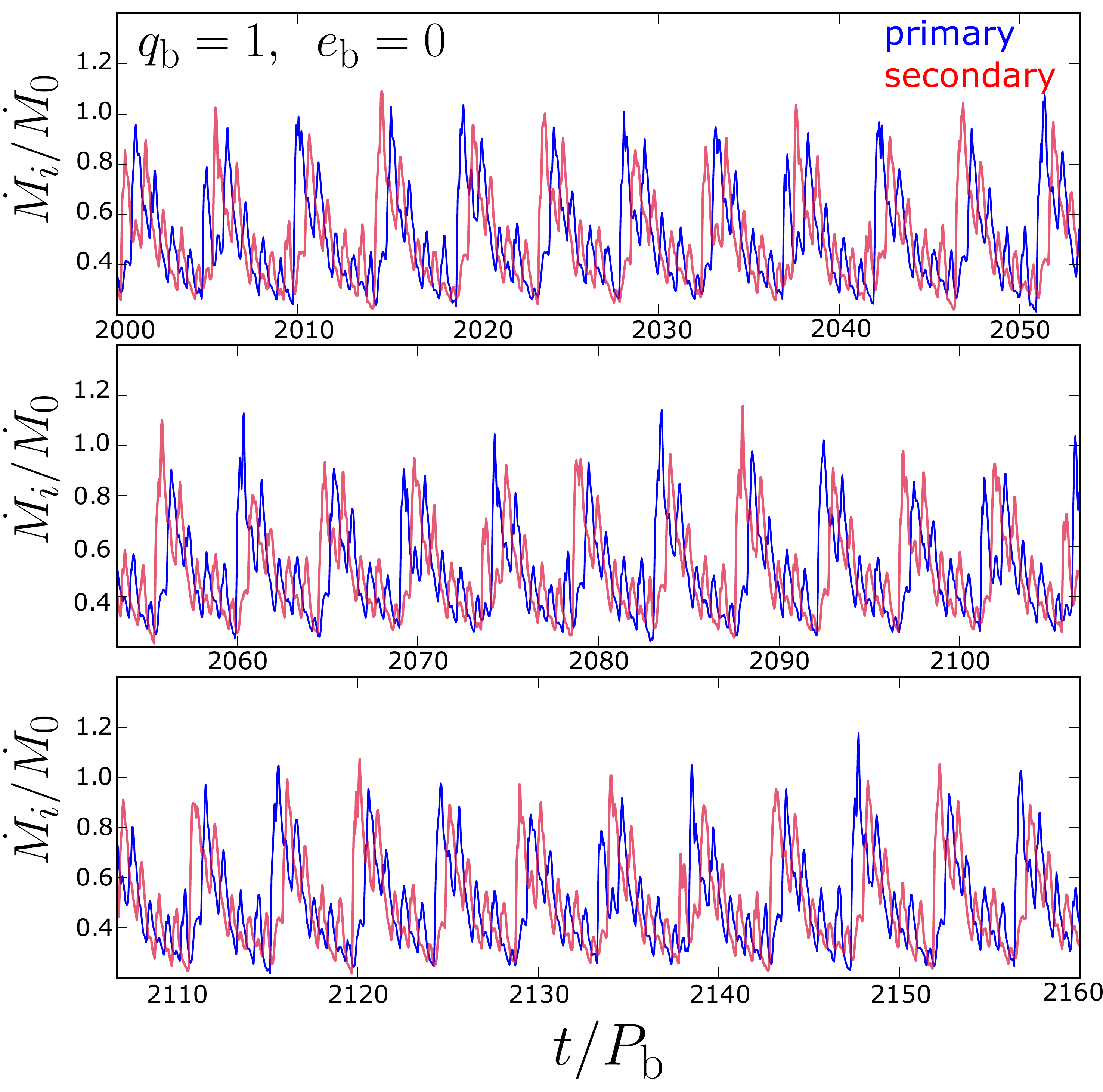}
\end{minipage}
\begin{minipage}[l]{0.495\textwidth}
\includegraphics[width=0.99\textwidth]{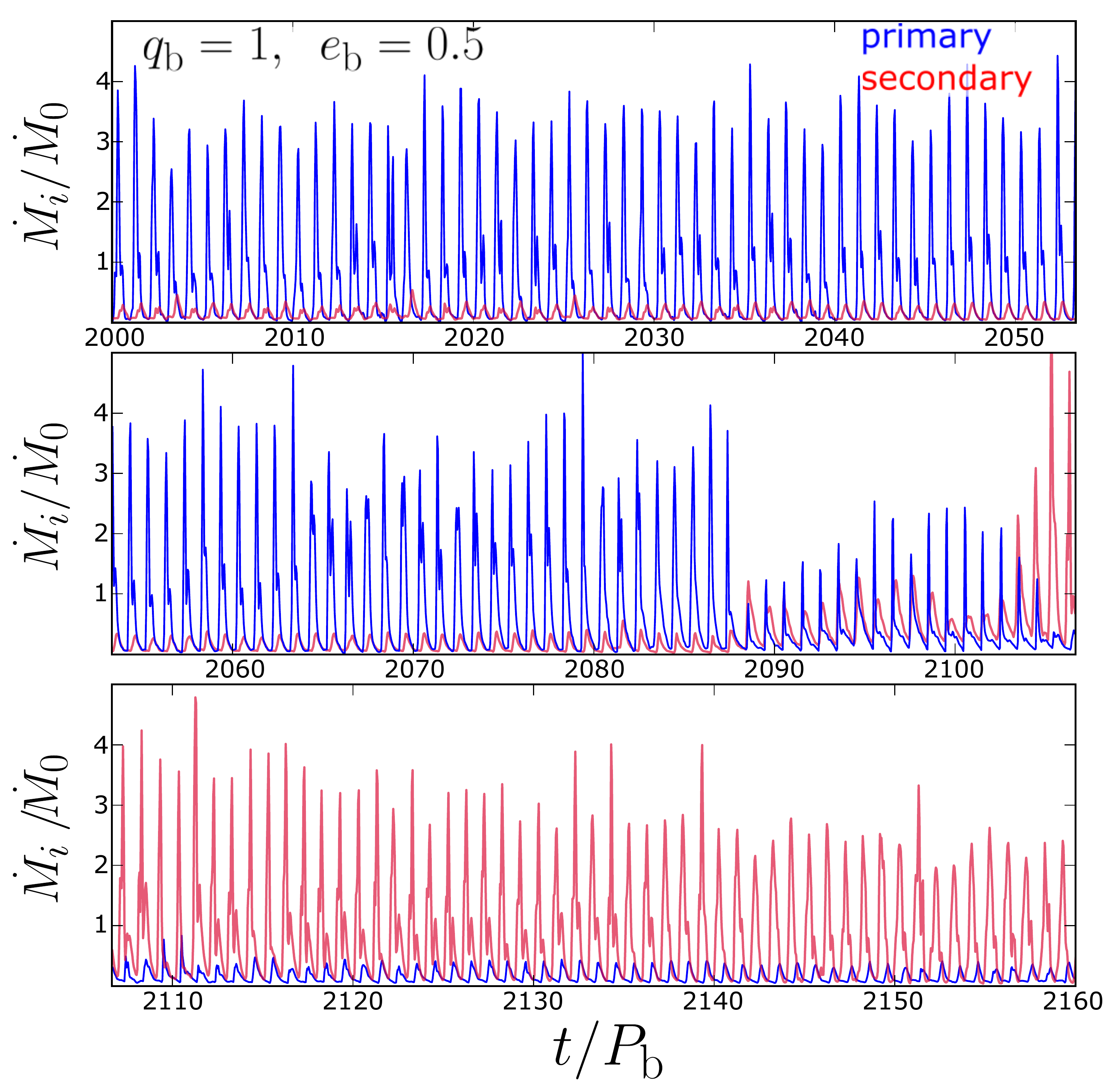}
\end{minipage}
\caption{
Accretion rates onto the primary and secondary members of the binary, $\dot M_1$ and $\dot M_2$ in
blue and red, respectively, for an equal-mass binary ($q_{\rm b}=1$). The  left panels show
the case of $e_{\rm b}=0$ and the right panels $e_{\rm b} = 0.5$.
The eccentric binary experiences a symmetry breaking, with one star accreting between
10 and 20 times more mass than its companion. This trend, however, is reversed at $t\sim 2110P_{\rm b}$
and then reversed back at $t\sim 2250P_{\rm b}$.
The individual accretion rates eventually average out to $\langle \dot{M}_1\rangle=\langle \dot{M}_2\rangle=0.5\dot{M}_0$.
Adapted from \citet{Munoz:2016} \copyright AAS. Reproduced with permission.
\label{fig:f4}}%
\end{figure*}

Binary accretion can vary on a timescale much longer the binary
orbital period (Figure~\ref{fig:cbd_timescales}), even when the mass supply rate $\dot {M}_0$ is
constant.  For an equal-mass circular binary, the accretion rates onto
individual stars are quite similar to each other, following the same
variable pattern in time, as illustrated in the left panels of Figure~\ref{fig:f4}. 
By contrast, for eccentric binaries, one of
the binary components can accrete at a rate 10-20 times larger than
its companion, even for mass ratio $q_{\rm b}=1$, as illustrated in the right panels
of Figure~\ref{fig:f4}. 

This ``symmetry breaking'' between the binary components, however, does not persist. Instead, it alternates over timescales of order $200P_{\rm b}$ (\citealp{Munoz:2016}; see also \citealp{Dunhill:2015}). 
This long-term quasi-periodicity can be attributed to a slowly precessing, eccentric
CBD. Indeed, the inner region of the disk generally develops eccentricity and 
precesses coherently (see below). The apsidal precession rate of an eccentric
ring (of radius $r$) around the binary 
is given by
\begin{equation}
  \dot\varpi_d\simeq {3\Omega_{\rm b}\over 4}{q_{\rm b}\over (1+q_{\rm b})^2}\left(1+{3\over 2}e_{\rm b}^2\right)
  \left({a_{\rm b}\over r}\right)^{7/2}.
\label{eq:dotvarpi}
\end{equation}
At $r\sim 3a_{\rm b}$ (the cavity radius), this corresponds to a period of 200-300$P_{\rm b}$.
Over longer timescales ($\gg 200P_{\rm b}$), the net accretion rates onto individual
binary components are the same (for $q_{\rm b}=1$).

This ``symmetry breaking'' behavior (even for $q_{\rm b}=1$) is interesting
in connection to AGNs around massive black hole binaries or
similar accretion phenomena around binary T Tauri stars.  It implies
that even when the two binary components are very similar in mass,
accretion may predominantly occur in one of components and last for
hundreds of orbits.
Interestingly, \citet{Tofflemire:2019} has reported
evidence for preferential accretion onto the primary of the 
 T Tauri Binary TWA 3A, in contradiction with expectations that accretion should be preferential onto the secondary (see Section~\ref{sec:unequal_masses} below). Having an eccentricity of $e_{\rm b}=0.67$ \citep{Tofflemire:2017b}, TWA 3A could be an example of this intriguing behavior.

\subsection{Disk Eccentricity and Precession}\label{sec:disk_eccentricity}

\begin{figure*}[h]
\centering
\includegraphics[width=\textwidth]{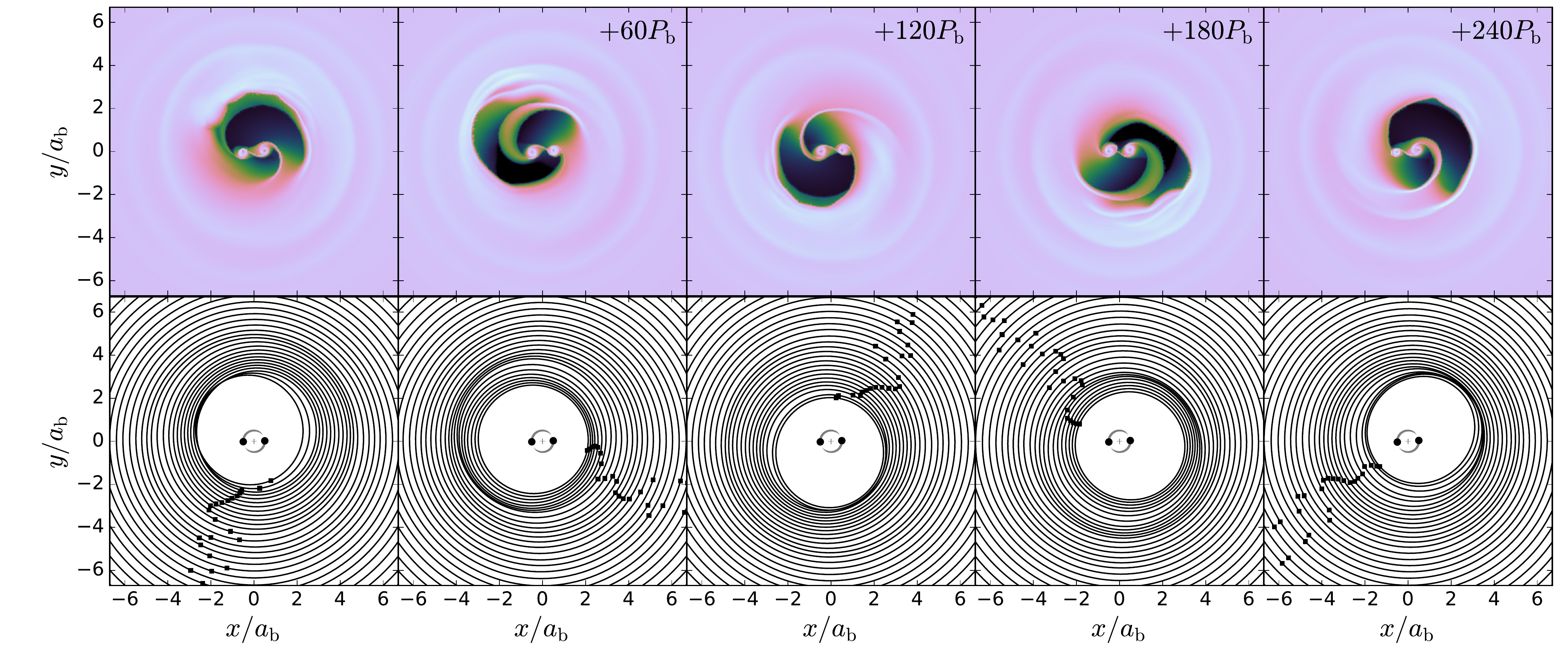}%
\caption{Evolution of CBDs over secular timescales. Top panels: surface density (logarithmic scale) in the vicinity
of a binary with $q_{\rm b}=1$, $e_{\rm b}=0$ (disk parameters are $h=\alpha=0.1$)
in intervals of 60 binary orbits
once steady-state has been achieved. The gas morphology is consistent throughout the panels, except for the orientation
of the central cavity, which evolves in tandem with the disk eccentricity.
Bottom panels: barycentric elliptical ``orbits'' corresponding to gas eccentricity   binned in semi-major axis.
These ``orbits'' change in time, exhibiting prograde apsidal precession, as evidenced by the advancement of the longitude of pericenter $\varpi_{\rm d}$,
 depicted by solid black squares; the orientation of the ellipses is roughly coherent ($\varpi_{\rm d}$ is approximately equal for all radii) out to
a distance of $\sim 10 a_{\rm b}$ from the barycenter.  Adapted from \citet{Munoz:2020b}  \copyright AAS. Reproduced with permission.
\label{fig:eccentric_disks}
}
\end{figure*}

Numerical simulations have long shown that
eccentricity can develop in the innermost region of the CBD
\citep{MacFadyen:2008,Miranda:2017}. While this is most noticeable by the formation of a lopsided tidal cavity \citep[e.g.][]{Kley:2006,Thun:2017,Ragusa:2020}, it responds to a phenomenon of much wider extent. Indeed, disk eccentricity $e_{\rm d}$ can be significant ($\gtrsim0.01$) out to radii of $\sim10-15a_{\rm b}$ \citep{Miranda:2017,Munoz:2020b}. 
Typically, this inner region undergoes coherent apsidal precession,
with the rate given by the appropriate spatial average of Eq.~(\ref{eq:dotvarpi}) (see below). 
But in some cases, it can also become apsidally locked relative to the binary's eccentricity vector, as found in the $0.2\lesssim e_{\rm b}\lesssim 0.4$ equal-mass binary
simulations of \citet{Miranda:2017} and in the unequal-mass eccentric binary 
simulations of \citet{Siwek:2022}.
An example of the coherent precession of a CBD and its correspondence to a lopsided cavity is illustrated in Figure~\ref{fig:eccentric_disks}.

The coherence of inner CBD precession requires efficient 
communication between different disk regions (recall that the apsidal
precession rate of a test particle around the binary depends strongly on $r$; see Eq.~\ref{eq:dotvarpi}). In the absence of disk self-gravity, this communication is achieved by gas pressure, which induces disk precession that balances the forced precession (Eq.~\ref{eq:dotvarpi})
\citep{Goodchild:2006, Teyssandier:2016}.
The eccentricity profile of the CBD actually corresponds to the eigenmode of a Sch\"rodinger-like eccentricity equation of \citet{Goodchild:2006} \citep[see also][]{Shi:2012,Teyssandier:2016,Lee:2019a,Lee:2019b,Munoz:2020b} and the global disk precession frequency is the associated eigenfrequency. The mode is trapped between two turning points that result from the sharply truncated CBD; as a result, most of the mode ``lives'' in the immediate vicinity of the cavity, thus obeying the local precession law (with the mode frequency approximately equal to 
$\dot{\varpi}_{\rm d}$ evaluated at the cavity radius) \citep{Munoz:2020b}. Interestingly, the mode is so strongly confined, that pressure-induced precession plays a minor role.

While the eccentricity profile and precession frequency can be understood via a linear analysis, the actual amplitude of the eccentricity eigenmode depends on the details of eccentricity growth, damping, and saturation. The growth of eccentricity likely results from a combination of the
direct ``hydraulic pumping'' from high Mach number streamers colliding directly with the inner edge of the disk as argued by \citet{Shi:2012} and the
tidal excitation via eccentric Lindblad resonances (ELRs)
\citep[e.g.,][]{Hirose:1990,Lubow:1991a, Lubow:1991b}.
ELRs is a parametric instability arising from the modulation of the disk particle's
epicyclic motion by the binary potential.
As discussed in Section 2, 
the gravitational potential from the binary on the disk can be
decomposed into many harmonic components, each having a pattern rotation frequency
$\omega_{\rm p}=\omega_{mn}=N\Omega_{\rm b}/m$, where $m, N=1,2,\cdots$ (and $n=N-m$). In the presence of this rotating potential, the epicyclic frequency $\kappa$
of the a disk fluid element attains a modulation term, proportional to
$\cos m(\omega_{\rm p}-\Omega)t$ (where $\Omega$ is the angular frequency of the disk). A parametric resonance occurs when $m(\omega_{\rm p}-\Omega)\simeq 2\kappa\simeq 2\Omega$, i.e., 
\begin{equation}
{\Omega(r_{\rm ELR})\over\Omega_{\rm b}}={N\over m+2}.
\end{equation}
The strength of the forcing depends on $e_{\rm b}$, with the components $m=2, N=1$ (which has $\Phi\propto e_{\rm b}$), and $m=2, N=2$
[with $\Phi\propto (1-5e_{\rm b}^2/2)$]
being the most important.

\begin{textbox}
\section{Eccentric Lindblad/Corotation Resonances and Mean-Motion Resonances}
The term "ELRs" are often used to describe Lindblad resonances that are associated eccentrcities. ELRs and "eccentric corotation resonances" are related 
to mean-motion resonances in celestial mechanics 
\citep{Murray:1999}
as follows. Consider the perturbation with a pattern rotation frequency $\omega_{mn}=(m\Omega_{\rm b}+n\kappa_{\rm b})/m$ (see Eq.~\ref{eq:omega_mn}).
A general Lindblad/corotation resonance occurs when
\begin{equation}
m(\omega_{mn}-\Omega)=n'\kappa,    
\end{equation}
where $n'=0$ gives the CR, and $n'=\pm 1$ gives the (linear) LRs, and $|n'|>1$ would involve nonlinear resonances. Using $\kappa_{\rm b}=\Omega_{\rm b}-\dot\varpi_{\rm b}$ and $\kappa=\Omega-\dot\varpi$, where $\dot\varpi_{\rm b}$ and $\dot\varpi$ are the apsidal precession rates of the binary and the disk, respectively, the above resonance condition becomes
\begin{equation}
(m+n)\Omega_{\rm b}-n\dot\varpi_{\rm b}-(m+n')\Omega+n'\dot\varpi=0.    
\end{equation}
This describes the $(m+n):(m+n')$ mean-motion resonance, and the strength of the interaction potential is $\propto e_b^{|n|} e^{|n'|}$.
\end{textbox}

For a binary with finite eccentricity, the secular (orbital-averaged) interaction can also drive the disk eccentricity. If we define the complex eccentricity
${\cal E}_{\rm d}(r,t)=e_{\rm d}(r,t)\exp[i\varpi_{\rm d}(r,t)]$ for the disk
particle (at radius $r$) and ${\cal E}_{\rm b}=e_{\rm b}\exp(i\varpi_{\rm b})$
for the binary, then the secular contribution to the time evolution of ${\cal E}_d$ is given by 
\begin{equation}
\left({\der{\cal E}_d\over \der t}\right)_{\rm sec}=i\omega_{\rm db}
{\cal E}_d -i\nu_{\rm db}{\cal E}_{\rm b}.
\end{equation}
The apsidal precession rate $\omega_{\rm db}=\dot\varpi_d$ 
(see Eq.~\ref{eq:dotvarpi}) is driven primarily by the quadrupole potential of the binary, while the eccentricity forcing rate $\nu_{\rm db}$ by the octupole potential and is given by \citep[e.g.,][]{Moriwaki:2004,Miranda:2017}
\begin{equation}
\nu_{\rm db}\simeq {15\,\Omega_{\rm b}\over 16}{q_{\rm b}(1-q_{\rm b})\over (1+q_{\rm b})^3}\left(1+{3\over 4}e_{\rm b}^2\right)
\left({a_{\rm b}\over r}\right)^{9/2}.
\label{eq:nudb}
\end{equation}
In the presence of an eccentricty-damping force (and neglecting 
other hydrodynamical effects), this secular forcing tends to drive
the disk toward a ``forced'' eccentricity, given by
${\cal E}_d=(\nu_{\rm db}/\omega_{\rm db}){\cal E}_{\rm b}$; this forced eccentricity is apsidally aligned with the binary eccentricity.

Thus, the eccentrcity evolution of the CBD around an eccentric binary can be quite complex, driven by the secular and resonant forcings from the binary and hydrodynamical effects.
Simulations show that the disk generally exhibits apsidal precession, but the precession rate may not be constant and the line of apses can vary with $r$. In analogy to test particle orbit, the CBD should evolve according to coexisting ``free'' and ``forced'' modes \citep[e.g., see][]{Lubow:2022}.  In some cases, the CBD can lock onto the binary and stop precessing altogether \citep{Miranda:2017,Siwek:2022}.

\subsection{Angular Momentum Transfer and Orbital Evolution: Circular Binaries}\label{sec:angular_momentum_transfer}

We now discuss the important problem of long-term angular momentum exchange between the disk and the binary. Without accretion, the binary always loses angular momentum to the CBD through gravitational (Lindblad) torque. With accretion, the net torque becomes uncertain.
As discussed above, the mass flow rate across the disk, $\dot M(r,t)$, and the mass accretion rate
onto the binary $\dot M_{\rm b}(t)=\dot M_1(t)+\dot M_2(t)$, are all highly variable. For an extended disk with a constant supply rate $\dot M_0$ at $r=r_{\rm out}\gg a_{\rm b}$,
a quasi-steady state is eventually reached, where the time-averaged mass flow rate
$\langle \dot M(r,t)\rangle=\langle \dot M_{\rm b}(t)\rangle=\dot M_0$. In general, the net torque on the binary can be obtained in two ways:
\begin{itemize}
\item[{\scriptsize$\blacksquare$}]  First, it can be computed directly as the sum of the gravitational torque from all the gas
plus the accretion torque (due to momentum of the accreting gas onto each binary component), i.e.
\begin{equation}\label{eq:angular_momentum_change_1}
  \langle\dot J_{\rm b} \rangle=  \langle \dot L_{\rm b}\rangle_{\rm grav}+\langle \dot L_{\rm b}\rangle_{\rm acc}+
\langle \dot S_1\rangle_{\rm acc}+
\langle \dot S_2\rangle_{\rm acc}.
\end{equation}
When the size of the binary component is much less than $a_{\rm b}$, the spin torques
$\langle\dot S_1\rangle$ and $\langle\dot S_2\rangle$ are negligible, and the total torque $\langle\dot J_{\rm b}\rangle$
acts on the binary orbit, i.e. $\langle\dot J_{\rm b} \rangle\simeq            \langle \dot L_{\rm b}\rangle=\langle \dot L_{\rm b}\rangle_{\rm grav}+\langle \dot L_{\rm b}\rangle_{\rm acc}$.

\item[{\scriptsize$\blacksquare$}]  Second, the angular momentum flow rate in the circumbinary disk (at radius $r$ in the disk)
  can be computed as \citep{Miranda:2017}
\begin{equation}
\dot J(r,t)=\dot J_{\rm d,adv}-\dot J_{\rm d,visc}-\dot J_{\rm d,grav},
\end{equation}
where $\dot J_{\rm d,adv}$ is the inward angular momentum advection rate,
$\dot J_{\rm d,visc}$ is the outward viscous angular momentum transfer rate,
and $\dot J_{\rm d,grav}$
is the torque from the binary acting on the gas exterior to radius $r$.
\end{itemize}
When the disk reaches a quasi-steady state, the time average $\langle \dot J(r,t)\rangle$ is
independent of $r$, and the circumbinary disk has two global constants: $\langle \dot M(r,t)\rangle                
=\dot M_{\rm b}=\dot M_0$ and $\langle \dot J(r,t)\rangle =\langle\dot J_{\rm b}\rangle$.

\begin{figure*}[h]%
\centering
\includegraphics[width=0.8\textwidth]{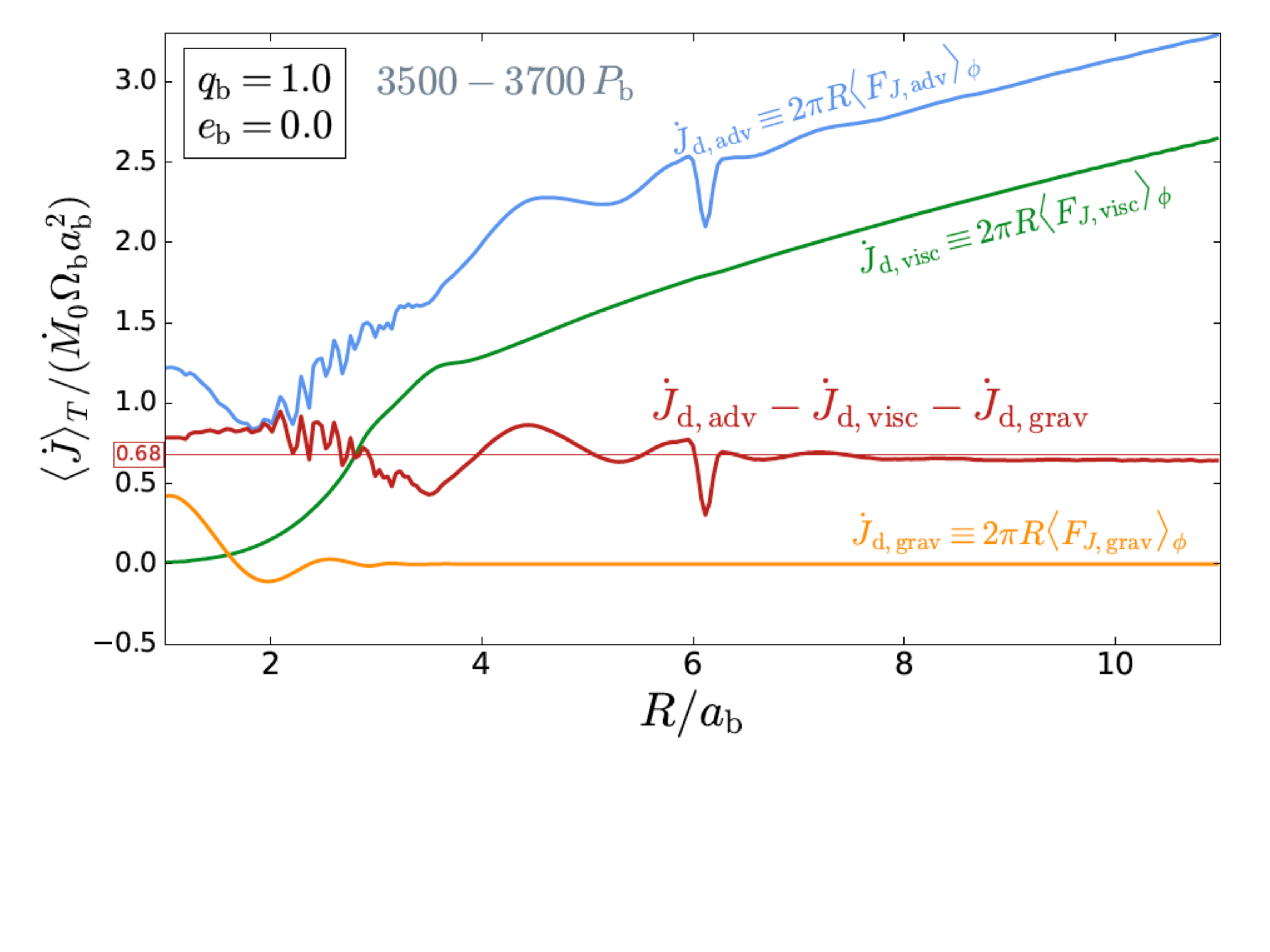}%
\vskip -1.3cm
\caption{Time-averaged angular momentum flow rate due to advection,
viscosity and gravitational torques in the CBD, for binary parameters
$q_{\rm b}=1$ and $e_{\rm b} = 0$. The total angular momentum flow rate across the disk
(red curve) is approximately constant, indicating
quasi-steady-state. For reference, the net torque 
on the binary $\langle \dot J_{\rm b}\rangle  = 0.68\dot M_0\Omega_{\rm b} a_{\rm b}^2$ is overlaid as a
straight red line. The ``blip'' at $r = 6a_{\rm b}$ and fluctuations at
$r\sim 2.5a_{\rm b}$ are artifacts of the mapping from the Voronoi cells (used in the AREPO code) onto a
regular polar grid.
From \citet{Munoz:2019}  \copyright AAS. Reproduced with permission.
}
\label{fig:angular_momentum_change}%
\end{figure*}

Using long-term AREPO simulations, \citet{Munoz:2019} demonstrated that the global quasi-steady
state can be achieved for circumbinary accretion, and $\langle\dot J_{\rm b} \rangle$ computed using both
methods are in agreement (see Figure~\ref{fig:angular_momentum_change}).  For equal-mass circular binaries (and with $h=0.1,\,\alpha=0.1$), the specific angular momentum
``eigenvalue'' (i.e., the angular momentum transferred to the binary per unit accreted mass) is
\begin{equation}
  l_0\equiv {\langle\dot J_{\rm b} \rangle\over \langle\dot M_{\rm b} \rangle}=0.68a_{\rm b}^2\Omega_{\rm b},\qquad
  (q_{\rm b}=1,\,\,e_{\rm b}=0)
\end{equation}
where $\Omega_{\rm b}=(GM_{\rm b}/a_{\rm b}^3)^{1/2}$ is the rotation rate of the binary.
The result was confirmed independently by \citet{Moody:2019} using the {\footnotesize Athena} code.
[A similar positive value of $l_0$ was first obtained by \citet{Miranda:2017} based on
PLUTO simulations with an excised cavity.]

For a binary accreting from a finite-sized disk/torus, a global
quasi-steady state obviously does not exist. Instead, 
the accretion proceeds in two phases:
an initial transient phase, corresponding to the filling of
the binary cavity, followed by a viscous pseudo-stationary phase
(achieved after the viscous time $r^2/\nu$ at a few  disk radii),
during which the torus viscously spreads and
accretes onto the binary. 
In the transient phase, the torque on the binary is negative since it is entirely gravitational (see Section~\ref{sec:2}).
\citet{Munoz:2020a} demonstrated that in the viscous phase, the net torque on the binary per unit accreted mass is close to $l_0$, the value derived for ``infinite'' disks.  
Because no global steady-state is required, this method
allows for a more efficient computation of the net torque on the binary
in non-steady situations (i.e., when $\langle\dot M_{\rm b}\rangle$ gradually evolves in time).

Using angular momentum conservation, $\der J_{\rm b}/\der t=
\langle\dot M_{\rm b}\rangle  l_0$ with 
$J_{\rm b}=\mu_{\rm b} a_{\rm b}^2\Omega_{\rm b}$, we have, for a $q_{\rm b}=1$ binary,
\begin{equation}\label{eq:migration_rate_circular}
{\dot a_{\rm b}\over a_{\rm b}}=  8\left({l_0\over a_{\rm b}^2\Omega_{\rm b}}-{3\over 8}\right){\langle\dot M_{\rm b}\rangle\over M_{\rm b}}.
\end{equation}  
Thus the binary orbit expands at the rate
${\dot a_{\rm b}/a_{\rm b}}\simeq  2.44\, {\langle \dot M_{\rm b}\rangle/M_{\rm b}}$. Note if we take account of the small amount of
spin torque $\dot S_{\rm b}=\dot S_1+\dot S_2$, the actual $\dot L_{\rm b}$ is a bit smaller than $\dot J_{\rm b}$,
and the resulting $\dot a_{\rm b}/a_{\rm b}$ is then smaller [\citet{Munoz:2019} found $\dot S_{\rm b}\simeq 0.028\dot M_{\rm b}
  a_{\rm b}^2\Omega_{\rm b}$ when assuming a ``stellar'' radius of $0.02a_{\rm b}$, and thus $\dot a_{\rm b}/a_{\rm b}\simeq
  2.2\dot M_{\rm b}/M_{\rm b}$.]

\paragraph{
Lessons from Numerical Studies of Long-term Binary Evolution.}
Until recently, it has always been thought  
that circumbinary accretion leads to binary orbital decay. \cite{Pringle:1991} suggested that the strong gravitational torque from the binary on the CBD prevents accretion, and therefore the binary loses angular momentum to the disk (see Section \ref{sec:2}). Gas accretion changes the picture completely. Only a few previous studies addressed the issue of angular momentum transfer in a quantitative way. Examples include \citet{MacFadyen:2008} and \citet{Shi:2012}, whose 
simulations excise the inner cavity. \citet{MacFadyen:2008}
considered $H/r = 0.1$ and a disk viscosity with $\alpha = 0.01$, and adopted a polar grid in the domain between $r_{\rm in} = a_{\rm b}$ and $r_{\rm out} = 100a_{\rm b}$.
They found a reduction of mass accretion onto the binary and the
dominance of the gravitational torque relative to advective torque
(therefore a negative net torque on the binary). However, with their
small $\alpha$ parameter the ``viscous relaxation'' radius at $t = 4000P_{\rm b}$ (the typical duration of
their runs) is only about $3a_{\rm b}$, and their surface density profile is
far from steady state even at $r\sim r_{\rm in}$. So it is likely
that the result of \citet{MacFadyen:2008} reflects a ``transient'' phase of their simulations. \citet{Shi:2012} obtained a positive value 
of $\dot J_{\rm b}$ in their 3D MHD simulations of CBDs (truncated at
$r_{\rm in} =0.8a_{\rm b}$). However, the duration of their simulations is only $\sim 100P_{\rm b}$,
and it is unlikely that a quasi-steady state is reached. Their value
of $l_0$, which is too small to cause orbital expansion, may not properly characterize the long-term evolution of the binary.
\citet{Tang:2017} carried out simulations of
accretion onto circular binaries using {\footnotesize DISCO} 
\citep{Duffell:2012,Duffell:2016}
with $h = 0.1$ and $\alpha = 0.1$. 
For the accretion presciption, they assumed that inside a
``sink'' radius (measured from each ``star''), the gas is depleted at a rate $d\Sigma/dt =-\Sigma/t_{\rm sink}$, with $t_{\rm sink}$ a free parameter. They claimed that the net torque on the binary is negative unless $t_{\rm sink}$ is much less than $P_{\rm b}$. This result is in
contradiction with \citet{Munoz:2019} and \citet{Moody:2019},
the latter adopted a similar accretion prescription and did not find
the same behavior as \citet{Tang:2017}. 
More recent works using {\footnotesize DISCO} \citep{Duffell:2020,Dittmann:2021,Dorazio:2021} and the cartesian-grid code {\footnotesize MARA3} \citep{Tiede:2020,Zrake:2021}
 have produced similar results as \citet{Munoz:2019,Munoz:2020a} and \citet{Moody:2019},
so it appears that different groups have reached
agreement (at least for disks with $h=0.1,\,\alpha=0.1$). See below for discussion on the numerical resolution requirement to obtain reliable net torques.

\paragraph{
Dependence on Disk Thickness and Viscosity.}
 Most of the recent simulations described thus far in this 
 review are two-dimensional, low-mass, viscous accretion disks with locally isothermal equation of state. In this case, the most important parameters are: $q_{\rm b}$, $e_{\rm b}$, $\alpha$ (or $\nu$) and $h$. Of these, $q_{\rm b}$ and $e_{\rm b}$ have been studied most extensively, but $h$ and $\alpha$ can be equally important.
 
\begin{figure*}[t]%
\centering
\begin{minipage}[l]{0.49\textwidth}
\includegraphics[width=\textwidth]{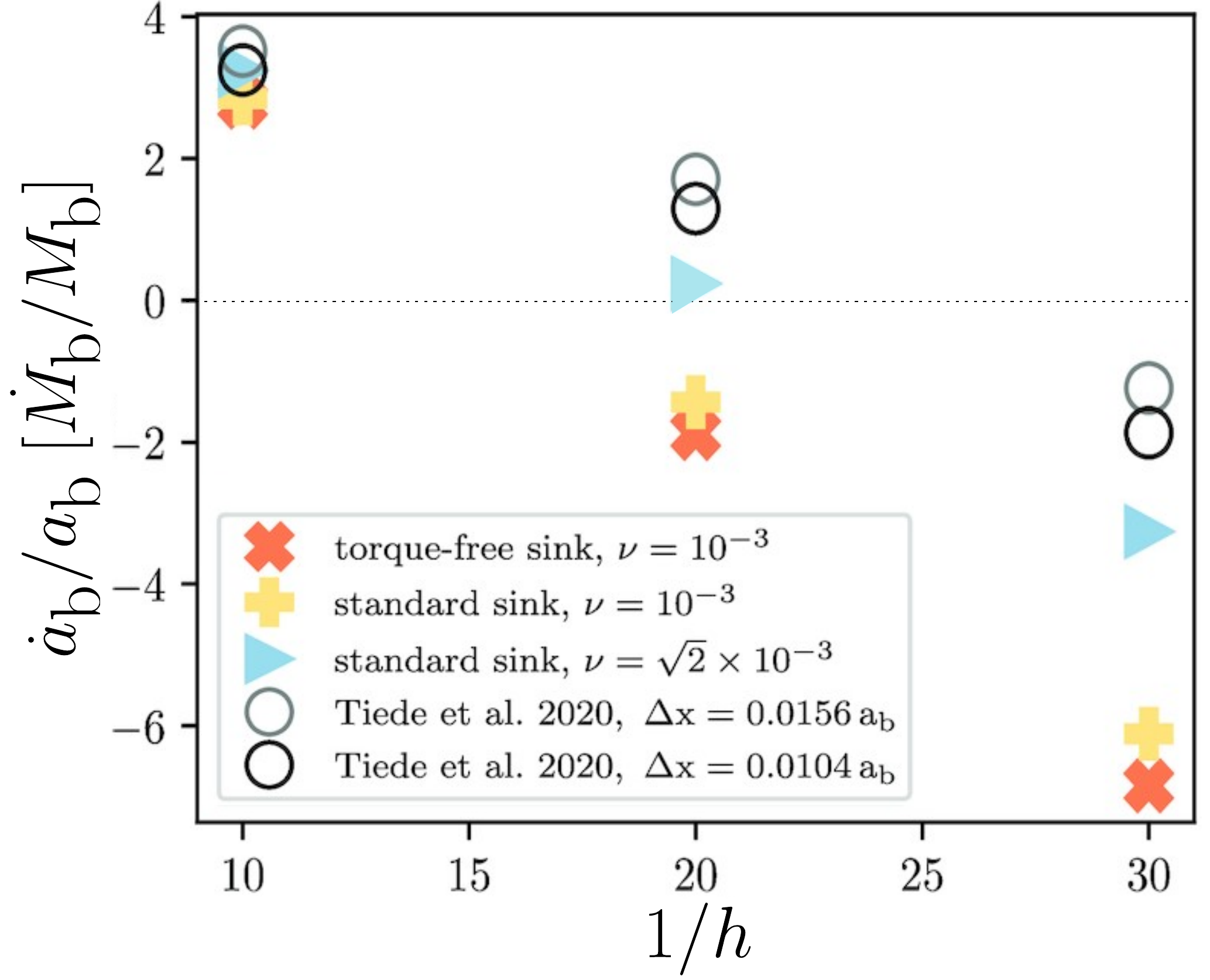}
\end{minipage}
\begin{minipage}[l]{0.49\textwidth}
\includegraphics[width=\textwidth]{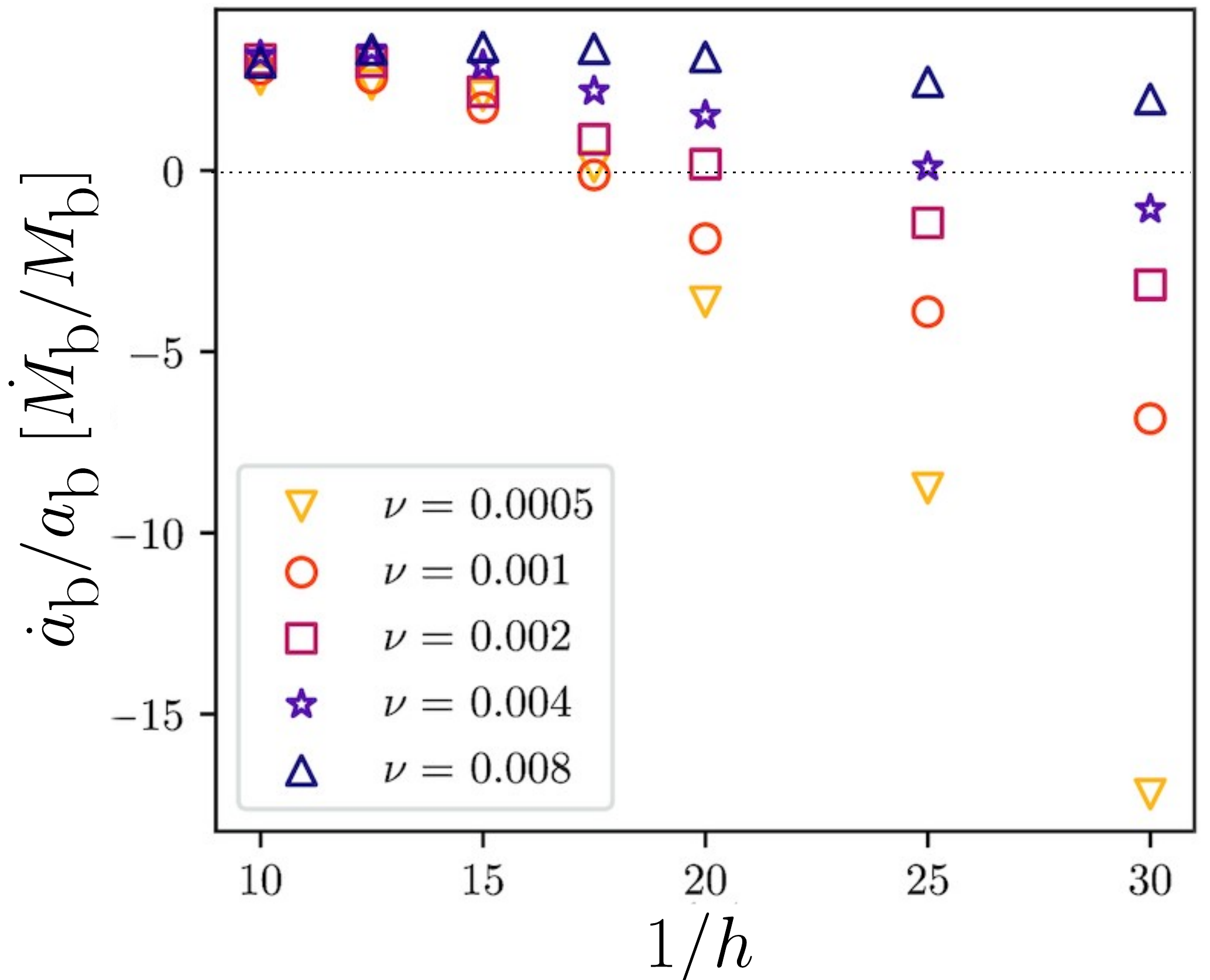}
\end{minipage}
\caption{Migration rate $\dot{a}_{\rm b}/a_{\rm b}$
in units of $\dot{M}_{\rm b}/M_{\rm b}$
as a function of disk aspect ratio $h$ and kinematic viscosity coefficient $\nu$. Left panel: numerical effects like insufficient resolution and sink prescription can affect the value of $l_0$, with some results reported by \citet{Tiede:2020} being non-converged, but the transition from outward to inward migration at 
$h\simeq0.04$ appears to be robust. Right panel:
the direction of migration  depends on $h$ and on the (globally constant) value of $\nu$, with the small-$\nu$ systems migrating inward more rapidly than the more viscous ones. This difference grows as $h$ is decreased.
Adapted from \citet{Dittmann:2022}.
\label{fig:disk_thickness}}%
\end{figure*}

An exploration of the dependence of $l_0$ and $\dot{a}_{\rm b}/a_{\rm b}$ on $h$
for equal-mass, circular binaries was carried out by \citet{Tiede:2020}.
Using the Godunov code {\footnotesize Mara3} (implemented in in Cartesian coordinates with static mesh refinement), they varied $h=\{0.02,0.033,0.05,0.1\}$, while fixing 
the kinematic viscosity $\nu=\sqrt{2}\times10^{-2}a_{\rm b}^2\Omega_{\rm b}$ (or $\alpha=10^{-2} h^{-2}$ at $r=2a_{\rm b}$). They found that
$l_0$ monotonically decreases with decreasing $h$ and that $l_0$ fall below $(3/8)a_{\rm b}^2\Omega_{\rm b}$
(see Equation~\ref{eq:migration_rate_circular}) 
when $h\lesssim 0.04$. This occurs in spite of the CBD mean cavity remaining roughly of the same size (since they use constant $\nu$ instead of constant $\alpha$, the truncation estimate of \citealp{Miranda:2015} would yield identical cavity sizes), albeit the streamers become more erratic, narrow, and denser than in thicker disks. The authors reported a dependence the results on resolution, which is likely due to the poorly resolved CSDs even with mesh refinement. \citet{Dittmann:2022} carried out follow-up work using a version of {\footnotesize DISCO}, and found that the \citet{Tiede:2020} results were likely unconverged but qualitatively confirmed the migration transition at $h=0.04$. \citet{Dittmann:2022} also explored the role of viscosity. While earlier works \citep{Munoz:2020a,Duffell:2020} found that the $h=0.1$ simulations exhibited little dependence on viscosity, when $h$ is varied, lower viscosity appears to further reduces $l_0$ with decreasing $h$ (see Figure~\ref{fig:disk_thickness}, right panel). The role played by disk thickness could introduce a crucial distinction between the young stellar binaries accreting from protostellar disks and massive BH binaries accreting from AGN disks. While the former systems can be well represented by disks with $h\sim 0.1$, the latter are better characterized by values of $h\sim10^{-2}$ or less \citep{Sirko:2003,Thompson:2005}.

Two difficulties must be addressed when discussing the roles of disk viscosity and aspect ratio. 
First, a low-viscosity disk simply takes longer to reach a quasi-steady state \citep{Miranda:2017,Munoz:2020a,Dittmann:2022}, and this is the likely the culprit of simulations reporting suppressed binary accretion at low viscosities \citep[e.g.,][]{Ragusa:2016}. 
To obtain reliable results, simulations often need an iterative reassessment of the initial condition \citep{Miranda:2017,Dempsey:2020a}. 
Second, the resolution requirements in the circum-single disks (CSDs) have not been rigorously addressed, and these can be very stringent. Insufficient resolution would dilute the strong spiral arms in the CSDs, which are the source of positive torques opposing the negative torque from the CBD and the streams. And the width of these spiral arms can be very sensitive to the choice of $h$. Borrowing a page from the theory of planet-disk interaction \citep[e.g., see][]{Papaloizou:2007}, we know that the one-sided Lindblad torque peaks at
an azimuthal number $\sim 1/(2h)$ \citep{Ward:1997}
and the width of a spiral arm is $\sim 4\pi r h^2$
 \citep[e.g.,][]{Masset:2008}, setting a minimum resolution requirement for the adequate torque calculation. While this resolution requirement can be easily met by modern simulations in the CBD region, it can become extremely difficult in the CSD region, where the number of resolution elements per radial interval around each accretor can be rather small. Moreover, in the frame of the moving accretor, the computational grid is never of a polar nature, and in many cases, is effectively Cartesian at the CSDs scales. Consequently, numerical diffusion can become very taxing in the CSDs, especially for low 
 values of $h$ where the spiral arms are narrow. Under-resolving these spiral arms has the unwanted consequence of reducing the positive torque stemming from this region, favoring the (well-resolved) negative torques from the CBD.

Recent SPH simulations carried out by \citet{Heath:2020} exhibit a drastically different torque reversal threshold of $h=0.2$. These simulations however, do not exhibit any substantial CSD formation, and are thus prone to underestimate the positive torque stemming from the CSD region. 
\citet{Franchinil:2022} showed that the SPH code 
{\footnotesize PHANTOM} can only reproduce the results \citet{Munoz:2019} if the CSD structure is properly resolved with the use of $10^7$ particles. On the other hand, they confirmed that the meshless particle-based code  {\footnotesize GIZMO} \citep{Hopkins:2015} can produce outward migration if the number of resolution elements is increased within the cavity region.

\begin{figure*}[t]%
\begin{minipage}[l]{0.5\textwidth}
\includegraphics[width=\textwidth]{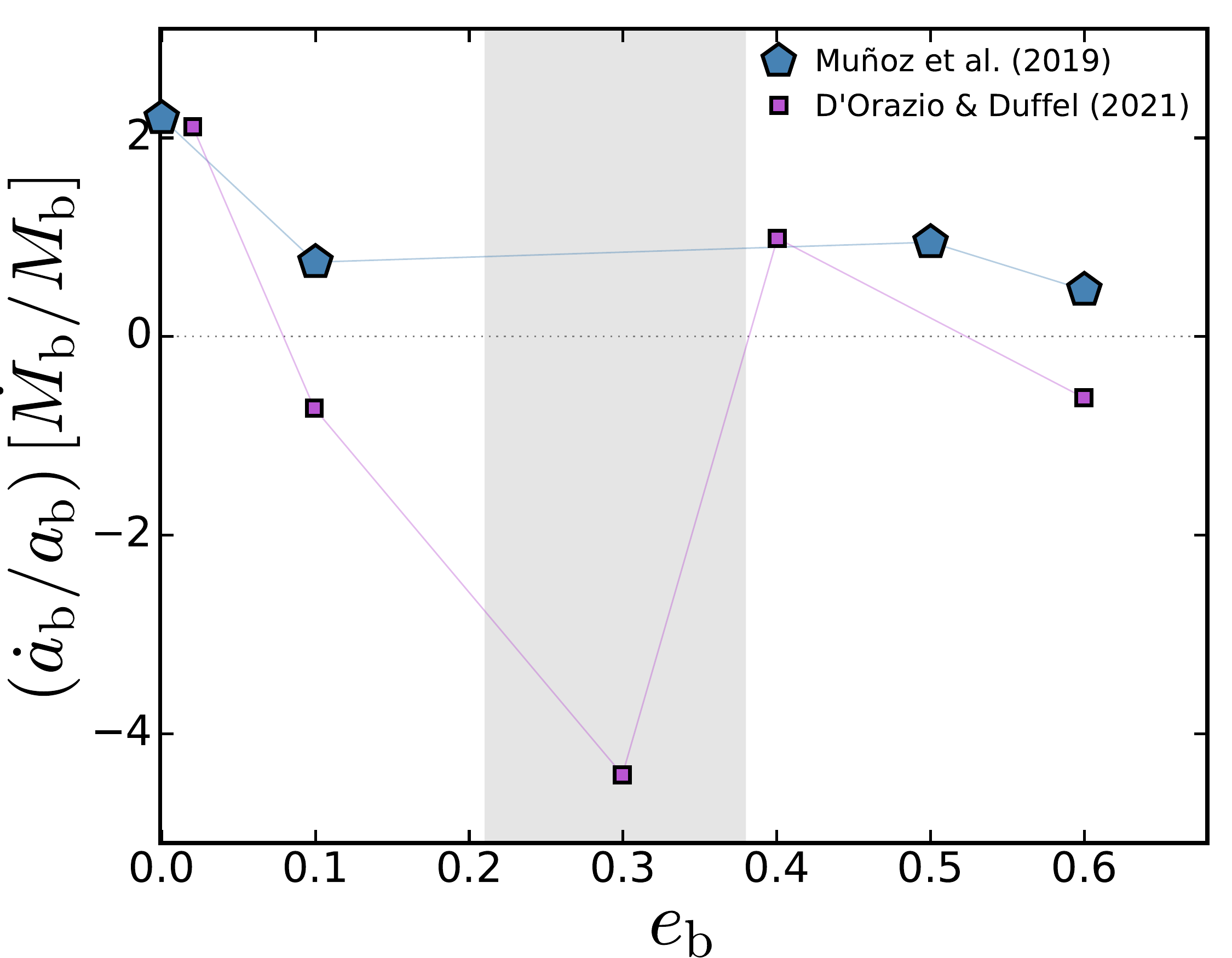}
\end{minipage}
\begin{minipage}[r]{0.5\textwidth}
\includegraphics[width=\textwidth]{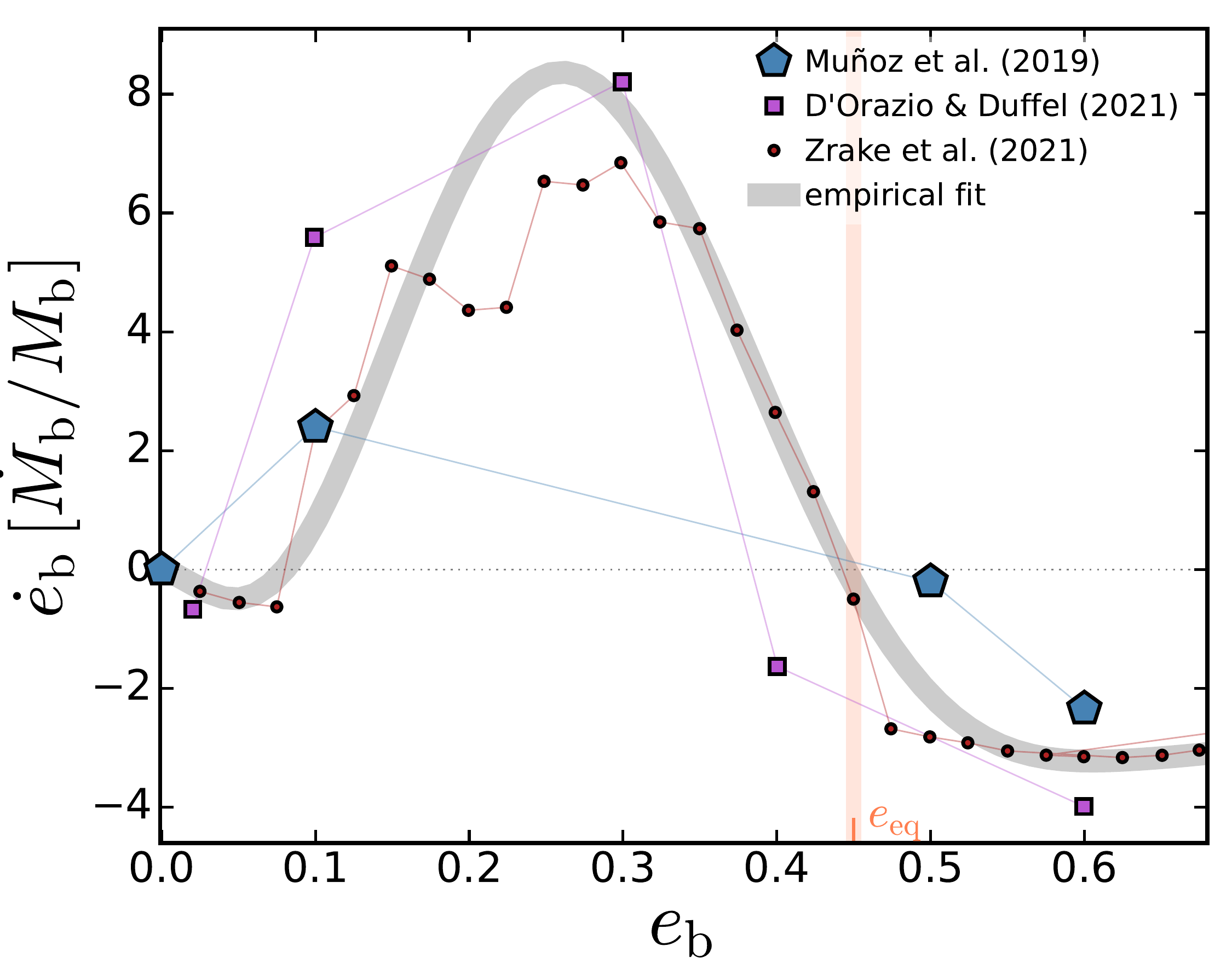}
\end{minipage}
\caption{Left panel:
secular rate of change of semi-major axis $(\dot {a}_{\rm b}/a_{\rm b})(\dot M_{\rm b}/M_{\rm b})=da_{\rm b}/d\ln M_{\rm b}$
for equal-mass binaries at different eccentricities and with the fiducial parameters $h=\alpha=0.1$.
Results from \citet{Munoz:2019} reported outward migration exclusively, while \citet{Dorazio:2021} reported a window (gray region) in which binaries could migrate inward concomitantly with growing in eccentricity.
Right panel: secular rate of change of eccentricity $\dot e_{\rm b}/(\dot M_{\rm b}/M_{\rm b})=de_{\rm b}/d\ln M_{\rm b}$
  of equal-mass binary undergoing circumbinary accretion as function of the binary eccentricity.
  Note that $\dot e_{\rm b}$ is negative for $e_{\rm b}\lesssim 0.08$, positive for larger $e_{\rm b}$, and becomes
  negative again for $e_{\rm b}\gtrsim 0.45$. Adapted from \citet{Zrake:2021}.}
\label{fig:eccentricity_evolution}
\end{figure*}

\subsection{Long-Term Orbital Evolution: Eccentric Binaries}
The secular binary migration rate of Equation~(\ref{eq:migration_rate_circular})
assumes that $e_{\rm b}=0$ at all times. 
For general eccentric binaries, \citet{Munoz:2019} devised a method to compute from simulations both
${\rm d} l_{\rm b}/{\rm d}t$ and ${\rm d}{\cal E}_{\rm b}/{\rm d}t$,
where ${\bf l}_{\rm b}={\bf r}_{\rm b}\times {\bf v}_{\rm b}$
(with ${\bf r}_{\rm b}={\bf r}_1-{\bf r}_2$ the binary separation vector, and 
${\bf v}_{\rm b}={\rm d}{\bf r}_{\rm b}/{\rm d}t$) and ${\cal E}_{\rm b}
=\tfrac{1}{2}{\bf v}_{\rm b}\cdot{\bf v}_{\rm b} -GM_{\rm b}/|{\bf r}_{\rm b}| =-GM_{\rm b}/(2a_{\rm b})$ 
are the specific orbital angular momentum and energy of the binary,
respectively. In particular, the energy transfer rate can be computed from
\begin{equation}
{{\der\cal E}_{\rm b}\over\der t}=-{G\dot M_{\rm b}\over r_{\rm b}}+{\bf v}_{\rm b}\cdot ({\bf f}_1-{\bf f}_2),
\end{equation}
where ${\bf f}_1={\bf f}_{\rm 1,grav}+{\bf f}_{\rm 1,acc}$ is the force per unit mass
on $M_1$ (from gravity and accretion), and similarly for ${\bf f}_2$. 
From these, one can obtain the orbital evolution rates
\begin{eqnarray}\label{eq:orbital_elements}
{\dot a_{\rm b}\over a_{\rm b}}&=&-{\dot{\cal E}_{\rm b}\over {\cal E}_{\rm b}}+{\dot M_{\rm b}\over M_{\rm b}},\\
 -{2e_{\rm b}\dot e_{\rm b}\over 1-e_{\rm b}^2}&=& 2{\dot l_{\rm b}\over l_{\rm b}}+{\dot{\cal E}_{\rm b}\over {\cal E}_{\rm b}}-2{\dot M_{\rm b}\over M_{\rm b}}.
\end{eqnarray}
The results of \citet{Munoz:2019} are shown as blue pentagons in Figure~\ref{fig:eccentricity_evolution}. They concluded that equal-mass binaries of different eccentricities
expand at the rate $\dot a_{\rm b}/a_{\rm b}\sim \dot M_{\rm b}/M_{\rm b}$ (left panel).
But while $\dot a_{\rm b}/a_{\rm b}$ is always positive, the non-monotonic feature of
$\dot a_{\rm b}/a_{\rm b}$ as a function of $e_{\rm b}$ is of interest. Related to this feature is the evolution of $e_{\rm b}$: 
\citet{Munoz:2019}
found that $\dot e_{\rm b}$ (right panel) is slightly negative for $e_{\rm b}\sim 0$, becomes positive
at $e_{\rm b}=0.1$, and becomes negative again at $e_{\rm b}\gtrsim 0.5$. Thus it appears that there is an
eccentricity ``attractor'' below $e_{\rm b}\sim 0.5$ and above  $e_{\rm b}\sim 0.3$. This behavior was confirmed by 
an independent study of \citet{Zrake:2021} (see Figure~\ref{fig:eccentricity_evolution}), who used the {\footnotesize MARA3} code to study accretion from finite-sized disks onto eccentric binaries finely sampling a range of values of $e_{\rm b}$,
and reported an equilibrium binary eccentricity of $e_{\rm b}\simeq 0.45$. These results are shown in the right panel of 
 Figure~\ref{fig:eccentric_binary} (red circles) alongside an associated fitting function (gray curve). The figure also includes the results of \citet{Dorazio:2021} for $\dot{a}_{\rm b}$ and $\dot{e}_{\rm b}$ (purple squares),  who found agreement with \citet{Munoz:2019} and \citet{Zrake:2021} for $\dot{e}_{\rm b}$, but reported negative $\dot{a}_{\rm b}$
 for intermediate values of $e_{\rm b}$. These authors argued that eccentric binaries close to
 $e_{\rm b}=0.4$ are able to migrate inward because of the non-axisymmetric distortion of the circumbinary cavity for those parameters.

\subsection{Accretion onto Unequal-Mass Binaries}\label{sec:unequal_masses}

Early SPH simulations of young stellar binaries accreting from gaseous environments 
\citep[e.g.,][]{Bate:1997b,Bate:2000,Bate:2002}
revealed that the accretion flow depends sensitively on the binary mass ratio $q_{\rm b}=M_2/M_1$, and that even over short timescales, the secondary grows in mass faster than the primary \citep[e.g.,][]{Bate:2000}. 
\citet{Farris:2014} conducted the first systematic study of circumbinary accretion as a function of $q_{\rm b}$ for circular binaries using an early version of the code {\footnotesize DISCO}. These simulations confirmed the existence of preferential accretion onto the secondary, and also found that the time-variability of accretion is a function of $q_{\rm b}$. Using a similar setup, \citet{Munoz:2020a} carried out {\footnotesize AREPO} simulations for $q_{\rm b}$ between 0.1 and 1 until a quasi-steady state was reached. Although they found some quantitative discrepancies with \citet{Farris:2014}, they confirmed that both preferential accretion and 
the accretion variability depend on $q_{\rm b}$.

\begin{figure*}[b]%
\centering
\includegraphics[width=0.75\textwidth]{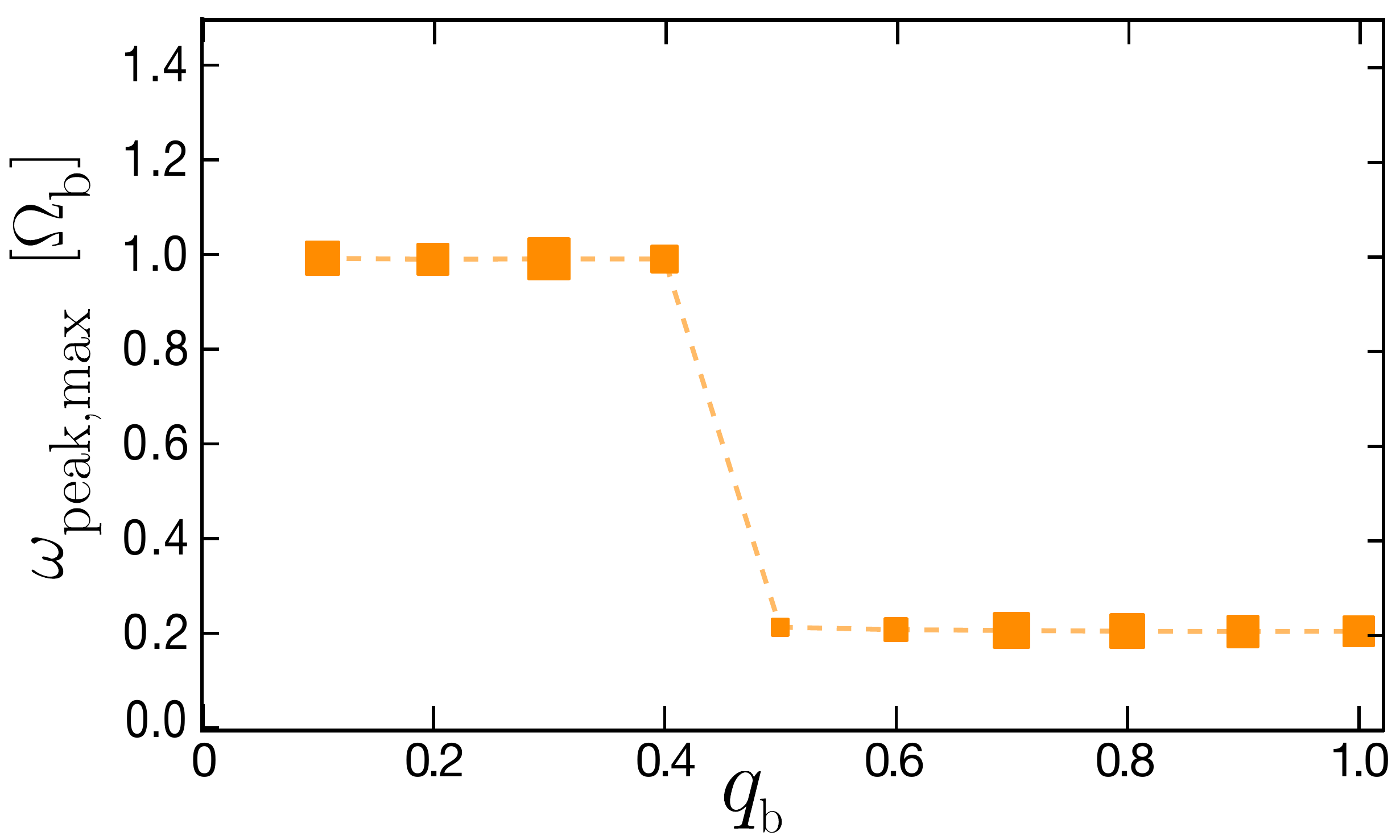}%
\caption{ Dominant frequency $\omega_{\rm peak,max}$ (in units of the binary orbital frequency $\Omega_{\rm b}$) from the spectral analysis of $\dot M_{\rm b}$ for different values of $q_{\rm b}=M_2/M_1$.  The dominant frequency is about $\Omega_{\rm b}/5$ for $q_{\rm b}\ge 0.5$ although its power (indicated by the size of the makers) decreases
 with decreasing $q_{\rm b}$. For $q_{\rm b}\le 0.4$, the dominant frequencies are $\Omega_{\rm b}$ and its harmonics.
 Adapted from \citet{Munoz:2020a}  \copyright AAS. Reproduced with permission.
}
\label{fig:accretion_modulation}%
\end{figure*}
Currently, a consensus has emerged across simulation studies of accreting circular binaries. For fiducial parameters $h=0.1$ and $\alpha=0.1$, the key findings are
\begin{itemize}
\item 
{\bf Accretion is primarily modulated at two frequencies: $\Omega_{\rm b}/5$ and  $\Omega_{\rm b}$.} For $q_{\rm b}\ge 0.5$, the accretion variability is dominated by the lower frequency mode (a period of about $5P_{\rm b}$); for $q_{\rm b}\le 0.4$, it is dominated by the higher frequency mode (a period of $P_{\rm b}$) (see 
Fig.~\ref{fig:accretion_modulation}).
The switch is associated to the disappearance of the orbiting ``lump'' at low binary mass ratios.
\item 
{\bf The secondary accretes more than the primary.} 
The long-term preferential accretion ratio, defined as
$\eta=\langle\dot M_2\rangle/\langle\dot M_{\rm b}\rangle$, is shown in Figure~\ref{fig:mdot_literature} from a collection of recent studies.
While some discrepancies remain, the monotonically decreasing trend of $\eta$ vs $q_{\rm b}$ 
is robust. The result of \citet{Munoz:2020a},
follows an approximately linear relation
\begin{equation}\label{eq:preferential_accretion}
\frac{\langle\dot {M}_2\rangle}{\langle \dot{M}_{\rm b}\rangle}
\simeq 0.5 +{4\over 9}(1-q_{\rm b}).
\end{equation}
\begin{figure*}[t]
\centering
\includegraphics[width=0.75\textwidth]{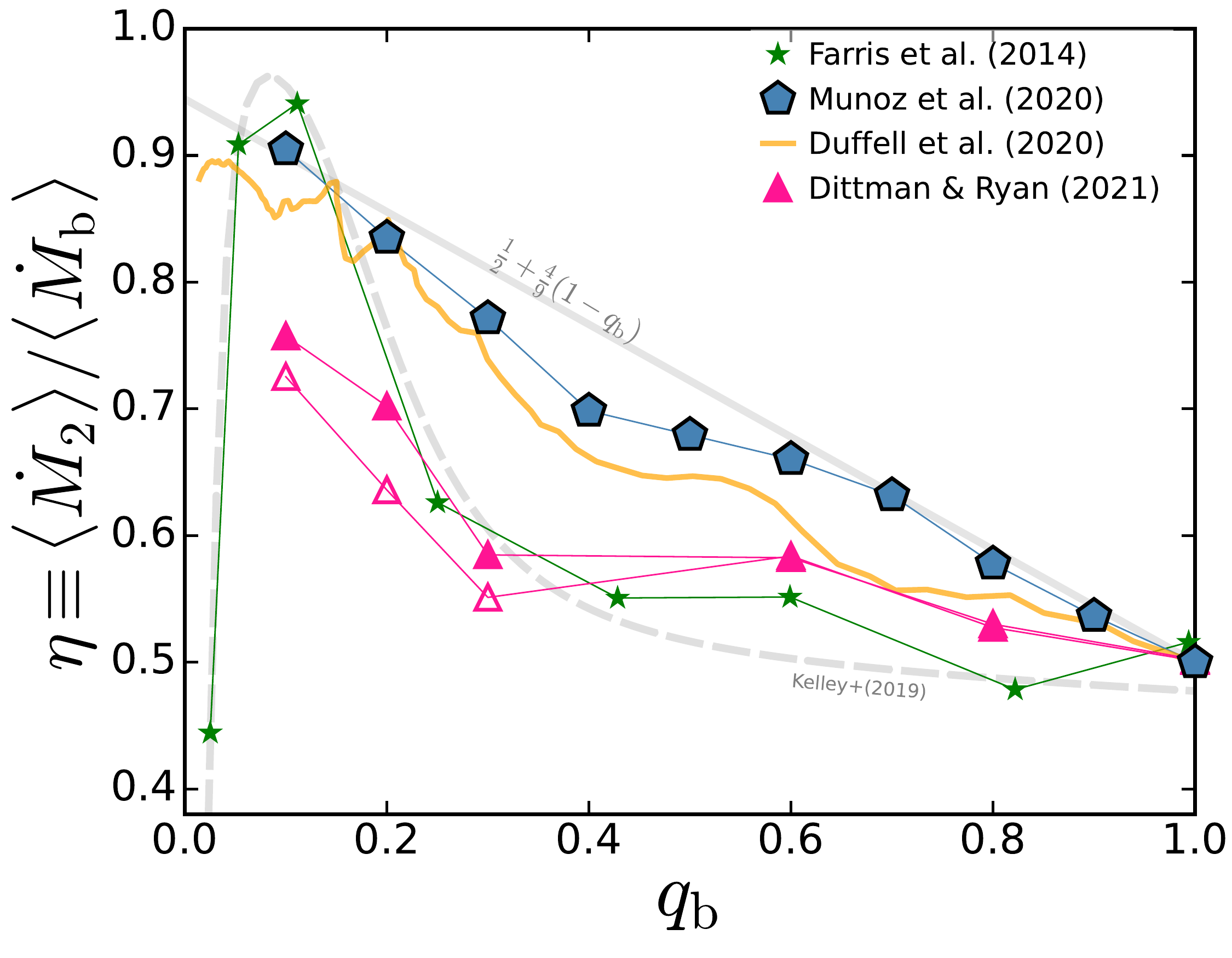}%
\caption{Preferential accretion ratio $\eta\equiv \langle\dot{M}_2\rangle/\langle\dot{M}_{\rm b}\rangle$
compiled from the literature
for $e_{\rm b}=0$ and $h=0.1$
(viscosity prescription is either constant $\alpha=0.1$ or constant $\nu=0.01a_{\rm b}^2\Omega_{\rm b}$). The data are from \citet{Farris:2014} ({\footnotesize DISCO}, $\alpha$-viscosity), \citet{Munoz:2020a} ({\footnotesize AREPO}, $\alpha$-viscosity),
\citet{Duffell:2020} ({\footnotesize DISCO}, constant $\nu$)
and \citet{Dittmann:2021} (modified {\footnotesize DISCO}, constant $\nu$; filled magenta triangles for standard sinks and empty triangles for torque-free sinks). The gray lines depict the \citet{Kelley:2019} fit to the Farris data and the linear relation of Eq.~(\ref{eq:preferential_accretion}). 
\label{fig:mdot_literature}}
\end{figure*}
\item {\bf The accretion angular momentum eigenvalue $l_0=\langle \dot J_{\rm b}\rangle /
\langle \dot M_{\rm b}\rangle$ stays around $0.68a_{\rm b}^2\Omega_{\rm b}$ to $0.8a_{\rm b}^2\Omega_{\rm b}$ for
$q_{\rm b}$ in the range 0.1 to 1}. Figure~\ref{fig:l0_literature} collects
the results reported by \citet{Munoz:2020a} and \citet{Dittmann:2021}, which are in remarkable agreement. \citet{Dittmann:2021} further showed that $l_0$ depends weakly on the sink prescription, provided that the sink region is small enough.
Note that this positive eigenvalue does no guarantee binary expansion, as some accreted angular momentum goes into equalizing the binary's mass ratio. For $\eta$  given by Eq.~(\ref{eq:preferential_accretion}),  the critical eigenvalue above which orbital expansion occurs
is
\begin{equation}\label{eq:critical_eigenvalue}
l_{0,{\rm crit}}=\frac{a_{\rm b}^2\Omega_{\rm b}}{2(1+q_{\rm b})^2}
\bigg[1+q_{\rm b}+q_{\rm b}^2
+\frac{8}{9}(1-q_{\rm b})^2(1+q_{\rm b})
\!\bigg],
\end{equation}
which reduces to $l_{0,{\rm crit}}=(3/8)a_{\rm b}^2\Omega_{\rm b}$ for $q_{\rm b}=1$.
Equation~(\ref{eq:critical_eigenvalue}) is shown in Fig.~\ref{fig:l0_literature} as a gray line, hinting at binary contraction for $q_{\rm b}\approx0.1$.
\begin{figure*}[t]
\centering
\includegraphics[width=0.75\textwidth]{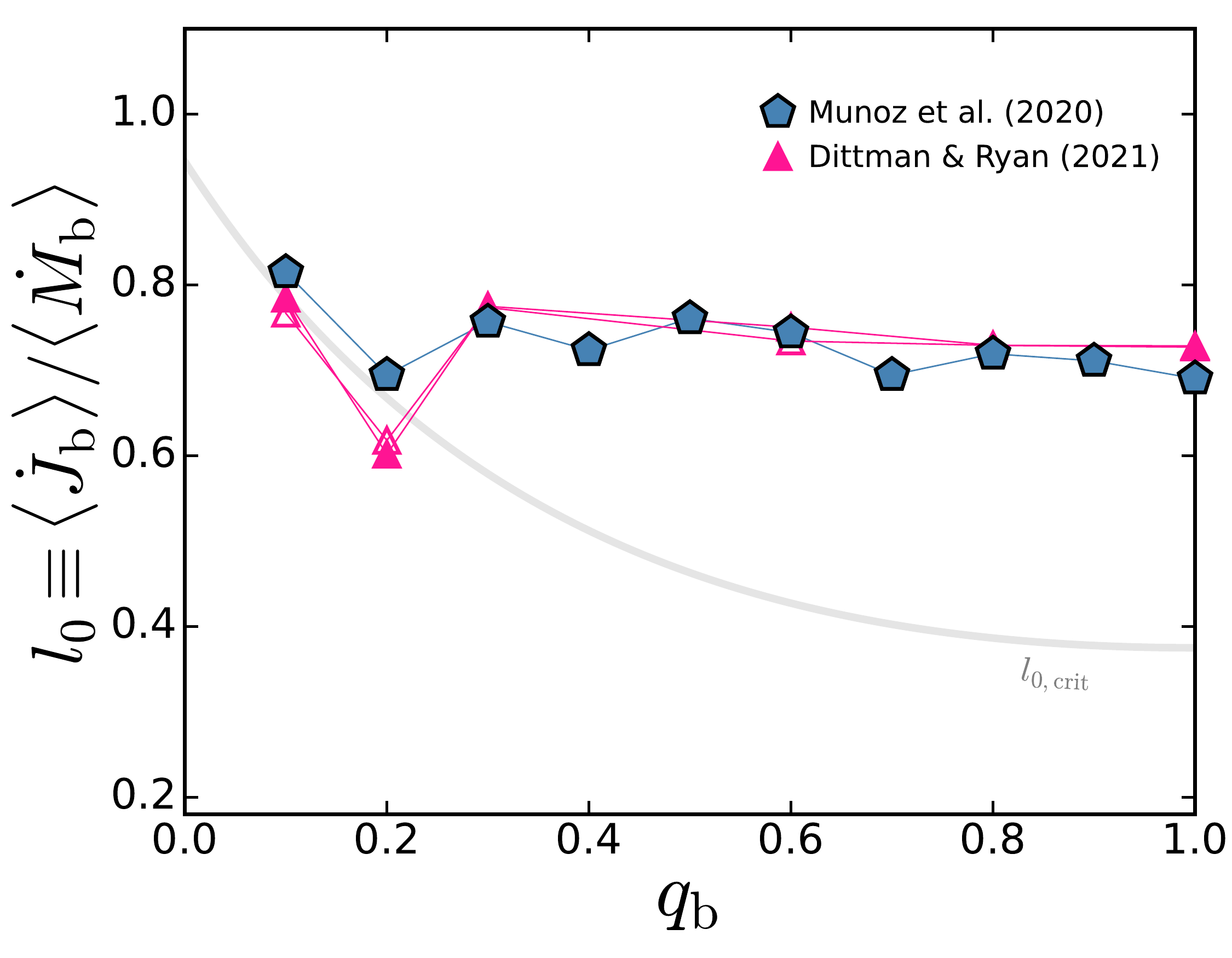}%
\caption{Angular momentum eigenvalue $l_0$ as a function of the binary mass ratio $q_{\rm b}$ compiled from the literature for $e_{\rm b}=0$ and $h=0.1$
(the viscosity prescription is either constant $\alpha=0.1$ or constant $\nu=0.01a_{\rm b}^2\Omega_{\rm b}$). The symbols are the same as in Fig.~\ref{fig:mdot_literature}. 
Note that \citet{Duffell:2020} did not report $l_0$, but only the gravitational torque $T_{\rm grav}$ on the binary; one can obtain the corresponding $l_0$ from 
$T_{\rm grav}$ and $\eta$ by neglecting 
the spin torques and accretion-induced specific torque.
}
\label{fig:l0_literature}
\end{figure*}
\item {\bf The binary orbit expands at a nearly constant rate $\langle\dot a_{\rm b}\rangle/a_{\rm b}\simeq 2 \langle\dot M_{\rm b}\rangle/M_{\rm b}$ for $0.3\le q_{\rm b}\le 1$; this rate
becomes significantly smaller for $q_{\rm b}\le 0.2$ and changes sign for $q_{\rm b}\lesssim 0.1$.} The left panel of Figure~\ref{fig:binary_migration} shows the binary migration rate as computed by \citet{Munoz:2020a}.
As $l_{0,{\rm crit}}$ grows for small $q_{\rm b}$ (Fig.~\ref{fig:l0_literature}), 
$\langle\dot{a}_{\rm b}\rangle$ approaches zero, hinting at a reversal of binary migration. \citet{Duffell:2020} 
probed this transition by running {\footnotesize DISCO} simulations that dynamically update the value
of $q_{\rm b}$, covering values down to $q_{\rm b}=10^{-3}$. They found that the net gravitational torque becomes negative 
for $q_{\rm b}\lesssim 0.05$, implying inward migration.
Recent results by \citet{Dempsey:2021} for $q_{\rm b}\ll 1$ suggest that this torque reversal transition 
depends on the dimensionless parameter\footnote{The parameter $K'$ has been found to accurately predict the gap width in simulations of planet-disk interaction \citep[i.e., when $q_{\rm b}\ll1$; see][]{Kanagawa:2016,Dempsey:2020a}.}
$K'\equiv q_{\rm b}^2/(\alpha h^3)$
and is associated with the disk becoming eccentric. For  $h=0.05$ and $\alpha=10^{-3}$, \citet{Dempsey:2021} report that the gravitational torque goes from negative to positive when $q_{\rm b}\gtrsim 1.5\times10^{-3}$, or when $K'$ surpasses 20. For
$\alpha>0.03$ and $h=0.1$,  $K'\gtrsim20$ is equivalent to  $q_{\rm b}\gtrsim 2\times10^{-2}$, in agreement with \citet{Duffell:2020}.
  \end{itemize}

\begin{figure*}[t]%
\centering
\begin{minipage}[l]{0.8\textwidth}
\includegraphics[width=\textwidth]{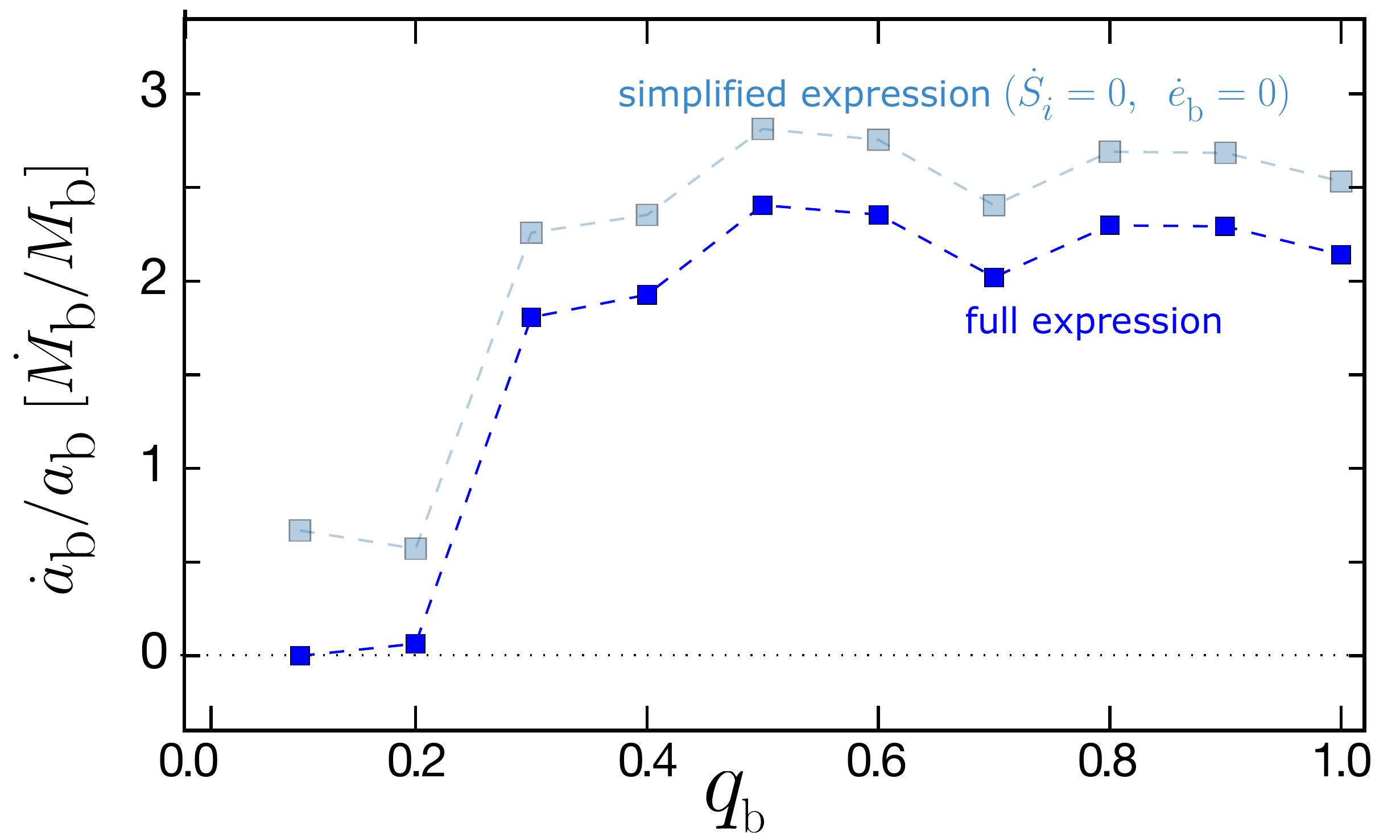}%
\end{minipage}
\begin{minipage}[r]{0.83\textwidth}
\includegraphics[width=\textwidth]{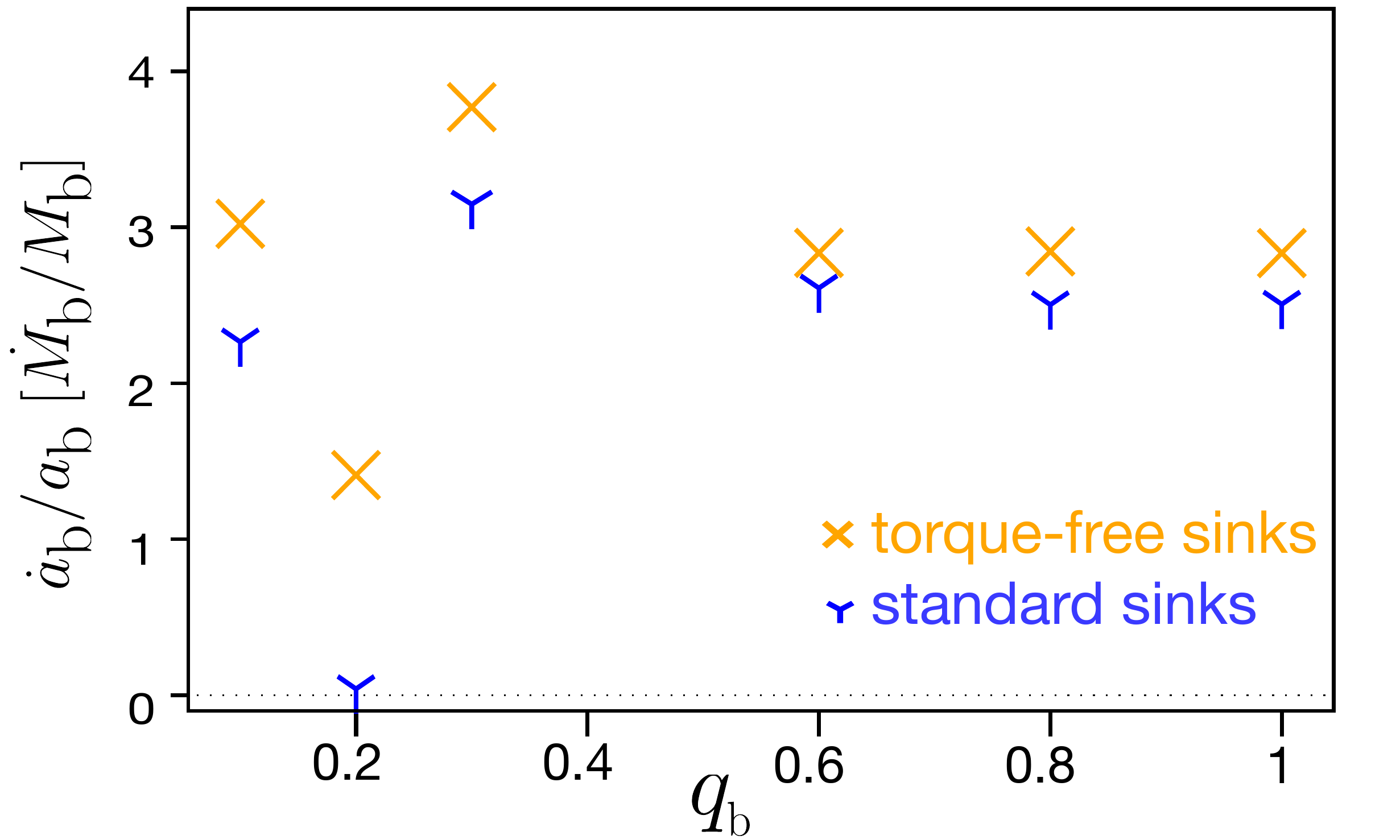}%
\end{minipage}
\vskip 0.5cm
\caption{Binary  orbital expansion rate
  $\dot a_{\rm b}/a_{\rm b}$ (in units of $\dot M_{\rm b}/M_{\rm b}$) for
$e_{\rm b}=0$ and $h=0.1$ (viscosity prescription is either constant $\alpha=0.1$ or constant $\nu=0.01a_{\rm b}^2\Omega_{\rm b}$). Left panel: results of \citet{Munoz:2020a}, who computed
$\langle\dot{a}_{\rm b}\rangle$ in two ways: from the energy transfer euqation (\ref{eq:orbital_elements}) (dark blue) and from angular momentum transfer
assuming $\langle\dot L_{\rm b}\rangle\simeq \langle\dot J_{\rm b}\rangle=l_0~\langle\dot M_{\rm b}\rangle$ 
(light blue). Right panel:
results of \citet{Dittmann:2021},who computed
$\langle\dot{a}_{\rm b}\rangle$ using two sink prescriptions: a standard method (blue symbols)
in which material and linear momentum are removed at the same rate (hence exerting a torque on the gas)
and the torque-free method of \citet{Dempsey:2020b} (yellow symbols), in which the azimuthal momentum of the gas is preserved after mass removal.
These results show that sink particles/mass removal algorithms play a minor role when resolution is sufficiently high, or sinks sufficiently small.
}
\label{fig:binary_migration}%
\end{figure*}

\subsection{Other Complications}\label{sec:complications}

In the previous subsections, we have reviewed the key results from
CBD simulations using idealized equation of state (locally isothermal with constant disk aspect ratio $h$) and viscosity prescription ($\alpha$ viscosity or constant $\nu$), and assuming small (local) disk mass 
(without disk self-gravity). These assumptions may not always apply to realistic disks.

The locally isothermal equation of state
is known not to conserve the angular momentum flux carried by density waves \citep{Lin:2011,Miranda:2019}, with a resulting modification of the torque profile in the disk. The effect on the integrated torque, however, is negligible for rapidly cooling disks, and amounts to a correction of 
$\sim 30\%$ in slowly cooling/adiabatic disks \citep{Miranda:2020}. Thus, the net contribution of this effect to the migration of binaries is likely minor, except for the borderline cases.

The locally isothermal equation of state
assumes an efficient cooling process that instantly brings disk temperature back to equilibrium. Without resorting to radiative transfer, the simplest way to assess the role of gas thermodynamics is to allow the disk temperature to relax towards the equilibrium
value over some cooling timescale $t_{\rm cool}$ (usually 
parameterized using $\beta=\Omega_K t_{\rm cool}$, where $\Omega_K$ is the local Kepler frequency). \citet{Wang:2022a,Wang:2022b}
carried out {\footnotesize ATHENA++} simulations using this $\beta$-cooling prescription, and found that 
with a longer cooling time, the accretion variability is
gradually suppressed and the morphology of the inner disk becomes more symmetric. They also found that the accretion angular momentum eigenvalue $l_0$ generally decreases with increasing $\beta$, such that an equal-mass, circular binary (with equilibrium $h\sim 0.1$) undergoes orbital decay for $\beta\gtrsim 0.2$, although the details depend on the viscosity prescription. 

A more complex behavior is expected to arise from 3D
magnetohydrodynamical (MHD) simulations, which are needed to capture viscous dissipation self-consistently. Such MHD simulations for circumbinary accretion are typically initialized with narrow, finite-supply tori, and are run for a small number of orbits. Thus, they are useful for studying short-term accretion variability, but may not be adequate for probing the long-term flow behavior and the secular evolution of the central binary.
Early MHD simulations of CBDs around quasi-circular binaries \citep[][using a version of the finite difference code {\footnotesize ZEUS}]{Shi:2012} and GRMHD \citep[][ using HARM3D]{Noble:2012}
were carried out excising the central binary for a run time of 80 binary orbits. These show the similar accretion modulation behavior as 2D viscous hydrodynamics, despite the development of MHD turbulence in the disk. 
More recent efforts have been focused on the general relativistic aspects of CBD evolution, important for black-hole binaries close to merger (with separations of 10-20 gravitational radii). 
For example, GRMHD simulations using {\footnotesize HARM} and its variants have been
implemented with an approximate analytic metric of binary black holes 
in order to include the CSDs in the computational domain,
although a circular region at the binary barycenter is still excised
\citep{Bowen:2017,Bowen:2018,Combi:2021,Combi:2022}. These simulations
(typically run for $\sim 20$ orbits)
show additional time modulations that are interpreted as purely relativistic effects.

The simulations discussed in the previous subsections all assume that the local disk mass near the binary is much less than $M_{\rm b}$ (although the mass of an extended disk can be much larger). When this condition is not satisfied,
the flow/binary dynamics and evolution can be quite different. Such massive disks (or ``envelopes'') naturally occur immediately following galaxy mergers.
A massive black-hole binary embedded in such an envelope is expected to 
undergo orbital decay through dynamical friction
\citep[e.g.,][]{Escala:2005,Cuadra:2009,Franchini:2021}.
Massive CBDs or envelopes may also be important for the formation of close (au-scale) proto-stellar binaries (see Section~\ref{sec:stellar_binaries}).
If CBDs are massive enough and no substantial CSDs can form, the binary is bound to lose angular momentum to the gas. 
Numerical simulations show that small-mass proto-binaries
embedded in collapsing cores decay in separation quickly\citep[e.g.,][]{Bate:2000,Bate:2002}.
Aided by dynamical friction, the pair becomes bound at small separations and continue to grow until it becomes dynamically dominant. Only then a true CBD may form, at which point the binary may have already reached au-scale separations.
More recent simulations, including the effects of magnetic fields, find qualitatively similar results, except for a further accelerated orbital decay attributed to magnetic braking of the gas envelope
\citep[e.g,][]{Zhao:2013,Kuruwita:2020}. 
For example, in the simulations of \citet{Kuruwita:2020}, the binary pair decreases in separarion from 300 au to 10 au in only $\sim 30$ radial incursions, at which point it becomes a Keplerian bound pair and its orbital decay is slowed down. In this scenario, orbital decay occurs early in the star formation process, and circumbinary accretion represents the late stage of binary formation.

\section{APPLICATIONS OF CIRCUMBINARY ACCRETION}\label{sec:4}

\subsection{Massive Black Hole Binaries and the Final Parsec Problem}\label{sec:mbhb}

In the $\Lambda$CDM paradigm of cosmology, MBH encounters are expected to occur as a natural consequence of hierarchical structure growth. 
If most galaxies host a MBH in their nuclei
\citep{Magorrian:1998,Kormendy:2013}, then
MBH binaries (MBHBs) should be an inevitable consequence
of galaxy mergers \citep{Begelman:1980}.
 Indeed, the discovery of dual AGNs with kiloparsec separations confirm that galactic mergers can form wide MBH pairs \citep[see][for a review]{DeRosa:2019}.
 Likewise, since galaxy mergers funnel large amounts of gas toward the center of the merger remnant \citep[e.g.,][]{Hopkins:2006}, the formation of gaseous CBDs should occur alongside that of MBHBs. 
 It is generally assumed that, if MBHBs form, they evolve toward merger one way or another \citep[for a review, see][]{Colpi:2014}. Such mergers are of great  importance for low-frequency GW astronomy in the 
 LISA band \citep{Haehnelt:1994,AmaroSeoane:2017} and the Pulsar Timing Array (PTA) band 
 \citep{Burke-Spolaor:2019}.

Following galaxy mergers, dynamical friction
by gas/stars and stellar “loss-cone” scatterings would serve to reduce the separations of MBHBs to about one parsec, while gravitational radiation would become effective
(and make the BHs merge) only if their separations are less than about 0.01 parsec
\citep[e.g.,][]{Begelman:1980,Polnarev:1994}.
Binary orbital evolution in the intermediate separation regime is a long-standing problem known as the “final parsec problem”
\citep{Milosavljevic:2003a,Milosavljevic:2003b}.

CBDs have been posited to be a gas-dynamical solution
to this final parsec problem \citep[stellar-dynamical solutions, on the other hand, have been thoroughly studied in the literature; e.g., ][]{Yu:2002b,Vasiliev:2015,Gualandris:2017}. In the  CBD hypothesis, a MBHB embedded in a viscous accretion disk is forced to migrate inward along with the accretion flow, much like massive planet embedded in a protoplanetary disk would \citep{Gould:2000}.  Most if not all cosmological-scale studies of MBHB coalescence that consider the influence of gas assume CBDs always promote inwardly driven migration \citep{Haiman:2009,Kelley:2017a,Kelley:2017b,Kelley:2018,Volonteri:2020,Volonteri:2022}.
The studies of CBD accretion of recent years put this standard assumption into question, potentially introducing a major caveat to the expected detection rates for LISA \citep[e.g.,][]{Sesana:2005} and PTAs  \citep[e.g.,][]{Sesana:2008}. 

Despite the encouraging discovery of wide dual AGN \citep[e.g.][]{Goulding:2019}, the direct observation of spatially resolved, sub-pc MBHBs will remain all but impossible for the foreseeable future (with the possible exception of very long baseline interferometric observations of nearby galaxies; \citealp{Burke-Spolaor:2011}). 
Thus, until a direct 
GW measurement of the inspiral and merger of two MBHs is made, indirect detection methods will be needed to probe the ellusive sub-pc regime.

\paragraph{Indirect Signatures of Ongoing Circumbinary Accretion onto Massive Black-Hole Binaries}
Spatially resolving MBHBs at sub-parsec separations is
difficult. The most compact, directly-imaged MBHB on record 
is the dual AGN in the radio galaxy 0402+379, with a projected separation 7.3 pc \citep{Rodriguez:2006}. For more compact sources, photometric and spectroscopic techniques can be used to infer the presence of MBHBs. Circumbinary accretion may prove crucial for upcoming multi-messenger efforts to identify such compact MBHBs
in the form of periodically varying electromagnetic (EM) counterparts
\citep[e.g.,][]{Bogdanovic:2022,Charisi:2022}.  The sough-after periodic signals in time-variable AGN include (i) time-variable kinematic offsets of the broad emission lines
\citep{Gaskell:1996},
(ii) Doppler boosted broad-band emission from the circum-secondary 
accretion disk \citep{Dorazio:2015},
(iii) periodic self-lensing flares \citep{Dorazion:2018,Hu:2020},
and 
(iv) broad-band photometric variability due to pulsed/modulated accretion \citep[e.g.,][]{Tang:2018}.
In practice, the kinematics approach is of limited use at sub-pc because of the confusion and truncation of the distinct broad line regions at these separations
\citep{Kelley:2021}, making the photometric approach a more viable and data-rich technique \citep[e.g.,][]{Charisi:2022}. Indeed, systematic searches for periodic signals in the Catalina Real-time Transient Survey \citep{Graham:2015b}, the Palomar Transient Factory \citep{Charisi:2016}, 
the Panoramic Survey Telescope and Rapid Response System \citep{Liu_T:2019}
have revealed 
hundreds of MBHB candidates. Measuring photometric variability is not exempt from systematic uncertainties \citep{Graham:2015b,Liu_T:2016,Witt:2022}, but has the potential of being a powerful multi-messenger tool in combination with simultaneous GW detections \citep{Charisi:2022}. At sub-pc separations, MBHBs
emit GWs in the the frequencies suitable for PTAs, 
which are forecast to provide individual binary detections within the next decade
\citep{Mingarelli:2017,Kelley:2018}. 
In this regime, the binary orbit is still essentially Newtonian, and the associated GW waveforms are well understood. 
Likewise, the known photometric variability mechanisms (accretion/boosting/lensing) are all modulated on timescales commensurate with the binary orbit. Simultaneous GW/photometric detection of these objects would not only conclusively prove their existence, but could also allow for constrains on the their orbital parameters.

\paragraph{Imprints of Circumbinary Accretion on Massive Black-Hole Binary Orbits}
In addition to detecting circumbinary accretion ``in action'' in the form of the photometric variability of individual sources, CBD physics can also leave an imprint in the population of MBHBs responsible for the stochastic gravitational wave background (GWB) \citep[e.g.,][]{Zimmerman:1980,Thorne:1987,Rajagopal:1995,Phinney:2001},
which is detectable by the method of pulsar timing \citep{Sazhin:1978,Detweiler:1979,Bertotti:1983}. The GWB spectrum can be computed from the GWs emitted from MBHB mergers across cosmic scales \citep{Phinney:2001}, although the uncertainties in its amplitude and spectral slope depend on the astrophysics of MBHB assembly \citep{Sesana:2008} and on whether or not binary coalescence is delayed, suppressed, or biased toward a specific range in binary mass $M_{\rm b}$  \citep[e.g.,][]{Shannon:2015,Middleton:2018}.
Moreover, recent calculations also conclude that the GWB can depend on the eccentricity distribution of the binaries \citep{Kelley:2017b}, and on the preferential accretion ratio \citep{Siwek:2020}. Consequently, the physics of circumbinary accretion can have a major impact in the GWB.

\subsection{Binary Star formation}\label{sec:stellar_binaries}
The existence of tidally cleared disks 
\citep[e.g.,][]{Dutrey:1994,Mathieu:1997,Jensen:1997,Carr:2001,Ireland:2008,Czekala:2021}
or rotationally supported structures \citep{Tobin:2016,Maureira:2020}
around  young stellar binaries,
plus the confirmation that such binaries may accrete in a modulated fashion \citep{Basri:1997,Jensen:2007,Muzerolle:2013,Tofflemire:2017a,Tofflemire:2017b,Tofflemire:2019}, have underscored the importance of binary-disk interaction in early 
stellar evolution. These stellar CBDs also contain information about circumbinary planet formation (see Section~\ref{sec:circumbinary_planets} below) and can serve as laboratories and proxies for the circumbinary physics taking place in galactic nuclei  (Section~\ref{sec:mbhb}).

Tidal truncation and pulsed accretion in young binaries (already discussed in Section 3) are the most straightforward smoking guns for circumbinary accretion. But other important clues of past circumbinary accretion may be found in the evolved population as well. These include binary mass ratios and the over-abundance of stellar twins, and migration history of compact and wide binaries.

\paragraph{The Over-Abundance of Stellar Twins}
Over the last few decades, several studies have consistently shown that 
stellar binaries exhibit an over-abundance of twins (with mass ratio $q_{\rm b}>0.95$;
\citealp{Lucy:1979,Tokovinin:2000,Halbwachs:2003,Lucy:2006,Raghavan:2010,Moe:2013,Moe:2017}).
This finding appears to be in agreement with some early hydrodynamical simulations which showed that, in binaries accreting from rotating gas, the secondary grew faster that the primary (\citealp{Bate:2000}, but see also \citealp{Artymowicz:1983,Bonnell:1992}).
Indeed, high resolution CBD simulations that systematically compute the ratio of accretion rates in circular binaries find 
that the secondary accretes more than the primary and 
the ratio $\dot{M}_{2}/(\dot{M}_1+\dot{M}_2)$ decreases monotonically with increasing $q_{\rm b}$
(see Figure~\ref{fig:mdot_literature}).

The twin excess was originally reported to be most significant at short orbital periods $P_{\rm b}\lesssim20$~d or $a_{\rm b}\lesssim0.4$~au,
(\citealp{Moe:2017}; see also \citealp{Tokovinin:2000,Raghavan:2010}), which was interpreted as being
consistent with the once-held idea that accreting binaries always migrate inward \citep[see discussion in][]{Raghavan:2010}. In recent years, however, astrometric measurements 
(using {\it Gaia} data) have revealed that the twin excess fraction is roughly constant 
for projected binary separations between 0.01 au and 10000 au \citep{ElBadry:2019}.
This finding 
contradicts the assumption that equal-masses and compact orbits are two sides of the same coin.
Instead, the large separation of twins may indicate that outward migration due to CBDs has occured.
But even if CBDs promote softer orbits in addition to equal masses, they alone may not 
produce binary orbits as wide as $\sim 10^4$au.  \citet{ElBadry:2019} conjectures that twins form first with separations $a_{\rm b}\lesssim100$ au while accreting from circumbinary disks, and are subsequently widened by dynamical interactions 
in their parental clusters. The fact that these binaries also tend to be eccentric \citep{Hwang:2022}
adds credence to the hypothesis that these orbits have been subject to strong perturbations {\it after} 
gas has dissipated.

\paragraph{Formation of Close Binaries}
In light of the emerging paradigm that 
binaries accreting from warm disks tend to expand (see Section 3), 
the origin of 1-10 au-scale binaries remains a puzzle. 
Numerical simulations suggest that binary stars initially form with large separations ($\gg 10^2$~au)
\citep[see][]{Offner:2022}.
The difficulty lies in the impossibility of fragmenting a disk at small separations where cooling is inefficient \citep{Matzner:2005,Rafikov:2005,Boley:2006,Whitworth:2006,Stamatellos:2008,Cossins:2009,Kratter:2010}, and in the impossibility of directly fragmenting a core when these are stable against non-axisymmetric perturbations (and because of the size of the first hydrostatic core is already a few au in size; \citealp{Bate:1998}). 
Naturally, a significant inward migration would be needed to produce close binaries.

A likely solution to this conundrum is that the required migration takes place 
at the early (Class 0-I) stage of star formation, when the newly fragmented binaries
are still embedded in massive envelopes \citep[see][]{Tokovinin:2020}.
As discussed in Section~\ref{sec:complications}, some simulations of 3D collapse with embedded binaries do show binary orbital decay at rates faster than 
the Lindblad torques of the associated CBDs.
In some cases, these ``binaries'' decay because they are not a bound pair initially, and their orbital evolution is dominated by dynamical friction and gas accretion from the envelope.

\subsection{Planets Around Binaries}\label{sec:circumbinary_planets}

One of the most exciting results in exoplanetary science in recent years
was the discovery of circumbinary planets (CBPs).
The Kepler mission has discovered 13 CBPs around 11 eclipsing binaries
\citep[e.g.,][]{Doyle:2011,Welsh:2018,Socia:2020}
and the TESS mission
has so far (Oct. 2022) detected two CBP systems
\citep{Kostov:2020,Kostov:2021}.
Most of these are large planets (with radius between $0.3R_J$ and $R_J$) in a sub-au, nearly co-planar orbit around the host binary.
A handful of CBP systems have been discovered using
gravitational microlensing \citep[OGLE-2007-BLG-349,][]{Bennett:2016},
and direct imaging
\citep[e.g., HD~106906][]{Bailey:2014,Lagrange:2016,Rodet:2017}.
In addition, a number of CBP systems have been inferred using eclipse timing variations (ETVs)
of binaries \citep[e.g., NN Serpentis][]{Qian:2009,Beuermann:2010}, although the validity of these planets remains uncertain.

Recent studies suggest that the occurrence rate of large, Kepler-like
CBPs is comparable to that of similar-mass planets in single-star
systems \citep[$\sim 10\%$][]{Armstrong:2014,Martin_D:2014,Martin_D:2019},
indicating that planet formation in circumbinary disks is a robust process.

Close-in planetary orbits around a binary are known to be dynamically unstable \citep[e.g.,][]{Dvorak:1989,Pilat-Lohinger:2003,Doolin:2011}.
For binaries with $q_{\rm b}\sim 1$, the critical planetary semi-major axis $a_{\rm crit}$ is about a few times $a_{\rm b}$. \citet{Holman:1999} provides an approximate expression of $a_{\rm crit}$ based on simulation of circular, co-planar circumbinary particles:
\begin{equation}
a_{\rm crit}=\Bigl( 1.6+ 5.1e_{\rm b}-2.22e_{\rm b}^2+4.12\mu_2
-4.27e_{\rm b}\mu_2-5.09\mu_2^2+4.61e_{\rm b}^2\mu_2^2\Bigr)\, a_{\rm b},
\end{equation}
where $\mu_2=M_2/M_{\rm b}$. 
A striking feature of the CBPs discovered by {\it Kepler} is that many of them lie close to the instability boundary (e.g. Kepler-16b has a semi-major axis $a_p= 1.09a_{\rm crit}$, and Kepler-34b has $a_p=1.14a_{\rm crit}$). This feature cannot be explained by the selection bias of transit observations \citep{Li:2016}.
It has been commonly interpreted as evidence for planetary migration,
since the circumbinary cavity acts as a ``trap'' in which migrating planets can be  ``parked'' \citep[e.g.,][]{Masset:2006,Pierens:2008,Pierens:2013,Kley:2014,Kley:2019,Penzlin:2021,Coleman:2022}.
The precise stopping location of the planet (for a given binary) depends on various disk parameters (such as viscosity and disk aspect ratio), which affect the intrinsic cavity size, and on the planetary mass -- A massive planet opens a gap in the disk and circularizes the inner cavity, and thus tends to migrate closer to the binary.
Current hydrodynamical simulations struggle to reproduce systems such as 
Kepler-34b, an eccentric ($e_p=0.18$) planet orbiting very close to an eccentric binary ($e_{\rm b}=0.52$), because the central cavity is large and has a significant eccentricity, causing the planet to park too far away \citep[e.g.,][]{Pierens:2013,Penzlin:2021}.
The eccentricity of the inner CBD (see Section 3.3)
can also leave an imprint on embedded CBPs. In particular, low-mass planets can 
inherit the eccentricity of the gas disk and precess with it in a state of apsidal corotation \citep{Thun:2018,Penzlin:2021}.
Clearly, the close-in CBPs provide clues on the intricate 
interplays of binary-disk and planet-disk interactions.

In-situ formation of CBPs at close-in locations are generally difficult
because of the large impact velocities between planetesimals
driven by the perturbations on the planetesimal orbits from the binary potential and non-axisymmetric density features within the CBD. Such large impact velocities 
lead to corrosive collisions and prevent the buildup of large planetary bodies
\citep{Scholl:2007,Marzari:2008,Meschiari:2012,Paardekooper:2012}.
The motion of planetesimals, however, can depend on the CBD as much as on the central binary.  \citet{Rafikov:2013} and \citet{Silsbee:2015} found that the (secular) planetesimal dynamics is affected by 
the gravity and gas drag from a precessing eccentric CBD. These effects 
(for sufficiently massive CBDs) may suppress the excitation from the binary, and couple to it resonantly, defining specific regions of the CBD where planetesimals can grow into planets.
Gas drag is fundamental in permitting planet formation under external eccentricity perturbations because it can lead to  coherent eccentric orbits \citep{Silsbee:2015} and/or coherent ``most circular'' closed orbits
 \citep{Pichardo:2005,Lithwick:2008,Youdin:2012,Bromley:2015}. Planetesimals in such orbits would still have low relative velocities, meaning that growth via collisions would not be hampered by the global eccentricity. 
 Overall, these studies suggest that close-in CBPs formed at relatively large distances 
 ($\gtrsim 10a_{\rm b}$ and move to their current orbits through disk-driven
migration.

\subsection{Post-Main-Sequence Binaries}\label{sec:post-MS}

The final evolution of 0.8-$8 M_\odot$ stars involves a rapid transition from the Asymptotic Giant Branch (AGB) over the post-AGB phase towards the planetary nebulae  stage. Many post-AGB stars are in binary systems (with a main-sequence companion). It is now well-established that most post-AGN binaries (with periods from 100 to a few thousands days) are surrounded by circumbinary gas-dust disks; the observational evidence comes from the disk-like spectral energy distribution of the systems, IR imaging and CO interferometry that resolves the Keplerian velocity field \citep[e.g.,][]{vanWinckel:2009,vanWinckel:2018,Kluska:2022}. Many 
of these binaries also have a bipolar outflow/jet launched from the circumstellar disk around the secondary (main-sequence) companion \citep[e.g.,][]{Bollen:2021,Bollen:2022}. The circumstellar disk is likely fed from the CBD, although mass transfer from the post-AGB primary star cannot be ruled out. 

The origin of the CBDs around post-AGB binaries is somewhat unclear. 
These disks are formed as a result of binary interactions during the AGB phase. One possibility is that during the common-envelope (CE) stage, not all envelope is ejected, and a fraction of the mass falls back to the remnant binary in the form of a CBD \citep[e.g.,][]{Sandquist:1998,Kashi:2011}. One issue with this scenario is that CE evolution is expected to produce systems with periods ranging from hours to hundreds of days, and yet no CBD has been observed around systems with periods less than 100 days. Another possibility is mass loss through the outer Lagrangian point L$_2$ associated with the mass transfer or wind of the AGB star \citep[e.g.,][]{Shu:1979,deVal-Borro:2009,Pejcha:2016}, although it is unclear if stable (and long-lasting) CBD can be produced in this way.

CBDs may produce some dynamical effects on the post-AGB binaries
or other evolved binaries. Many post-AGB binaries with periods
100 to a few thousands days are observed to have significant eccentricities (up to 
$e_{\rm b}\simeq 0.6$) \citep{vanWinckel:2018}. This is surprising since such binaries are supposed to have circularized during the AGB phase. \citet{Dermine:2013} suggested that the eccentricities of the post-AGB binaries could be excited as a result of their gravitational interaction with the CBD. \citet{Antoniadis:2014} suggested that similar effect could explain the eccentricities of some binary millisecond pulsars with the white-dwarf companions.
The eccentricity-growth mechanism can be easily understood from gravitational binary-CBD interaction \citep[e.g.,][]{Goldreich:1980,Lubow:1996}. From Section \ref{sec:2}, we know that the binary loses angular momentum and energy through the OLR associated with the $(mn)$ potential component at the rates $\der J_{\rm b}/\der t=-T_{mn}$
and $\der E_{\rm b}/\der t=-\omega_{mn} T_{mn}$.
These imply that the binary eccentricity evolves
at the rate 
\begin{equation}
\dot e_{\rm b}={(1-e_{\rm b}^2)^{3/2} T_{mn}\over
e_{\rm b}J_{\rm b}}\left[(1-e_{\rm b}^2)^{1/2}-{\omega_{mn}\over\Omega_{\rm b}}\right].
\label{eq:edot}\end{equation}
For small but finite $e_{\rm b}$, the $m=2$, $n=-1$
potential component has $\omega_{mn}=\Omega_{\rm b}/3$, and $T_{mn}\propto e_{\rm b}^2$, leading to $\dot e_{\rm b}/e_{\rm b}> 0$.

It is important to note that equation (\ref{eq:edot}) relies on pure gravitational
interaction between the binary and the CBD.
This could apply in the "transient" stage before gas accretion onto the binary sets in [see, for example, the simulation of accretion from a finite-sized CBD by \citet{Munoz:2020a}].
However, as discussed in Section \ref{sec:angular_momentum_transfer}, once the 
accretion starts (typically within a viscous time of the inner disk) and the system 
settles into a quasi-steady state, the secular evolution of the binary can be quite different from that predicted by the pure gravitational effect.
In addition, when applying to post-AGB binaries, it is not clear that the CBD has a sufficient mass to 
change the binary eccentricity in an appreciable way
even if the accretion effect is ignored
\citep{Rafikov:2016}. Obviously, 
more works are needed to understand the 
origin of the peculiar eccentricities of the post-MS binaries.

\section{MISALIGNED DISKS}\label{sec:5}

In the previous sections, we have focused on aligned disks, i.e.,
disks that are coplanar with their central binaries. Such alignments
may not be realized in many situations.  For example, in the current
theory of star formation, supersonic turbulence in molecular clouds
leads to the creation of clumps, which then condense and fragment into
stars and binaries \citep[e.g.,][]{Mckee:2007}. The gas that falls
on to the central protostellar core/binary and assembles on to the
disk at different times may rotate in different directions 
\citep{Bate:2003,Bate:2010,Offner:2010,Tsukamoto:2013,Fielding:2015,Takaishi:2020}.
In this scenario, it
is reasonable to expect a newly formed binary to be surrounded by a
highly misaligned CBD, which forms as a result of
continued gas accretion. Similar chaotic accretion may also occur in
the formation of massive black holes (MBHs) \citep[e.g.,][]{King:2006}.
In particular, MBH binaries at the centers of galaxies may experience accretion of
successive low-mass gas ``clouds'' with uncorrelated angular momenta,
which naturally lead to misaligned CBDs.

Observationally, most CBDs around young stellar binaries
are found to be aligned with their host binary orbital planes \citep{Czekala:2019}.
For example, the gas rich CBDs
AK Sco \citep{Czekala:2015} and DQ Tau
\citep{Czekala:2016}, and the debris disks $\alpha$ CrB and $\beta$ Tri \citep{Kennedy:2012b}, all have mutual disc–binary inclinations less than a few degrees. However, there are some notable exceptions.
For example, the pre-MS binary KH 15D possesses a low inclination (10-20$^\circ$)
precessing circumbinary ring or disk \citep{Chiang:2004,Winn:2004,Capelo:2012,Poon:2021}
(see \citealp{Zhu:2022} for two similar systems discovered by ZTF).
The disks (circumbinary and two
circumstellar) in the system IRS 43 are highly misaligned 
($\sim 60^\circ$) with each other and with the binary \citep{Brinch:2016}.
The gaseous CBD around the eccentric binary HD 98800B is nearly
polar (with a misalignment angle $\sim 90^\circ$; \citealp{Kennedy:2019}). 
The 6–10 Gyr old eccentric binary 99 Herculis has a nearly polar
($\sim 87^\circ$) debris ring around it \citep{Kennedy:2012a}.
The young ($\sim 1$~Myr) hierarchical triple star system GW Ori possesses a misaligned circumtriple disk with three broken dusty rings \citep{Krauss:2020,Bi:2020}.

\subsection{Disk Warping, Breaking and Alignment}
\label{sec:warp}

\def\hatl{\hat\boldmath{l}}

Consider a CBD surrounding a circular binary (with total mass $m_{\rm b}$, reduced mass $\mu_{\rm b}$, 
semi-major axis $a_{\rm b}$ and eccentricity $e_{\rm b}=0$). 
The orientation of the disk at radius $r$ (from the center of mass of the binary) is specified by the unit normal vector $\bl(r,t)$. 
Averaging over the binary orbital period and
the disk azimuthal direction, the binary imposes a torque per
unit mass on the disk element at radius $r$ given, to the leading order in $a_{\rm b}/r$, by
\begin{equation}
\boldsymbol{T}_{\rm b}=-r^2\Omega\, \omega_{\rm prec} (\blb\cdot\bl)(\blb\times\bl), 
\end{equation}
where $\Omega=(GM_{\rm b}/r^3)^{1/2}$ is the orbital frequency at radius $r$,
$\blb$ is the unit vector along the binary angular momentum axis, and 
\begin{equation}
\omega_{\rm prec}={3\mu_{\rm b}\over 4M_{\rm b}}\left({a_{\rm b}\over r}\right)^2 \Omega.
\end{equation}
characterizes the (nodal) precession rate of the disk mass element around the binary axis.
Since $\omega_{\rm prec}$ depends strongly on $r$, the disk would lose coherence if different parts of the disk
did not ``communicate'' with each other.  In reality, the combination of differential precession and 
internal fluid stresses can give rise to a coherently warped/twisted disk.

Theoretical studies of warped disks \citep{Papaloizou:1983,Papaloizou:1995}
have shown that there are
two dynamical regimes for the linear propagation of warps
in an accretion disk. For high-viscosity disks with
$\alpha\gtrsim h\equiv H/r$ (where $H$ is the half-thickness of the disk and
$\alpha$ is the Shakura–Sunyaev parameter such that the viscosity is $\nu =\alpha H^2\Omega$), 
the warp is communicated via angular momentum advection by the oscillatory internal flow whose amplitude is determined by the viscosity, and 
satisfies a diffusion-type equation. The 
corresponding diffusion coefficient $\nu_2$ measures the $r$-$z$ viscous stress and can differ from the usual 
viscosity $\nu$ (which measures the $r$-$\phi$ stress)\footnote{For Keplerian disks, resonance
between the epicyclic frequency and orbital frequency leads to $\nu_2/\nu=1/(2\alpha^2)\gg 1$;
However, the resonance can be easily ```detuned'' by small non-Keplerian effects such as general relativity, quadrupole potential from the binary \citep[see][]{Ogilvie:1999}.}.
\citet{Ogilvie:1999} has developed a fully nonlinear theory of diffusive warps that is in agreement with 3D numerical simulations \citep{Lodato:2010}. In such high-viscosity regime, we expect that an inclined CBD at large radii
transitions to alignment with the binary at small radii, with the characteristic transition radius (``warp radius'') $r_{\rm warp}$ 
determined by $\omega_{\rm prec}(r_{\rm warp})\simeq \nu_2/r_{\rm warp}^2$, giving 
\begin{equation}
{r_{\rm warp}\over a_{\rm b}}\simeq 14\left({4\mu_{\rm b}\over M_{\rm b}}{\nu\over\nu_2}{0.1\over\alpha}\right)^{1/2}{0.1\over h}.
\end{equation}
(This warp transition behavior is similar to the Bardeen-Petterson effect of accretion disks around spinning black holes; e.g., \citealp{Bardeen:1975,Kumar:1985,Scheuer:1996}).

Depending on the initial condition, a misaligned viscous CBD may
be susceptible to tearing (i.e. breaking up into two or more 
disconnected ``rings'') if the steady warped state cannot be attained. Such disk tearing was observed in the SPH simulations of \citet{Nixon:2013} when the CBD is initiated with a sufficiently large inclination angle $\theta$ with respect to the binary \citep[see also][]{Nealon:2020}. 
The disk breaking radius can be estimated by comparing the viscous warp torque (per unit mass)
$(\nu_2/r^2)r^2\Omega$ to the precessional torque from the binary, $|T_{\rm b}|\sim \omega_{\rm prec} r^2\Omega|\sin 2\theta|$, giving
\begin{equation}
r_{\rm break}\simeq r_{\rm warp}|\sin 2\theta|^{1/2}.
\end{equation}
Disk breaking requires sufficiently large $\theta$, for which $r_{\rm break}\sim r_{\rm warp}$. If the initial $\theta$ is small such that $r_{\rm break}$ lies
inside the inner radius of the CBD, disk breaking would not occur and we expect the disk to evolve into the steady warped state,
with a smooth transition between alignment (with the binary) at small radii to misalignment at large radii.

For low-viscosity disks with $\alpha\lesssim h$, a low-amplitude warp propagates as a
bending wave at about half the sound speed, $c_s/2$, provided that the disk is sufficiently Keplerian (i.e. the apsidal precession and nodal precession rates are less than $h\Omega$)
(\citealp{Lubow:2000}; see also \citealp{Lubow:2002,Ogilvie:2006}).
 Protoplanetary disks around young binary stars (with $\alpha\sim 10^{-4}$-$10^{-2}$, $h\gtrsim 0.05$ and $r/a_{\rm b}\gtrsim$ a few) generally satisfy these conditions \citep{Foucart:2013}.
The nonlinear behavior of low-viscosity warped disks is complicated and poorly understood owing to 
the resonant excitation of vertical "breathing" 
motions \citep{Fairbairn:2022} and a parametric instability associated with the inertial waves \citep{Gammie:2000,Ogilvie:2013a,Paardekooper:2019,Deng:2021,Deng:2022}.

Because of the efficient communication of warps by the bending waves,
an inclined protostellar disk around a binary is approximately flat 
-- this is the case provided that the bending wave crossing time $2r/c_s$
is shorter than the characteristic precession time $\omega_{\rm prec}^{-1}$, 
a condition that is well satisfied everywhere in the disk except the inner-most region \citep{Zanazzi:2018b}.
The interplay between the small (non-zero) disk twist/warp and viscous 
dissipation drives the long-term evolution of the disk inclination.
\citet{Foucart:2013,Foucart:2014} studied the warp and the dissipative torque that drives the inclination evolution of a CBD around a circular binary. \citet{Foucart:2013} considered an infinite disk and included the effect of accretion on to the binary, while \citet{Foucart:2014} considered a more realistic disk
of finite size and angular momentum, which can precess coherently
around the binary. They showed that under typical protoplanetary
conditions, both viscous torque associated with disk warp/twist and
accretion torque tend to damp the mutual disk–binary inclination
on time-scale much shorter than the disk lifetime (a few Myr). In
contrast, a circumstellar disk inside a binary can maintain large
misalignment with respect to the binary orbital plane over its entire lifetime 
\citep{Lubow:2000,Foucart:2014}.
Qualitatively, the key difference between CBDs and circumstellar disks is that in former/latter, the binary torque is exerted at the inner/outer region of the disk, which contains small/larger amount of angular momentum, leading to relatively
large/small warp and thus faster/slower viscous damping.
Overall, these results are consistent with the observations that most CBDs
are nearly coplanar with their host binaries \citep{Czekala:2019}, while circumstellar disks within young stellar binaries are often misaligned \citep[e.g.,][]{Jensen:2014,Ichikawa:2021}.

\subsection{Polar Alignment of Disks Around Eccentric Binaries}
\label{sec:polar}
Although disks around circular binaries tend to evolve toward alignment, 
recent works suggested that other outcomes may be possible for
disks around eccentric binaries. \citet{Aly:2015} carried out SPH simulations 
of disks around eccentric MBHBs  (which typically lie in the ``viscous'' regime of disk warps, with $\alpha\gtrsim h$), and showed that 
the disk may be driven into polar alignment (i.e., the disk plane is perpendicular to the binary plane). \citet{Martin:2017} found numerically that a circumbinary protoplanetary disk (typically in the bending-wave regime, with $\alpha \lesssim 
h$) inclined to an eccentric ($e_{\rm b} = 0.5$) binary by $60^\circ$ will evolve to a
polar configuration. 
This dynamical outcome arises from the combined influences of the gravitational torque on
the disk from the eccentric binary and viscous torque from disk warping.

To understand the possibility of polar alignment, it is useful to consider the secular dynamics of a circular test particle around an eccentric binary
\citep[e.g.,][]{Farago:2010,Li:2014}.
\begin{equation}
\bTb = -r^2\Omega\, \omega_{\rm prec} \left[ (1-\eb^2)(\bl \cdot \blb) \blb \times \bl - 5 (\bl \cdot \beb) \beb 
\times \bl \right],
\label{eq:bTb}
\end{equation}
where $\beb$ is the binary eccentricity vector.  In the absence of hydrodynamical 
forces, the time evolution of the test particle's orbital angular momentum vector $\bl$ is governed by 
${\der \bl}/{\der t} = \bTb/(r^2\Omega)$. It is clear that the evolution of $\bl$ has four possible fixed points 
(where $\der\bl/\der t=0$): $\bl=\pm\blb$ and $\bl=\pm\eb/e_{\rm b}$. To examine the stability of the fixed points, it
is useful to analyse the trajectory of $\bl(t)$ using the 
the ``energy'' curves. The equation of motion for $\bl(t)$ admits an integral of motion
\begin{equation}
\Lambda = (1-\eb^2)(\bl \cdot \blb)^2 - 5 (\bl \cdot \beb)^2,
\label{eq:Lam}
\end{equation}
which is simply related to the quadrupole interaction energy (double-averaged over the binary and test-particle orbits) by 
\citep[e.g.,][]{Tremaine:2009,Liu:2015}
\begin{equation}
\Phi_{\rm quad} = \frac{G \mu_{\rm b} a_{\rm b}^2}{8r^3}(1-6 \eb^2 -3\Lambda).
\label{eq:Phi_quad}
\end{equation}

\begin{figure*}
\centering
\includegraphics[width=0.95\textwidth]{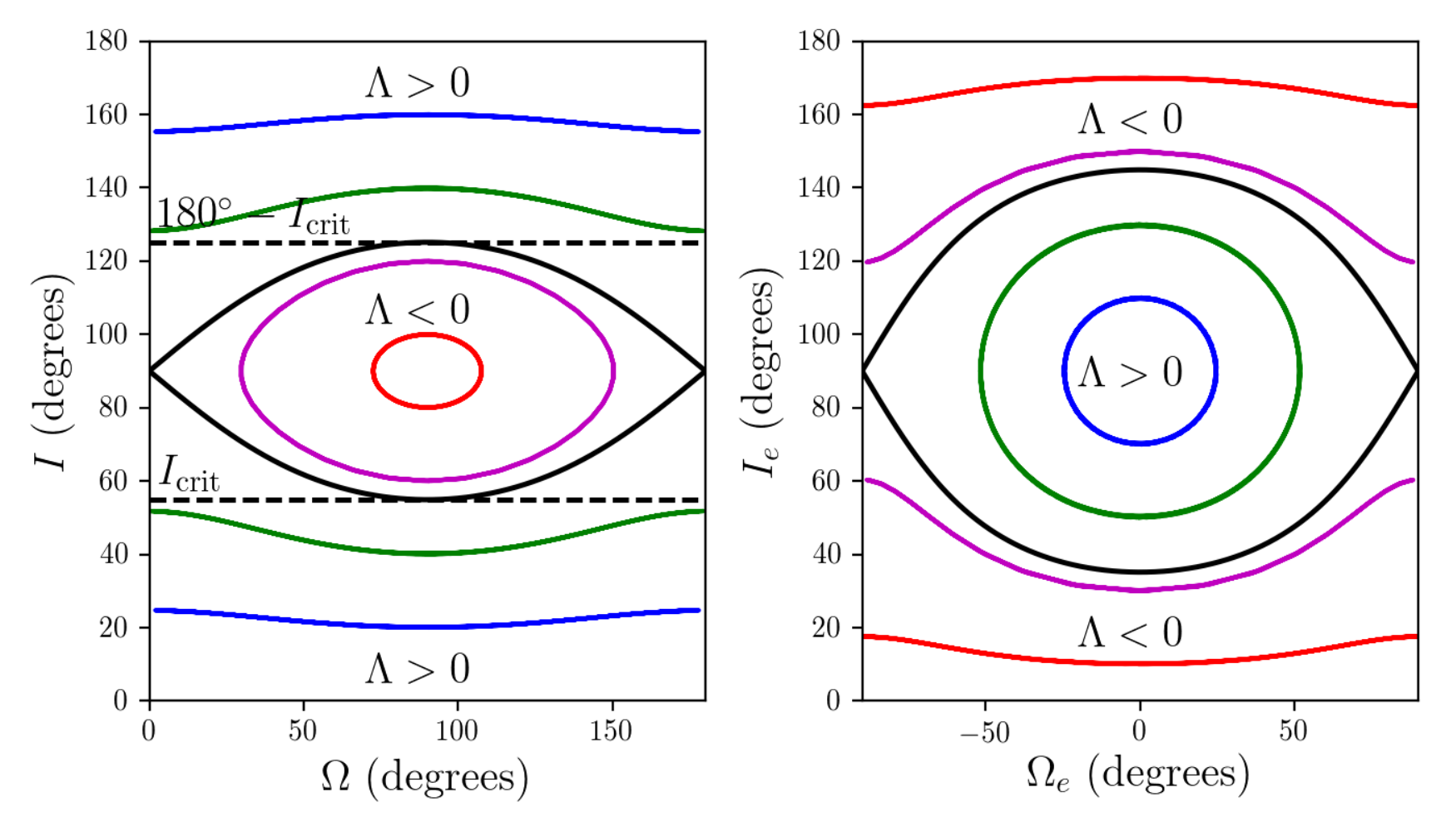}
\caption{ Test particle trajectories around an eccentric binary in the $I-\Omega$ and $I_e-\Omega_e$
  planes, with the binary eccentricity $e_{\rm b} =0.3$.  The inclination angles $I$ (between $\bl$ and $\blb$)
  and $I_e$ (between $\bl$ and $\beb$), and the nodal angles $\Omega,\,\Omega_e$ (not to be confused with the
  angular frequency) are defined by 
  $\bl= (\sin I \sin \Omega, -\sin I \cos \Omega, \cos I)= (\cos I_e, \sin I_e \sin \Omega_e, \sin I_e 
  \cos \Omega_e)$ in the Cartesian coordinate system where $\blb = \bm{\hat z}$ and $\beb = e_{\rm b} \bm{\hat x}$. 
  When $\Lambda > 0$, $\bl$ precesses around $\blb$, with $I \sim
  \text{constant}$ and $\Omega$ circulating its full range of values                                        
  ($0^\circ$-$360^\circ$).  When $\Lambda < 0$, $\bl$ precesses around $\beb$, with $I_e \sim
  \text{constant}$ and $\Omega_e$ circulating its full range of values                                      
  ($-180^\circ$-$180^\circ$). The black lines denote the $\Lambda = 0$                                        
  separatrix.  The other curves have $\Lambda = 0.751$ (blue), $\Lambda =
  0.348$ (green), $\Lambda = -0.110$ (magenta), $\Lambda = -0.409$ (red).
  Only $\Omega$ and $\Omega_e$ in the range $[0^\circ,180^\circ]$ and
  $[-90^\circ,90^\circ]$ are shown. From \citet{Zanazzi:2018a}.
\label{fig:cons}}
\end{figure*}

Figure~\ref{fig:cons} shows the test particle trajectories (constant$-\Lambda$ curves) 
in the $I-\Omega$ (left panel) and $I_e-\Omega_e$ (right panel) planes for 
$e_{\rm b} = 0.3$.  The critical separatrix $\Lambda = 0$ is displayed in black in both plots.  When $\Lambda > 0$,
$\bl$ precesses around $\blb$ with $I \sim \text{constant}$ and
$\Omega$ circulating the full range ($0-360^\circ$). When $\Lambda <0$, $\bl$ precesses around
$\beb$ with $I_e \sim \text{constant}$ and $\Omega_e$ circulating the
full range ($0-360^\circ$). Thus, the test particle angular momentum axis $\bl$ transitions from
precession around $\blb$ for $\Lambda > 0$ to precession around $\beb$
for $\Lambda < 0$.  A necessary condition for $\bl$ to precess around $\beb$ is 
$I_{\rm crit} < I < 180^\circ - I_{\rm crit}$, where 
\begin{equation}
I_{\rm crit} = \cos^{-1} \sqrt{ \frac{5 \eb^2}{1 + 4 \eb^2} }
= \tan^{-1} \sqrt{ \frac{1-\eb^2}{5\eb^2} }.
\label{eq:Icrit}
\end{equation}

\citet{Zanazzi:2018a} carried out a detailed theoretical analysis 
of the dynamics of inclined, warped disks around eccentric binaries 
and their long-term evolution driven by viscous torques 
\citep[see also][]{Lubow:2018}. For disks with 
$H/r\gtrsim \alpha$ (as appropriate for protoplanetary disks), 
bending wave propagation effectively couples different regions of the disk, making it precess as a quasi-rigid body. Zanazzi \& Lai showed 
explicitly that the dissipative torque (associated with disk warp/twist)
tends to drive the disk to one of two states, depending on the initial sign of $\Lambda$:  For $\Lambda>0$, the disk angular momentum axis
$\bl_d$ aligns (or anti-aligns) with the binary orbital angular momentum vector $\blb$; for $\Lambda<0$, $\bl_d$ aligns with the binary 
eccentricity vector (polar alignment).  
They also showed that when the disk has a non-negligible angular momentum compared to the binary, the system’s fixed points are modified and the disk may then 
evolve to a state of near polar alignment, with the inclination somewhat less than $90^\circ$ \citep[see also][]{Martin:2019}.
Note that $\Lambda$ depends on both $I$ (the disk-binary inclination) and $\Omega$ (the longitude of ascending node of the disk).  Thus
for a given $e_{\rm b}$, the direction of inclination evolution depends not only on the initial $I(0)$, but also on the initial $\Omega(0)$.
The time-scale of evolution of the disk–binary inclination
angle is approximately given by (assuming the disk surface density profile $\Sigma\propto 1/r$)
\begin{equation}
\tau_{\rm b} \sim  10^4 \left( \frac{0.01}{\alpha} \right) 
\left( \frac{h}{0.1} \right)^2 \left( \frac{r_{\rm in}}{2a_{\rm b}} \right)^4
\left( \frac{M_{\rm b}}{4\mu_{\rm b}}\right)^2 \left(\frac{2M_\odot}{M_{\rm b}}
\right)^{1/2} \left( \frac{r_{\rm out}}{100 \,\text{AU}}
\right)^{3/2} \text{yr}
\label{eq:cgb_scale}
\end{equation}
where $r_{\rm in}, r_{\rm out}$ are the inner and outer radii of the disk. Thus 
$\tau_{\rm b}$ is generally less than a few Myr, the lifetime of proto-planetary disks. This suggests that highly inclined disks may exist around eccentric binaries.

This theoretical expectation was recently confirmed by the observation of the young ($\sim 10$~Myr)
protoplanetary system HD 98800. Using ALMA observations of dust and CO emissions,
\citet{Kennedy:2019}
showed that the inner binary BaBb (with $a_{\rm b}\simeq 1$~AU, $e_{\rm b}=0.785$) 
in the system (which is a ``2+2'' hierarchical quadruple system with two inner binaries orbiting each 
other with a semi-major axis of 54~AU) is surrounded by a CBD in a polar-aligned configuration. The old circumbinary debris disk nearly perpendicular to the orbital plane of 
the binary 99 Herculis \citep{Kennedy:2012a} may also have gone through such polar-alignment
process early in its lifetime when the gas was present \citep[e.g.,][]{Smallwood:2020}.

\begin{figure*}
\centering
\includegraphics[width=0.95\textwidth]{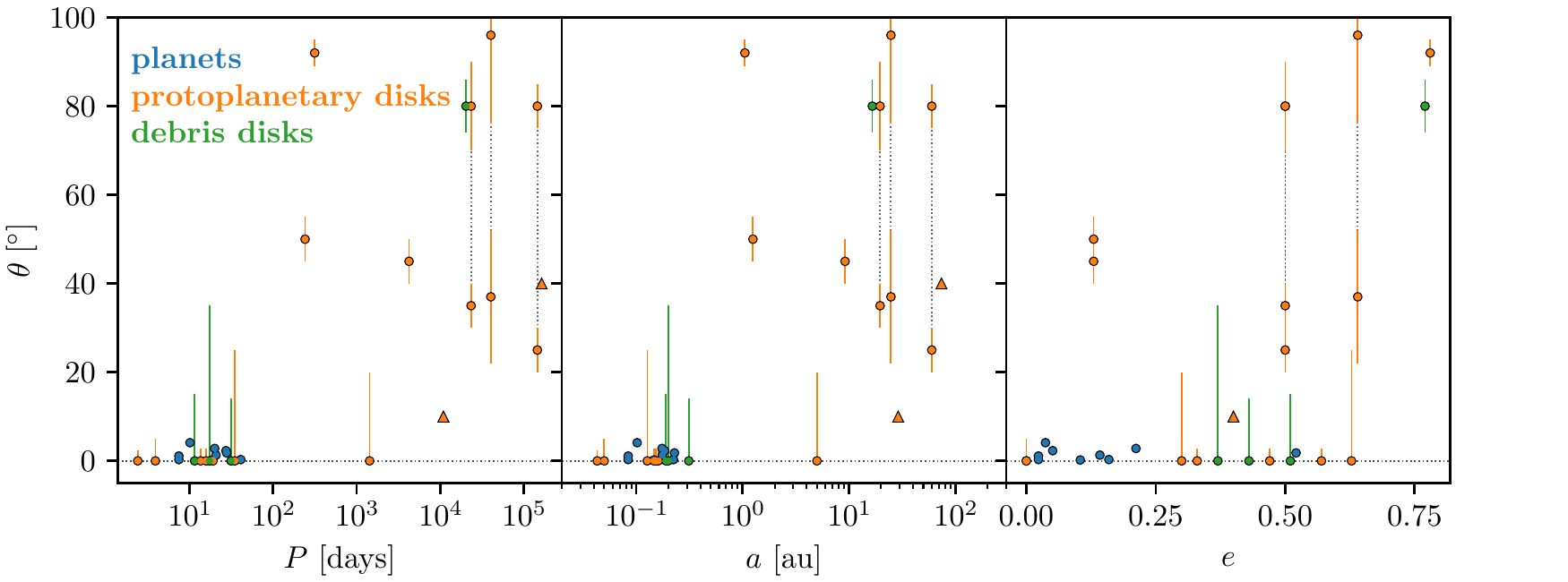}
\caption{\emph{Left}: The mutual inclinations of {\it Kepler} circumbinary planets and all circumbinary
 protoplanetary and debris disks, as a function of binary orbital period. The triangles represent the lower 
 limits on $\theta$ for R~CrA and IRS 43. 
 \emph{Center}: Mutual inclination as a function of semi-major axis. \emph{Right}: Mutual inclination as a function 
of binary eccentricity. Note that the binary eccentricity is unknown for IRS\,43, and so it is not plotted in the eccentricity panel). The two points at $e=0.13$ correspond to GW~Ori A-B and AB-C, which collectively host a 
circumternary protoplanetary disk. Long-period, eccentric binaries are more likely to host CBDs with significant mutual inclinations. From \citet{Czekala:2019}.}
\label{fig:cezkala}
\end{figure*}

Figure \ref{fig:cezkala} \citep[from][]{Czekala:2019} shows the 
inclinations of {\it Kepler} circumbinary planets and 
circumbinary protoplanetary and debris disks, as a function of binary orbital period, semi-major axis and
eccentricity.   A clear trend emerges: All CBDs orbiting binaries
with period less than 30 days (semi-major axis less than 0.4~AU) 
and/or eccentricity less than 0.2 are consistent with being
coplanar, while disks orbiting longer period and/or more eccentric binaries exhibit a wide range of mutual inclinations, from coplanar to polar. The origin of this ``critical'' orbital period is unclear.
But the trend is consistent with the general idea that stellar binaries form/fragment at large separations and migrate inwards to small separations in massive disks/envelopes, during which the binary and disk become aligned (see Section~\ref{sec:stellar_binaries}).

The near coplarity of CBDs around 
short-period binaries implies that planets formed in such disks should be similarly coplanar. This is consistent with the finding that {\it Kepler} circumbinary planets, orbiting binaries with $P_{\rm b} <40$~days, have small mutual
inclinations, and indicates that the planet occurrence
rate around such binaries is similar to that for single stars
\citep[see also][]{Armstrong:2014,Li:2016,Martin_D:2019}.
Beyond $P_{\rm b} > 40$~days, however, the existence misaligned CBDs suggests 
that circumbinary planets around eccentric binaries 
may be found to have a broad distribution of mutual
inclinations, with a possible concentration of polar-aligned
systems.

\section{BINARIES EMBEDDED IN ``BIG'' DISKS}\label{sec:6}

A special type of ``circumbinary accretion'' has gained significant interest 
in recent years.  This concerns stellar binaries embedded in AGN disks around supermassive black holes (SMBHs).

The detection of gravitational waves from the merging binary black holes (BHs) by the LIGO/Virgo collaboration
\citep{LIGO:2021}
has motivated many theoretical studies of the formation channels of the BH binaries.  In addition to the isolated binary evolution channel 
\citep[e.g.,][]{Lipunov:1997,Belczynski:2016},
there are several flavors of dynamical channels, including strong gravitational scatterings in dense star clusters 
\citep[e.g.,][]{Portegies:2000,Kremer:2019},
more gentle ``tertiary-induced mergers'' (often via Lidov-Kozai mechanism) that take place either in stellar triple/quadrupole systems
\citep[e.g.,][]{Silsbee:2017,Liu:2018,Liu:2019a}
or in nuclear clusters dominated by a central SMBH 
\citep[e.g.][]{Antonini:2012,Liu:2019b,Liu:2021},
and (hydro)dynamical interactions of binaries in AGN disks 
\citep[e.g.,][]{Bartos:2017,Stone:2017,McKernan:2018,Tagawa:2020}.
It has also been suggested that intermediate-mass BHs (IMBHs) may be formed efficiently in AGN disks
via accretions or mergers of stars and compact objects
\citep[e.g.,][]{McKernan:2012,McKernan:2014}. If these IMBHs are themselves members of binaries,
they could be susceptible to ``evection resonances'' driven by the SMBH's tidal field \citep{Munoz:2022},
which could then accelerate their GW-driven inspiral and produce an eccentric waveform \citep[see also][]{Bhaskar:2022}.

In the AGN disk scenario, an important question concerns how single stellar-mass BHs can be captured into bound binaries and merge. \citet{Li_J:2022b}
showed that when the gas effect is negligible, two BHs in tightly-packed orbits around a SMBH become bound to each other in rare, very close encounters due to gravitational wave emission, 
leading to highly eccentric BH mergers in the LIGO band.  
At sufficiently high gas densities, the gas drags on the BHs
can facilitate the capture process
\citep[e.g.,][]{Tagawa:2020,Li_J:2022a}.

Once a bound BH binary forms in the AGN disk, an important question concerns how
the binary evolves in the presence of the surrounding gas. It is tempting to consider this as a ``circumbinary accretion'' problem \citep[e.g.,][]{Stone:2017}, with the 
background AGN disk feeding gas onto the CBD around the binary. For example, 
one might use the modified Bondi-Hoyle-Lyttleton accretion formula 
\citep[e.g.,][]{Edgar:2004}
to estimate the mass supply rate from the background AGN disk (with gas density 
$\rho_{\rm bg}$ and sound speed $c_{\rm s,bg}$) onto the binary,
$\dot m_{\rm b}\sim \pi \rho_{\rm bg} v_{\rm eff} r_{\rm acc}{\rm min}(r_{\rm acc},H)$, where $H$ is the disk scale-height and $r_{\rm acc}$ is the accretion
radius $r_{\rm acc}\sim Gm_{\rm b}/v_{\rm eff}^2$, with $v_{\rm eff}^2\sim c_{\rm s,bg}^2+v_H^2$
(the second term accounts for the ``Hill velocity'', the velocity shear across the Hill radius).
However, such an estimate could be quite misleading as the strong velocity shear and flow angular momentum can 
significantly reduce the accretion rate compared to the Bondi-Hoyle-Lyttleton estimate \citep{Li_R:2022a}. In fact, even when the binary is replaced by a single object, the accretion rate can be much smaller than the Bondi-Hoyle-Lyttleton estimate because of the strong upstream velocity shear. Numerical simulations
show that the accretion rate generally depends on the physical size of the accretor, indicating that the physics near the accretor can strongly influence the accretion flow \citep[see][]{Xu:2019}.
Thus, it is not clear that the results from circumbinary accretion (Section 3) can be directly adapted and applied to the problem of ``Binaries embedded in big disks''.

The hydrodynamical evolution of binaries in AGN disks has been studied numerically by a handful of works so far. \citet{Baruteau:2011}
carried out global simulations in 2D isothermal disks and found that a massive (gap-opening) prograde, equal-mass binary is hardened by dynamical friction from the lagging spiral tails trailing each binary component inside the Hill radius.  
\citet{Li_YP:2021}
used a similar global setup and found that adequately resolved 
circum-single disk (CSD) regions in fact lead to expanding binaries.
\citet{Li_YP:2022}
further found that the temperature structure of the CSDs plays an important role in the evolution of the binary; when CSDs attain a sufficiently high temperature, binaries contract rather than expand.

Resolving CSDs around each binary component is therefore important, but
is computationally demanding in global disk simulations. A useful approach
is to use a co-rotating local disk (“shearing-box”) model, where the
global cylindrical geometry of the disk is mapped onto local Cartesian
coordinates centered at the binary's center of mass which rotates around the 
SMBH. This is the approach adopted by \citet{Li_R:2022a,Li_R:2022b}
and \citet{Dempsey:2022}.
Note that although there are multiple length scales and velocity scales associated with
the problem, only a few dimensionless ratios are important. For example, the relevant 
length scales are:
\begin{itemize}
\item[{\scriptsize$\blacksquare$}]  Binary semi-major axis $a_{\rm b}$;
\item[{\scriptsize$\blacksquare$}]  Hill radius $R_{\rm H} \equiv R (M_{\rm b}/M)^{1/3} \equiv R q^{1/3}$, where $M_{\rm b}$ is the total mass of the binary, $M$ is the mass of the SMBH, and $R$ is the orbital radius of the binary around the SMBH
(note that this definition of Hill radius differs from the standard one 
$R_{\rm H}' = R_{\rm H}/3^{1/3}$);
\item[{\scriptsize$\blacksquare$}]  Bondi radius $R_{\rm B} = G M_{\rm b} / c_{\rm s,bg}^2$, where $c_{\rm s,bg}$ is the sound speed of the background gas (far from the binary);
\item[{\scriptsize$\blacksquare$}]  Scale height of the background disk $H = c_{\rm s,bg} / \Omega_{\rm K}$, with $\Omega_K=(GM/R^3)^{1/2}$.
\end{itemize}
However, their ratios depend on only two dimensionless parameters:
\begin{align}
\label{eq:R_H_over_H}
  \frac{R_{\rm H}}{H} &= \left( \frac{R_{\rm B}}{H} \right)^{1/3} = \left( \frac{q}{h^3} \right)^{1/3},  \\
  \label{eq:R_H_over_a}
  \frac{R_{\rm H}}{a_{\rm b}} &= \lambda.
\end{align}
Similarly, the relevant velocity scales are $c_{\rm s,bg}$, $v_{\rm b}=(GM_{\rm b}/a_{\rm b})^{1/2}$, the velocity shear across the binary $V_{\rm s} = (3/2)\Omega_{\rm K} a_{\rm b}$, and $|\Delta V_{\rm K}|\sim h^2V_K$ (the deviation of the background gas velocity around the SMBH from the Keplerian velocity). The first three are related by the same dimensionless parameters:
\begin{align}
  \frac{v_{\rm b}}{c_{\rm s,\infty}} &= \lambda^{1/2} \left(\frac{q}{h^3} \right)^{1/3}, \\
  \frac{V_{\rm s}}{c_{\rm s,\infty}} &= \frac{3}{2\lambda} \left(\frac{q}{h^3} \right)^{1/3}.
\end{align}
For thin discs ($h \ll 1$), $|\Delta V_{\rm K}|$ is very subsonic ($|\Delta V_{\rm K}|/c_{\rm s,bg} \sim h \ll 1$) and is typically much smaller than $V_{\rm s}$.
Therefore, the flow dynamics and the binary orbital evolution,
when appropriately scaled, depend on various physical quantities only 
through two dimensionless parameters: $q/h^3=(R_{\rm H}/H)^3$ and $\lambda$ (in addition to 
other obvious parameters such as the binary eccentricity and mass ratio). Note that the parameter $q/(3h^3)$ is the ratio of the binary mass $M_{\rm b}$ to the so-called thermal mass, $3h^3 M$.  When $R_{\rm H}/H\gtrsim 1$ (which also implies $R_{\rm B}\gtrsim R_{\rm H}$), the flow onto the Hill sphere is quasi-2D, we expect a CSD to form around each binary component; 2D shearing-sheet simulations are appropriate in this regime.
When $R_{\rm H}/H\lesssim 1$, vertical flows onto the Hill sphere are important, 3D simulations are needed to accurately capture the flow structure \citep{Dempsey:2022}. Current models of AGN disks \citep{Sirko:2003,Thompson:2005} contain regions with a wide range of $R_{\rm H}/H$ ratio (for typical $M_{\rm b}/M$), from $\lesssim 1$ to $\gg 1$ (see Fig.~1 of \citet{Dempsey:2022}).  The dynamical stability of the binary requires $\lambda\gtrsim 2$; even for $\lambda\gg 1$, the flow structure still depends on $R_{\rm H}/H$ and can be quite different from that around isolated binaries.

\begin{figure*}
\centering
\includegraphics[width=\textwidth]{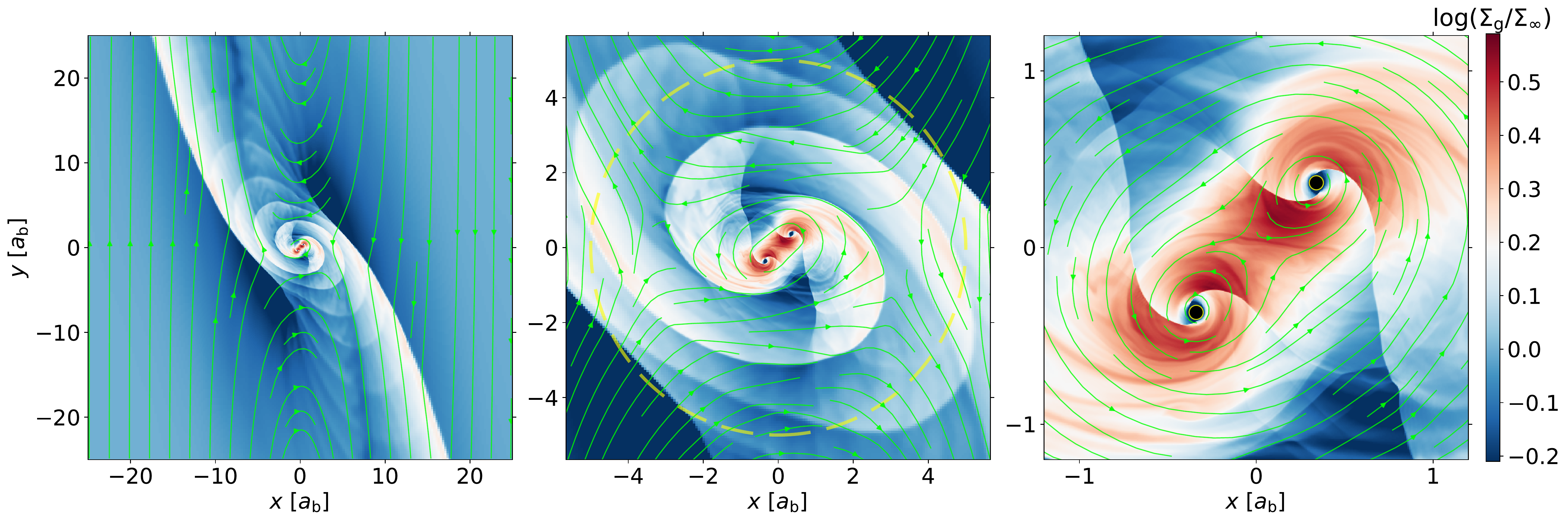}
\caption{A snapshot of the flow property around an equal-mass, circular binary embedded in a ``big'' (AGN) disk, where the mesh is refined progressively towards the binary (zooming in from left to right). The parameters are $\lambda=R_H/a_{\rm b}=5$, $q/h^3=1$ and $\gamma=1.6$. The shearing-box frame is centered at the center of mass of the binary, and the SMBH is to the left.
The background background flow (far from the binary) in the shearing box frame is given by
${\bm V}_{\rm w} = {\bm V}_{\rm sh} + \Delta{\bm V}_{\rm K}= -(3/2)\Omega_{\rm K} x \bm{\hat{y}} +\Delta V_{\rm K}\bm{\hat{y}}$,
where $\bm{V}_{\rm sh}(x)$ denotes the Keplerian shear, $\bm{\Delta V}_{\rm K}$ is the deviation from Keplerian velocity, with 
$\Delta V_{\rm K}\sim -h^2V_{\rm K}$.
The color specifies the gas surface density (with $\Sigma_\infty=\Sigma_{\rm bg}$ the background disk density far from the binary). The green streamlines show the detailed flow structure, and the yellow dashed circles in the middle panel with a radius of $R_H$ denotes the Hill radius of the binary. 
Adapted from the left column of Fig.~3 in \citet{Li_R:2022a}  \copyright AAS. Reproduced with permission.}
\label{fig:rixin}
\end{figure*}

Figure \ref{fig:rixin} illustrates the flow structure of an equal-mass circular binary
embedded in the AGN disk, from the 2D shearing-box simulation of
\citet{Li_R:2022a} with $\lambda=5$, $q/h^3=1$ and the gas satisfying 
the $\gamma$-law equation of state ($\gamma=1.6$).
Grand spirals originated from the circumbinary flows extend
to large distances along the shear flow directions; they
can be considered as half bow shocks of the binary accretion with upstream
flow gradients due to the shear. Some of these features are also found in the global simulations of \citet{Baruteau:2011} and \citet{Li_YP:2021}.
There are also horseshoe flows and the inner/outer shear flows around the binary. Such flow structures are similar to those observed in previous studies
of a single accretor 
\citep[e.g.,][]{Fung:2015,Zhu:2016,Kuwahara:2019,Bailey:2021}.
Due to the fast orbital velocity $v_{\rm b}$, the flow close to the binary is much more dynamic. Both CSDs contain two spiral shocks that drive accretion throughout the disk. Each CSD is then encompassed and attached by a small 
half bow shock, the tail of which is slingshot away along each grand spiral once a binary orbit; the propagation of such waves is visible in 
Fig.~\ref{fig:rixin}. 
More examples of the flow properties (which depend
on $\lambda,\,q/h^3,\,\gamma,\, M_1/M_2$ and $e_{\rm b}$) can be found
in \citet{Li_R:2022a,Li_R:2022b} and \citet{Dempsey:2022}
(the latter considered 3D simulations with isothermal equation of state). Overall, the flow structure is qualitatively different from that of isolated ``circumbinary accretion'' discussed in Section 3.

\citet{Li_R:2022a,Li_R:2022b}
have used a suite of high-resolution 2D shearing-box simulations of binaries embedded in AGN disks to determine the accretion dynamics and secular evolution of the binary, covering a range of values for the parameters $\lambda,\,q/h^3,\,\gamma,\, 
M_1/M_2$ and $e_{\rm b}$. 
The flows close to the binary (including the CSDs) are
generally found to be more massive with decreasing $\gamma$ and with increasing $q/h^3$ and $\lambda$; they are also hotter and more turbulent with increasing $\gamma$, $q/h^3$ and $\lambda$.
The time-averaged accretion rate $\langle \dot{M}_{\rm b} \rangle$ (in units of $\Sigma_{\rm bg} a_{\rm b} v_{\rm b}$) 
depends on the physical size of the accretor (see above), and
monotonically decreases with $\gamma$ and increases with $q/h^3$.
In general, circular comparable-mass binaries contract if the EOS is far from isothermal, with an orbital decay rate of a few times the mass doubling rate. When the EOS is close to isothermal ($\gamma=1$), the binary orbit expands
[however, retrograde binaries always experience orbital decay; see also \cite{Li_YP:2021}].
Eccentric binaries tend to experience eccentricity damping. Prograde binaries with higher eccentricities or smaller mass ratios tend to have slower orbital decay rates, with some extreme cases exhibiting orbital expansion. Note that some of the quantitative results may be modified by 3D effects when $q/h^3\lesssim$ a few \citep{Dempsey:2022}.
The accretion flows are highly variable, and the dominant variability frequency is the apparent binary orbital frequency (in the rotating frame around the central massive BH) for circular
binaries but gradually shifts to the radial epicyclic frequency as the binary eccentricity increases. These calculations also  
suggest that the hardening timescales of the binaries are much shorter than their migration timescales in the AGN disk, for all reasonable binary and disk parameters.
Overall, these studies show that the dynamics of binaries embedded in AGN disks is quite different from that of isolated binaries in their own CBDs. Obviously, current simulations are still idealized, and future works will be needed to assess the effects of additional physics (e.g. magnetic fields, radiation and accretion feedbacks) on the evolution of BH binaries in AGN disks.

\section{SUMMARY AND FUTURE PROSPECTS}\label{sec:7}

Accretion disks have long played a central role in many areas of astronomy, from 
bright X-ray sources associated with accreting compact objects in the Galaxy, 
to outflows and jets associated with active galactic nuclei, to various complex processes and phenomena associated with star and planet formation. When the central accreting object is replaced by a binary, a new set of dynamical behaviors become important, including Lindblad torques and cavity opening, accretion variabilities on different timescales,
eccentricity driving and precession of the inner disk, preferential accretion and long-term binary evolution, and, in the case of misaligned accretion, disk warping and breaking, and secular evolution towards alignment or polar alignment, etc.  All these dynamical behaviors have direct or indirect observational manifestations.

Since circumbinary accretion involves intrinsically 
2D or 3D phenomena, numerical simulations are crucial to unravel its dynamical behaviors and obtain  quantitative answers to some of the key questions,
with semi-analytic theory providing complementary insights. These simulations are challenging (compared to ``normal'' accretion disk simulations) because of
the wide range of spatial 
(from $\gg a_{\rm b}$ for the circumbinary region to $\ll a_{\rm b}$ for the circum-single disks) 
and time scales (from $\ll P_{\rm b}$ to $\gg P_{\rm b}$)
involved. For example, to determine the secular effect of accretion on the binary evolution, sufficiently long simulations must be carried out in order to average out the highly dynamical flow behaviors.
As a result, systematic numerical simulations have so far focused on idealized 2D setups, with simple equation of state (usually locally isothermal) and viscosity prescription.
Some of key findings include:
\begin{itemize}
\item[{\scriptsize$\blacksquare$}]  Circumbinary accretion is highly dynamical, and the accretion variability is dominated by the period of $P_{\rm b}$
or $\sim 5P_{\rm b}$, depending the binary eccentricity and mass ratio.
\item[{\scriptsize$\blacksquare$}]  The inner region ($\lesssim 10 a_{\rm b}$) of the circumbinary disk can develop coherent eccentric structure, which may modulate the accretion and affect the physical processes (such as planet migration) taking place in the disk.
\item[{\scriptsize$\blacksquare$}]  Over long timescales, a binary undergoing accretion evolves towards equal masses between the components. 
\item[{\scriptsize$\blacksquare$}]  While the gravitational torque between the binary and circumbinary disk tends to drive binary orbital decay, once
the inner disk region is sufficiently relaxed and accretion sets in, the net angular momentum transfer between the binary and the disk depends on the competition between the accretion and gravitational torques.  Circumbinary accretion does not necessarily lead to binary orbit decay, as commonly assumed; the secular orbital evolution depends on the binary parameters (such as the mass ratio and eccentricity) and the thermodynamic properties of the accreting gas.
In general, the binary orbital decay/expansion rate is of order the mass doubling rate, i.e.
$\dot a_{\rm b}/a_{\rm b}\sim \pm \dot M_{\rm b}/M_{\rm b}$.
\end{itemize}
Much work remains to go beyond these ``idealized'' simulations in order to systematically evaluate the roles of gas thermodynamics and magnetic fields (which can generate turbulent viscosity and outflows).
Note that the above results apply to the regime where the local disk mass near the binary
is much less than $M_{\rm b}$. 
When this condition is not satisfied, the flow and binary dynamics can be quite different. For example, in the early stages of young binary stars and massive black-hole binaries, 
the local disk/envelope surrounding the binary
can be much more massive than the binary, and the disk self-gravity is important -- such massive disks can drive rapid binary orbital decay. Eventually, the local disk "thins out", and we are back to the "proper" circumbinary accretion regime that determines the late-stage binary evolution.

On the observational front, the study of accreting proto-stellar binaries and the detailed characterization of ``mature'' binary stars (e.g. using GAIA) can shed light on the circumbinary accretion process in the aftermath of star formation. Similarly, the observation of accreting massive black-hole binaries (e.g. through variable lightcurves) and the future detection of low-frequency 
gravitational waves from such binaries (by the Pulsar Timing Arrays and by the space interferometers such as LISA) will help constrain the assembly history of massive black holes and the role of circumbinary accretion.

Circumbinary disk can often be misaligned with the orbital plane of the central binary, as expected in realistic scenarios of star formation and massive black hole evolution. Such misaligned accretion gives rise several new dynamical features:
\begin{itemize}
\item[{\scriptsize$\blacksquare$}]  A misaligned disk is warped, and the ``degree'' of the warp depends on the internal hydrodynamical stresses of the disk (viscosity and bending waves). Under some conditions, a highly warped disk may break up into two or more separate ``rings''.
\item[{\scriptsize$\blacksquare$}]  Over long (viscous) timescales, a misaligned disk around a low-eccentricity binary tends to evolve toward alignment driven by viscous dissipation. When the binary eccentricity is significant, the circumbinary disk can evolve toward ``polar alignment'', with the disk plane perpendicular to the binary plane.
\end{itemize}
Numerical simulations of misaligned disks are challenging because they require capturing the large-scale disk warps while resolving 
small-scale dissipations. Much remains to be done in the future, e.g.
to evaluate the possibility and condition of disk breaking.
Observationally, polar-aligned circumbinary disks have already been detected, and future characterization of misaligned disks would help constrain the binary formation/migration process. Will highly misaligned planets be detected in the future?

Binaries embedded in a ``big'' disk present a special type of 
circumbinary accretion. The mass supply/accretion onto such a binary depends on the intricate coupling between the small-scale
(within the binary) and large-scale flow dynamics. The orbital evolution of the binary depends on several dimensionless parameters
as well as the gas thermodynamics. It has been suggested that 
some merging black-hole binaries detected by LIGO/VIRGO are produced by binaries embedded in AGN disks. Future observations,
including more merger events and related electromagnetic counterparts,
may provide a more definitive answer.

\section*{DISCLOSURE STATEMENT}
The authors are not aware of any affiliations, memberships, funding, or financial holdings that
might be perceived as affecting the objectivity of this review. 

\section*{ACKNOWLEDGMENTS}
The writing of this review has been motivated by several seminars/colloquia that DL gave on this topic in the last few years,
most recently at KITP, UCB, Caltech, UIUC, KIAA and TDLI.
We thanks our collaborators, including
Xuening Bai, Francois Foucart, Lars Hernquist, Katlin Kratter, Jiaru Li, Rixin Li, Yoram Lithwick, Adam Dempsey, Ryan Miranda, Volker Springel, Haiyang Wang and J.J. Zanazzi, for their contributions and insights. We also thank Adam Dempsey, Julian Krolik, Gordon Ogilvie, Eliot Quataert and Noam Soker for useful comments on an early version of this article.
This work has been supported in part by the NSF grant AST-17152
and NASA grant 80NSSC19K0444, and by Cornell University.

\bibliographystyle{ar-style2}

\end{document}